\documentclass[12pt,preprint]{aastex}
\usepackage{epsfig}
\usepackage{natbib}
\usepackage{graphicx}
\usepackage{slashbox}
\usepackage{multirow}
\usepackage{lscape}
\usepackage{mathrsfs,amssymb}
\usepackage{amsmath}
\usepackage{subfigure}
\usepackage{amssymb}
\usepackage{multirow}
\usepackage{tabularx}
\usepackage{rotating}
\usepackage{longtable}
\usepackage{amsmath}
\usepackage{ulem}
\usepackage{lineno}
\usepackage{graphicx}
\usepackage{amssymb}
\usepackage{amsmath}
\usepackage{hyperref}
\usepackage{varwidth}
\defcitealias{Schlegel1998}{SFD}
\hypersetup{
  colorlinks = true,
  urlcolor = blue,
  linkcolor = blue,
  citecolor = blue
}

\newcommand{\Laro}{$L_\textrm{3.3}$}

\newcommand       \km           {\,{\rm km}}

\newcommand       \Mpc         {\,{\rm Mpc}}
\newcommand       \s            {\,{\rm s}}

\newcommand       \yr       {\,{\rm yr}}

\newcommand       \simlt        {\lesssim}
\newcommand       \simgt        {\gtrsim}

\newcommand       \mum          {\,{\rm \mu m}}

\newcommand       \Teff         {T_{\rm eff}}

\newcommand       \Msun         {\,{M_\odot}}
\newcommand       \Lsun         {\,{L_\odot}}
\newcommand       \mJy         {\,{\rm mJy}}

\newcommand       \LIR        {L_{\rm IR}}

\newcommand       \simali       {\sim\,}
\newcommand       \magni        {\,{\rm mag}}

\newcommand       \etaali       {\eta_{\rm ali}}
\newcommand       \alifrac      {\eta_{\rm ali}}

\newcommand       \be           {\begin{equation}}
\newcommand       \ee           {\end{equation}}
\countdef\decade=200
\advance\decade by \year
\countdef\hours=201
\advance\hours by \time
\divide\hours by 60
\countdef\mins=202
\advance\mins by \hours
\multiply\mins by 60
\multiply\hours by 100
\countdef\miltime=203
\advance\miltime by \hours
\advance\miltime by \time
\advance\miltime by -\mins

\shorttitle{Aromatics and Aliphatics in Local Star-Forming Galaxies}
\title{
\vspace*{-2.0em}
{\normalsize\rm {Accepted for publication in \it The Astrophysical Journal}}\\
\vspace*{1.0em}
  Aromatics and Aliphatics in Local Star-Forming Galaxies
  as Probed by AKARI
}
\author{Junhao Peng\altaffilmark{1},
             Xuejuan Yang\altaffilmark{1,2}
             and Aigen Li\altaffilmark{2}
             }
\altaffiltext{1}{Department of Physics,
                  Xiangtan University,
                  411105 Xiangtan, Hunan Province, China;
                       \sf{xjyang@xtu.edu.cn}}
\altaffiltext{2}{Department of Physics and Astronomy,
                        University of Missouri,
                        Columbia, MO 65211, USA;
                        {\sf lia@missouri.edu}}

\begin{document}

\begin{abstract}
Polycyclic aromatic hydrocarbon (PAH) molecules
are abundant and widespread in galaxies and their
infrared (IR) emission traces star formation.
PAH molecules in astronomical environments
often have aliphatic contents as revealed by
the detection of the 3.4$\mum$ aliphatic C--H
stretch, a weak satellite feature accompanying
the 3.3$\mum$ aromatic C--H stretch.
Here, we selected 102 local star-forming galaxies
from the AKARI archive, including 66 galaxies,
each of which hosts an active galactic nucleus (AGN).
We analyzed their AKARI near-IR spectra, which exhibit
pronounced 3.3$\mum$ aromatic and 3.4$\mum$
aliphatic C--H emission.
We also compiled their multi-wavelength photometric
data and performed a decompositional analysis of
their spectral energy distributions (SEDs)
from the ultraviolet (UV) to the far-IR
to derive the star formation rates (SFRs),
stellar masses, metallicities, and luminosity
of the galaxies.
We explored the 3.3$\mum$ PAH emission
luminosity ($L_{3.3}$) as a calibrator of the SFR
and found a close agreement with previous studies.
We also found that $L_{3.3}/L_{\rm IR}$
and $L_{3.4}/L_{\rm IR}$ exhibit a strong
dependence on metallicity, but remain nearly
constant above 12\,+\,log(O/H)\,$\simali$8.5,
where $\LIR$ is the total luminosity emitted by dust,
and $L_{3.4}$ is the luminosity
of the 3.4$\mum$ aliphatic emission.
We derived from $L_{3.4}/L_{3.3}$ the PAH aliphatic
fractions, defined as the fractions of carbon atoms
in aliphatic units, to be in the range of
$\simali$0.38\%--6.8\%,
with a median fraction of $\simali$3.1\%.
The  PAH aliphatic fractions are lower in AGN hosts
and show a weak negative correlation with
the SFR and $\LIR$,
suggesting that UV photons in regions
with AGN or strong star formation activities
may photodissociate the aliphatic structures
associated with PAH molecules.
% Our study examines the behavior of {\ali},
% offering valuable insights into the role of
%aliphatic PAHs in star-forming galaxies.
%
\end{abstract}




\section{Introduction} \label{sec:intro}
Polycyclic aromatic hydrocarbon (PAH) molecules
composed of fused aromatic benzene rings
are ubiquitous in the interstellar medium (ISM)
of the Milky Way and external galaxies,
both near and far \citep{Allamandola1985,Leger1984,Li2020,Tielens2008}.
% and are commonly believed to be the carriers of
%a series of infrared emission bands at 3.3, 6.2,
%7.7, 8.6, and 11.3 {\um} \citep{Allamandola1985, Leger1984}.
These molecules absorb ultraviolet (UV) photons
and re-emit the absorbed energy in the infrared (IR),
mostly through a series of aromatic IR emission bands
at 3.3, 6.2, 7.7, 8.6, and 11.3$\mum$. This process is
especially prominent in regions of active star formation,
making them valuable tracers of star formation activity
\citep[e.g.,][]{Pope2008}.
PAHs play a vital role in various astrophysical
and astrochemical processes. They dominate
the heating of the interstellar gas and the ionization
balance in molecular clouds, and are essential to
the evolution of galaxies \citep{Li2020}.

The aromatic emission bands of PAHs arise from
their vibrational modes. The shortest band
(in wavelength) at 3.3$\mum$ originates from
aromatic C--H stretching and is often accompanied
by a weaker satellite feature at 3.4$\mum$, which is
attributed to aliphatic C--H stretching \citep{Yang2017}.
This implies that astronomical PAHs also contain
an aliphatic component, e.g., aliphatic sidegroups
like methyl (-CH$_{3}$) may be attached as functional
groups to PAHs.
% Although the 3.4$\mum$ feature could alternatively
%originate from anharmonicity \citep{Barker1987}
%or superhydrogenation \citep{Bernstein1996, Yang2020},
%it is generally  considered to be an aliphatic feature.

The 3.3$\mum$ emission traces small, neutral PAHs
\citep{Draine2007,Draine2021} and can be employed
to diagnose active galactic nuclei (AGN) and starburst
activities in ultra-/luminous IR galaxies (U/LIRGs)
for identifying buried AGN \citep{Imanishi2010,Ichikawa2014}.
Based on statistical studies of a large number of
galaxies, the 3.3$\mum$ luminosity ($L_{3.3}$) was
found to closely correlate with the IR luminosity
in the wavelength range
of $\simali$8--1000$\mum$
\citep[e.g.,][]{Yamada2013}.
It has also been established that the 3.3$\mum$ emission
is a valid SFR tracer \citep[e.g.,][]{Lai2020,Lyu2025}.
%
%[***AL: add more references for L3.3 vs LIR, L3.3 vs SFR ***].
%

Due to its adjacency to the 3.3$\mum$ band
and its aliphatic origin, the 3.4$\mum$ emission
is routinely used to probe the aliphatic contents
of PAHs and how they vary with environments
\citep[e.g.,][]{LD2012,Yang2023}.
\citet{Mori2014} investigated Galactic H\,II regions
and discovered that the intensity ratio of
the aliphatic to aromatic bands decreases
with increasing PAH ionization, highlighting
the erosion of aliphatic structures
inside or at the  boundaries of ionized regions.
\citet{Lai2023} conducted a spatially resolved
study of PAH emission in the nearby active galaxy
NGC\,7469 and found that the aliphatic-to-aromatic
intensity ratio decreases toward the AGN,
suggesting that the aliphatic bonds are more
susceptible to photodestruction than the aromatic bonds.

To further explore the astrophysics of
the aliphatic and aromatic emission,
we utilize the near-IR spectra of
local star-forming galaxies obtained
with the IR Camera (IRC) on board AKARI,
which exhibit prominent 3.3 and 3.4$\mum$
emission in their AKARI/IRC spectra
\citep[e.g., see][]{Katayama2025}.
We compile the multi-wavelength photometric
data of these galaxies and perform
a decompositional analysis of their
spectral energy distributions (SEDs)
from the UV to the far-IR.
We examine the possible connections
between the aliphatic fraction of PAHs
and various galaxy properties,
such as the SFR, metallicity,
and stellar mass. %\footnote{%
%  In the appendix, we present
%  the spectral fitting results for our sample, t
%  ogether with the compiled multi-wavelength
%  photometric catalog from the UV to the far-IR,
%  as well as the galaxy properties derived from
%  the SED fitting.
This paper is organized as follows.
In \S\ref{sec:obs}, we describe the sample selection
and data reduction. In \S\ref{sec:measure},
we describe the PAH fitting and SED decomposition.
The results are presented in \S\ref{sec:result},
discussed in \S\ref{sec:discussion},
and finally summarized in \S\ref{sec:summary}.

Throughout this work, we adopt the cosmology
of a flat universe with
$H_{0} =70\km\s^{-1}\Mpc^{-1}$,
$\Omega_{m}=0.3$, and $\Omega_{\Lambda} = 0.7$.
Also, throughout this work, we take the assumption
of ascribing the emission bands at 3.3, 6.2, 7.7, 8.6, 11.3
and 12.7$\mum$---collectively known as
the ``unidentified infrared (IR) emission'' (UIE)
bands---to PAH molecules. While the PAH hypothesis
is popular, other candidate materials have also been
proposed as carriers of the UIE bands
\citep[e.g., see][]{Papoular1989, Jones1990, Sakata1990, Kwok2011, Cataldo2013}.
As a matter of fact, prior to the proposition of
the PAH model, \citet{Knacke1977} and
\citet{Duley1981} had already revealed
the aromatic nature of the interstellar
3.3$\mum$ band.\footnote{%
  \citet{Duley1981} also suggested
  the aliphatic nature of the 3.4$\mum$ band.
  }
\citet{Tokunaga2025} argued
that the PAH hypothesis as an explanation for
the UIE bands has a number of problems,
e.g., the exact wavelength of the astronomical
3.3$\mum$ band is inconsistent with PAH molecules
\citep{Tokunaga2021}.
If the UIE carriers have a considerable amount
of aliphatic content, substances with mixed
aromatic and aliphatic hybridization may be
more appropriately considered
as the UIE carriers \citep[e.g., see][]{Kwok2022}.
Possible candidates include nano-sized
hydrogenated amorphous carbon \citep[HAC;][]{Jones1990},
quenched carbonaceous composites \citep[QCC;][]{Sakata1990},
coal or kerogen \citep{Papoular1989},
and ``mixed aromatic aliphatic organic
nanoparticles'' \citep[MAONs;][]{Kwok2011, Kwok2013}.
Nevertheless, as the main focus of this
work is to explore the aromatic and aliphatic
contents of the UIE carriers in local star-forming
galaxies and their roles as a SFR probe
and an environmental indicator,
here we will confine ourselves to the PAH hypothesis.
After all, as will be seen later,
the aliphatic fractions of the UIE carriers
derived here are small, and the UIE carriers
are predominantly aromatic, consistent 
with previous studies. 

\section{Observations and Data Reduction} \label{sec:obs}
\subsection{AKARI/IRC Spectroscopy}
To obtain the near-IR spectra, we utilized spectra
observed with the AKARI/IRC grism spectroscopy
\citep{Murakami2007, Onaka2007}, which covers
the range 2.5--5.0$\mum$ with a resolution of
$R$\,$\simali$120.
The near-IR camera on the IRC is equipped with
a $1\arcmin\times1\arcmin$ aperture at its center,
designed to ensure that the target light is observed
without interference.
The NG (high-resolution grism)
%***[AL: what is NG?]***
is used in combination with this aperture.
Each pointing consists of eight or nine subframes,
a strategy that mitigates the impact of
cosmic ray contamination \citep{Onaka2007}.

The AKARI program was divided into three phases.
In Phases 1-2, liquid helium served as the coolant.
In Phase 3, also known as the post-helium phase,
liquid helium was depleted, and the sensitivity decreased.
We downloaded the raw data from the AKARI archive
website\footnote{%
  \url{https://data.darts.isas.jaxa.jp/pub/akari/}
   }
and then used the IDL package provided by the AKARI team
to perform data reduction. The toolkits used are
``IRC Spectroscopy Toolkit Version 20181203''
and ``IRC Spectroscopy Toolkit  for Phase 3
data Version 20181203'', which are available
on the AKARI support website.\footnote{%
  \url{https://www.ir.isas.jaxa.jp/AKARI/Observation/support/IRC/}
   }

The data reduction followed the standard pipeline.
In addition, we applied an appropriate adjustment
to the wavelength zero reference point.
We note that the quality of the spectroscopy flat for NG is very poor, which is not
enough for good flat fielding and may even reduce the signal-to-noise ratio
(SNR) of the spectrum.
Following the advice of the AKARI team, we enabled \verb|/no_slit_flat| in the
pipeline to skip flat fielding.
We extract 1D spectra from 2D spectral images by adopting a Gaussian spatial
profile assumption for each source, using a 2$\sigma$ aperture width
(6 to 14 pixels, corresponding to 8$\farcs$76 to 20$\farcs$44) to ensure $>$ 95$\%$
flux recovery while maintaining sufficient SNR.

Due to the depletion of the helium coolant,
Phase 3 data have been affected by contamination.
The data quality was enhanced by implementing
a correction method suggested by \citet{Lai2020}:
spatial profiles were constructed following the optimal
extraction algorithm from \citet{Horne1986},
with subsequent scaling applied to repair defective pixels.
For sources with multiple pointings, we performed
a weighted average of pointings with a comparable
SNR, while discarding those with an excessively low SNR.

\subsection{Multiwavelength Observations}
We compiled UV-to-IR multiwavelength photometry
for our sample from various survey catalogs
to perform SED decomposition and obtain reliable
estimates of galaxy properties.
To ensure the data are well-suited to our scientific objectives,
we exclusively utilized the photometry
corresponding to extended sources provided in the catalogs.

The {\it Galaxy Evolution Explorer}
\citep[GALEX;][]{Martin2005, Bianchi2011} provided
the far-UV (FUV) and near-UV (NUV) photometry
adopted in this study, which were derived from
the latest version of its catalog. \citep[GR6plus7;][]{Bianchi2017}.
For NGC\,1614, since it was not included
in the GALEX observations, we used the  UVW2, UVM2,
and UVW1 band photometry from the Swift
UV/Optical telescope \citep[UVOT;][]{Roming2005}
provided by \citet{Brown2014}.
We obtained optical photometry from catalogs
released by three ground-based telescopes:
SDSS DR18 \citep{Almeida2023},
Pan-STARRS1 DR2 \citep{Chambers2016, Waters2020,
Flewelling2020, Magnier2020, Magnier2020a, Magnier2020b},
and SkyMapper Southern Sky Survey DR4 \citep{Onken2024}.
For galaxies included in multiple catalogs,
we prioritized the SDSS cModel magnitudes
in the $u$, $g$, $r$, $i$, and $z$ bands.
%**[AL: cModel?]**
Based on the estimates given
in the tutorial\footnote{%
  \url{https://sdss.org/dr18/tutorials/conversions/}
   },
   corrections were made to the $u$ and $z$ bands:
$u_{\rm AB} = u_{\rm SDSS} - 0.04\magni$,
$z_{\rm AB} = z_{\rm SDSS} + 0.02\magni$.
When the SDSS data were unavailable,
Pan-STARRS Mean Kron magnitudes
in the $g$, $r$, $i$, $z$, and $y$ bands
were predominantly adopted,
with a correction factor of 100/90
applied to account for the missing
flux fraction \citep{Yamada2023}.
In other cases, we used the Petrosian magnitude
provided by Skymapper in the $u$, $v$, $g$, $r$,
$i$, and $z$ bands, applying zero-point corrections
following \citet{Casagrande2018}.

The near-IR data were obtained
using the $J$ (1.235$\mum$), $H$ (1.662$\mum$),
and $K_{s}$ (2.159$\mum$) filter data
from the extended catalog \citep{2MASS}
released by the {\it Two Micron All Sky Survey}
\citep[2MASS;][]{Skrutskie2006}.
For the mid-IR band, we use the ALLWISE \citep{ALLWISE}
or ALLSKY \citep{ALLSKY} catalogs
released by the {\it Wide-field IR Survey Explorer}
\citep[WISE;][]{Wright2010}.
Photometric corrections for the W1 (3.4$\mum$),
W2 (4.6$\mum$), W3 (12$\mum$), and W4 (22$\mum$)
bands were implemented
following the method of \citet{Wright2010},
which incorporated a $F_{\nu} \varpropto \nu^{-\alpha}$
power-law SED adjustment.
In addition, we also utilized the far-IR data
provided by \citet{Chu2017},
which were derived from
the {\it Herschel Space Observatory}
{\it Photodetector Array Camera and Spectrometer}
\citep[PACS;][]{Poglitsch2010} and
the {\it Spectral and Photometric Imaging Receiver}
\citep[SPIRE;][]{Griffin2010}.
For sources absent from the primary catalog,
we instead used data from the PACS Point Source Catalog
\citep{ppscteam2024, ppscteam2024a, ppscteam2024b}
and the SPIRE Point Source Catalog
\citep{nhsc2020, nhsc2020a, nhsc2020b}.
The PACS catalog provides photometry at 60, 100,
and 150$\mum$, while the SPIRE catalog includes
measurements at 250, 350, and 500$\mum$.
Tables~\ref{tabE:photometry1} and \ref{tabE:photometry2}
provide the photometric data we compiled.

We corrected the photometric data
for Galactic extinction using the same method
as \citet{Yamada2023}. We obtained the reddening
$E(B-V)$ for our sample from the dust map provided
by \citet[][hereafter \citetalias{Schlegel1998}]{Schlegel1998}
and then calculated the extinction $A_{\lambda}$
in magnitudes for FUV, NUV, UVW2, UVM2, UVW1,
$u$-$y$, $J$, $H$, $K_{s}$, W1, and W2 bands,
using $R_{\lambda} = A_{\lambda}/E(B-V)$.
For the extinction coefficient $R_{\lambda}$ in each band,
we adopted those from \citet{Bianchi2017} for GALEX,
\citet{Yi2023} for UVOT, \citet{Schlafly2011} for Pan-STARRS,
\citet{Wolf2018} for Skymapper,
and \citet{Yuan2013} for other telescopes.

\subsection{Sample Selection}
As shown in Table~\ref{tab:programs},
we used 1035 galaxies from 12 AKARI programs
as the parent sample.
To meet the scientific objectives of this study,
we imposed stringent detection criteria
that required robust ($>$\,3$\sigma$ significance)
identification of both the 3.3$\mum$ aromatic
and the 3.4$\mum$ aliphatic PAH features
in all targeted sources.
Based on the WISE photometry,
we further constrained our sample to sources
with a power-law spectral slope $\alpha\simgt2$
(where $F_{\nu} \varpropto \nu^{-\alpha}$).
Galaxies with such red colors are more likely to host
active star formation and/or AGNs.
Our final sample consists of 102 galaxies,
of which 66 are AGNs, with basic information
provided in Table~\ref{tabA:SEDresult}.

\section{Measurements} \label{sec:measure}

\subsection{SED Decomposition}

We used the CIGALE
\citep{Boquien2019, Burgarella2005, Noll2009} code
to perform UV-to-IR SED decomposition,
enabling us to derive various physical properties
of the galaxies, such as star formation rate,
stellar mass, luminosity, and more.
The analysis utilized CIGALE version 2022.1,
employing a multi-component framework
comprising stellar populations, nebular emission,
attenuation laws, dust emission, AGN templates,
and parametrized star formation history (SFH).

Table~\ref{tab:parameters} lists the models
employed in our SED decomposition
together with their adopted parameter values.
The delayed SFH model extended by \citet{Ciesla2017}
allows for instantaneous bursts and recent quenching
of star formation, thereby providing a high degree of flexibility.
The stellar population and nebular emission
were modeled using standard prescriptions,
while the attenuation law was treated with
the starburst attenuation curve of \citet{Calzetti2000},
which is well suited to the characteristics of our sample.
We additionally incorporated the AGN component,
applying different $f_{\mathrm{AGN}}$
(defined as the fraction of AGN IR luminosity
relative to the total IR luminosity)
for star-forming galaxies and AGN-host galaxies
in our sample. The parameters of the above modules
were mainly set with reference to recent studies of
similar samples \citep{Paspaliaris2021, Yamada2023}.
The dust emission model was adopted from
\citet{Draine2014}, which builds upon
the \citet{Draine2007} model.
Referring to the grid of \citet{Draine2007},
we set $q_{\rm pah}$ to range from 0.47 to 4.6,
considered all possible values of $U_{\mathrm{min}}$,
set $\alpha$ to either 2.0 or 2.5,
and assigned $\gamma$ a smaller value of 0.02
and a larger value of 0.15,
although 0.02 is suitable for most galaxies.

Given the absence of high-quality spectroscopic data
required to resolve metallicity-sensitive emission lines
in our input dataset, the metallicity could not be directly
constrained through SED fitting.
Following the approach of \citet{Shivaei2024},
we re-estimated the gas-phase metallicities
based on the fundamental metallicity relation (FMR)
between stellar mass, SFR, and metallicity
provided by \citet{Sanders2021}.

\begin{figure*}[h!]
\centering
\includegraphics[width=0.8\textwidth]{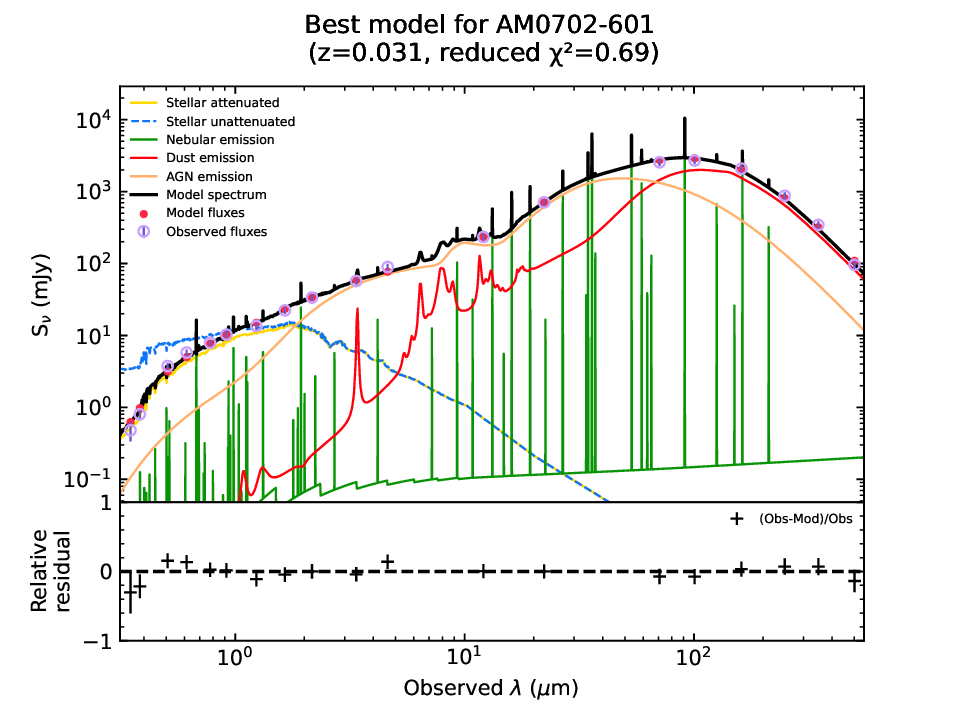}
\includegraphics[width=0.8\textwidth]{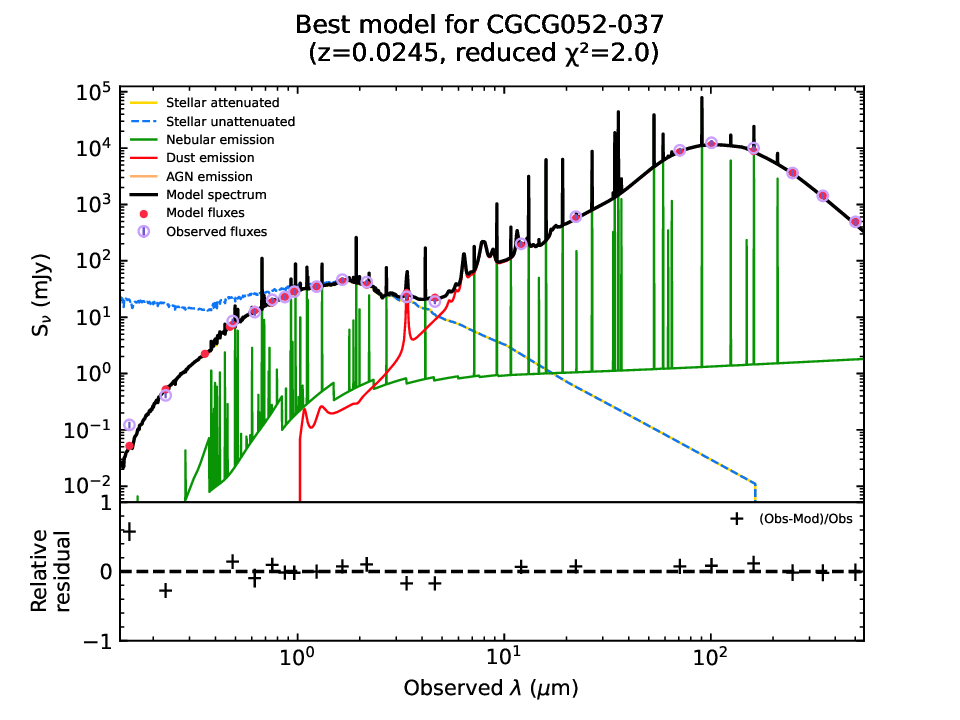}
\vspace{-0.2cm}
\caption{\label{fig:SEDexample}
Best-fit models for decomposing the UV-to-IR SEDs
of our sample galaxies.
The curves represent different model components
along with the total model spectrum.
Purple open circles and red filled circles
denote the observed and model-predicted
flux densities, respectively. The bottom panel
shows the residuals.
The complete figure set (102 subfigures)
is available in the online journal.
}
\end{figure*}

Once all models were built, CIGALE performed
SED fitting to the observational data and derived
posterior probability distributions for the physical
properties of the galaxies through Bayesian inference.
Figure~\ref{fig:SEDexample} shows the best-fit model.
The Bayesian estimates of the physical properties
for our sample can be found in Table~\ref{tabA:SEDresult}.

\subsection{Spectral Fitting}

To derive the PAH feature strengths of
our sample galaxies, we performed spectral
fitting on the AKARI spectra in the range of 2.6--3.8$\mum$.
Our model consists of a background continuum,
H$_{2}$O ice absorption, and multiple PAH components
modeled using Drude profiles, which can be expressed as:
\begin{small}
\begin{eqnarray}
F_\nu = \frac{(1 - e^{-\tau_{\lambda}})}
{\tau_{\lambda}}\left[f_0\lambda^{\alpha}
+\sum_{j}^{}\frac{P_j\times (2\gamma_j/\pi )}
{(\lambda -\lambda _{o,j}^2/\lambda )^2
+\gamma _{j}^2}\cdot\frac{\lambda^2}{c}\right] ~~,
\end{eqnarray}
\end{small}\par\noindent
where $\tau_\lambda$ is the optical depth
of H$_{2}$O ice; $f_0$ and $\alpha$ are
the coefficient and slope of the power law model
representing the thermal dust and stellar continuum,
respectively; $c$ is the speed of light;
$\lambda_{o,j}$ and $\gamma_{j}$
are the central wavelength and width of
the $j$-th Drude profile; $P_j$, the power
emitted from the $j$-th Drude profile
(in units of W\,m$^{-2}$), is obtained by
integrating the emission feature over wavelength:
\begin{eqnarray}
P_j = \int_{\lambda_{j}}^{} \Delta F_{\lambda}\,d\lambda ~~.
\end{eqnarray}
The broad 3.05$\mum$ H$_{2}$O ice absorption feature
frequently overlaps with the 3.3$\mum$ PAH emission,
making its inclusion in spectral fitting models essential.
We adopt the same configuration as \citet{Lai2020},
but allow the central wavelength of $\tau_\lambda$
to vary within the range of 3.0--3.1$\mum$,
instead of fixing it at 3.05$\mum$.
The absorption is assumed to be from a fully
mixed geometry, with the attenuation governed
by the optical depth $\tau_\lambda$ through
a scaling factor of
$\left\{1 - \exp\left(-\tau_{\lambda}\right)\right\}/\tau_{\lambda}$.
In addition to the relatively prominent 3.3 and
3.4$\mum$ PAH features, there are often some weak
subfeatures at longer wavelengths
(e.g., 3.47, 3.51, and 3.56$\mum$),
sometimes a broad plateau as well.
We fit these subfeatures using one to three
Drude profiles and assign the power emitted
from these subfeatures to $P_{3.4}$.

\begin{figure*}[h!]
\centering
\includegraphics[width=0.43\textwidth]{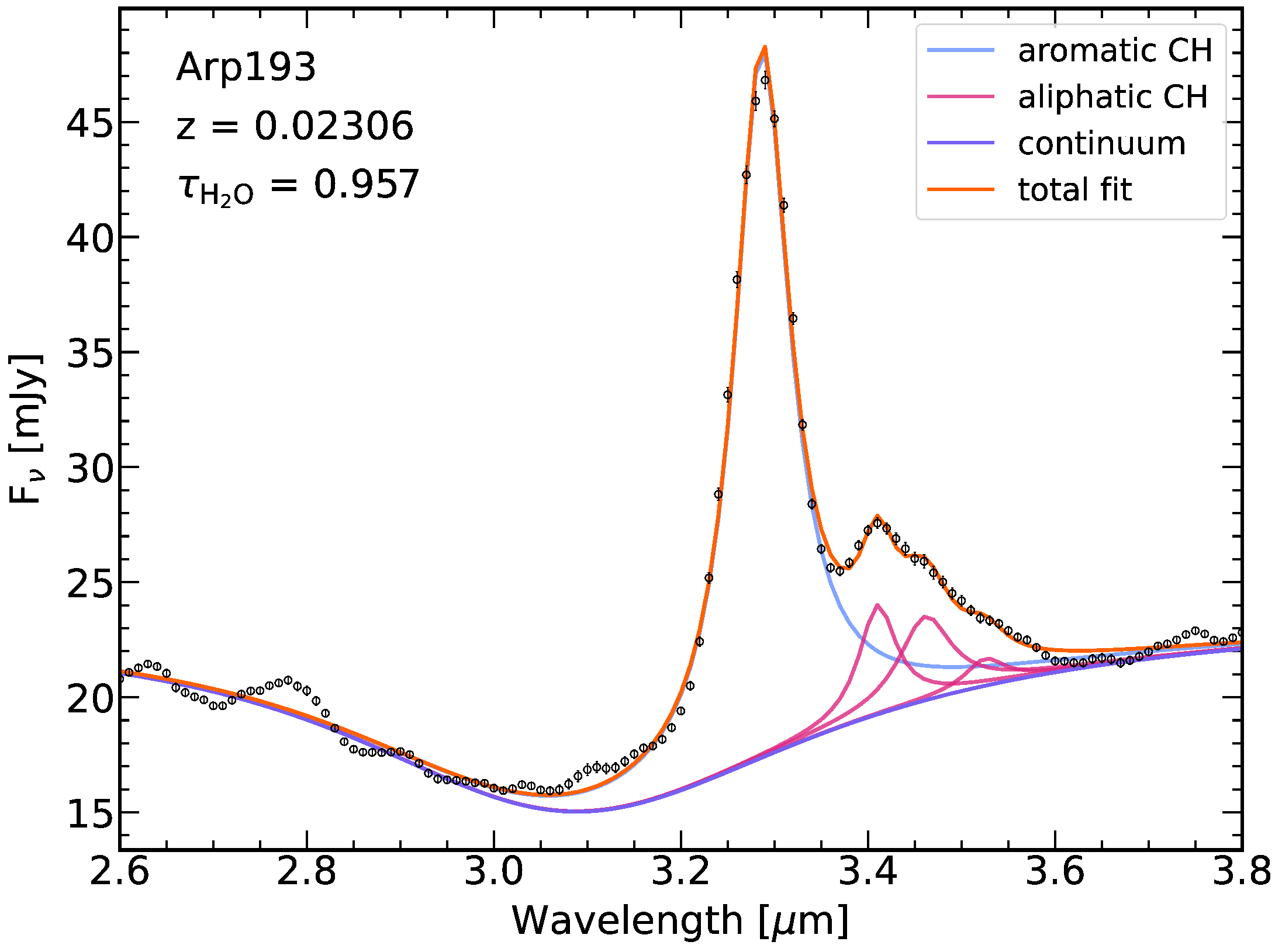}
\includegraphics[width=0.43\textwidth]{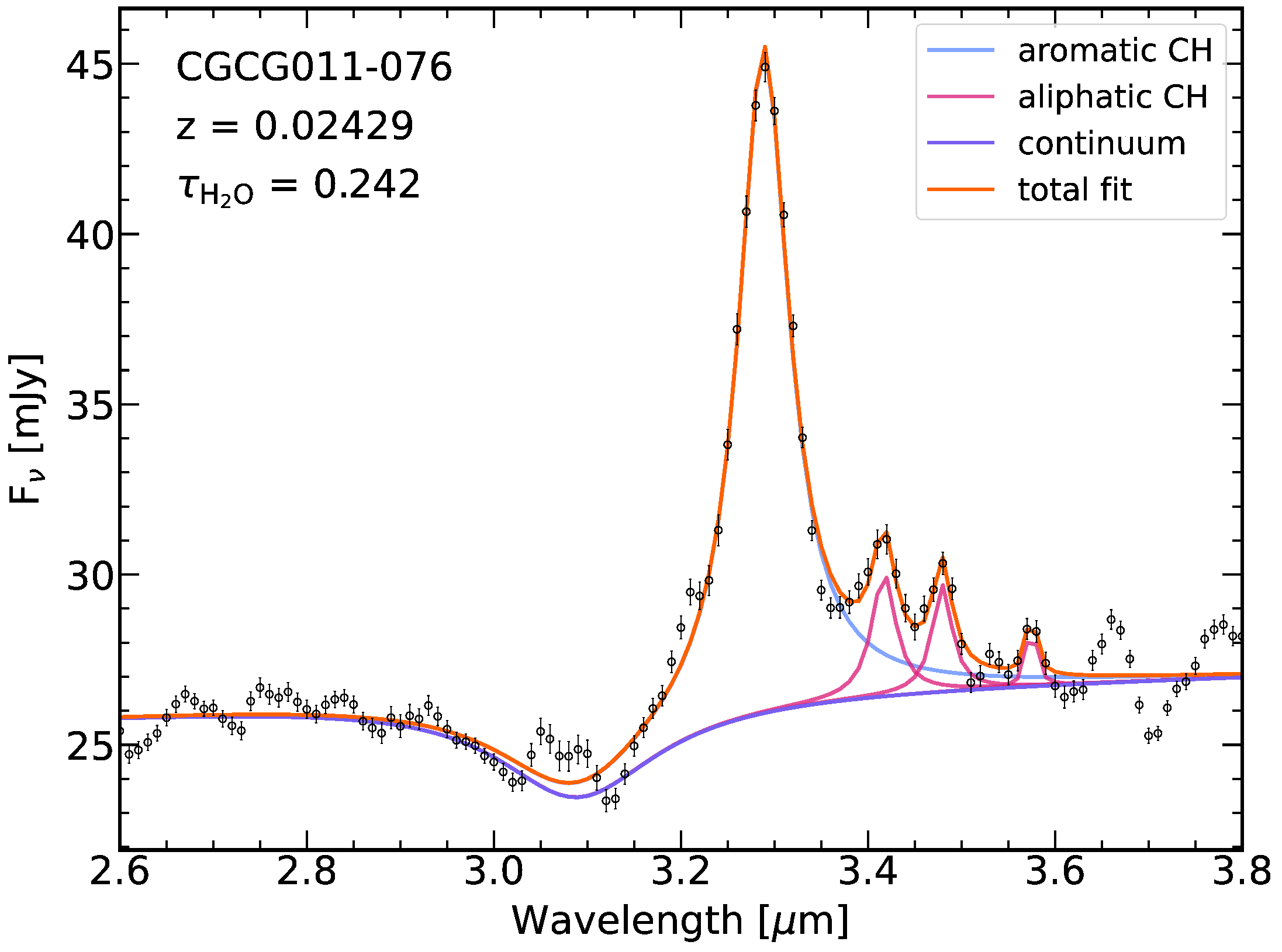}
\includegraphics[width=0.43\textwidth]{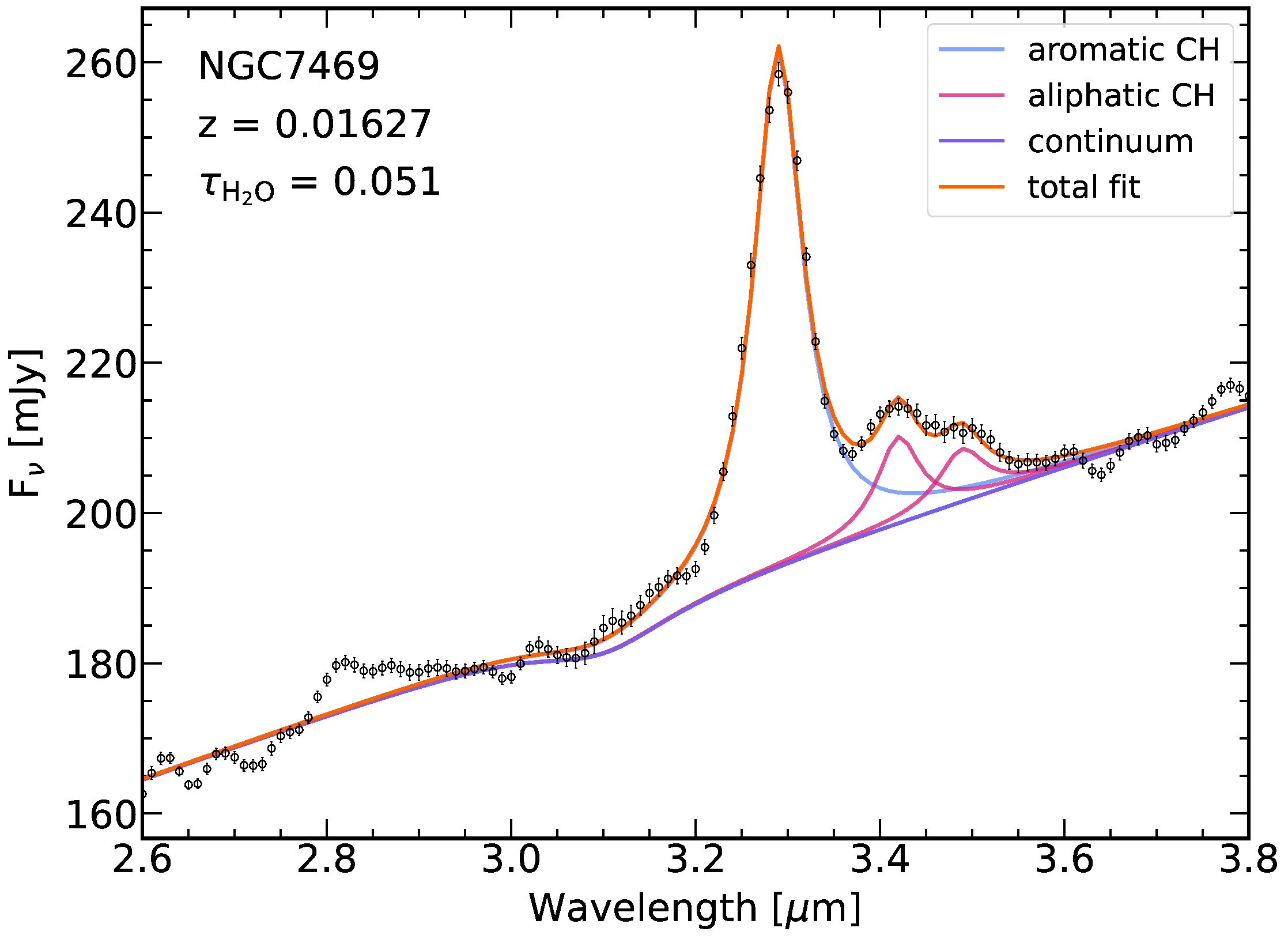}
\includegraphics[width=0.43\textwidth]{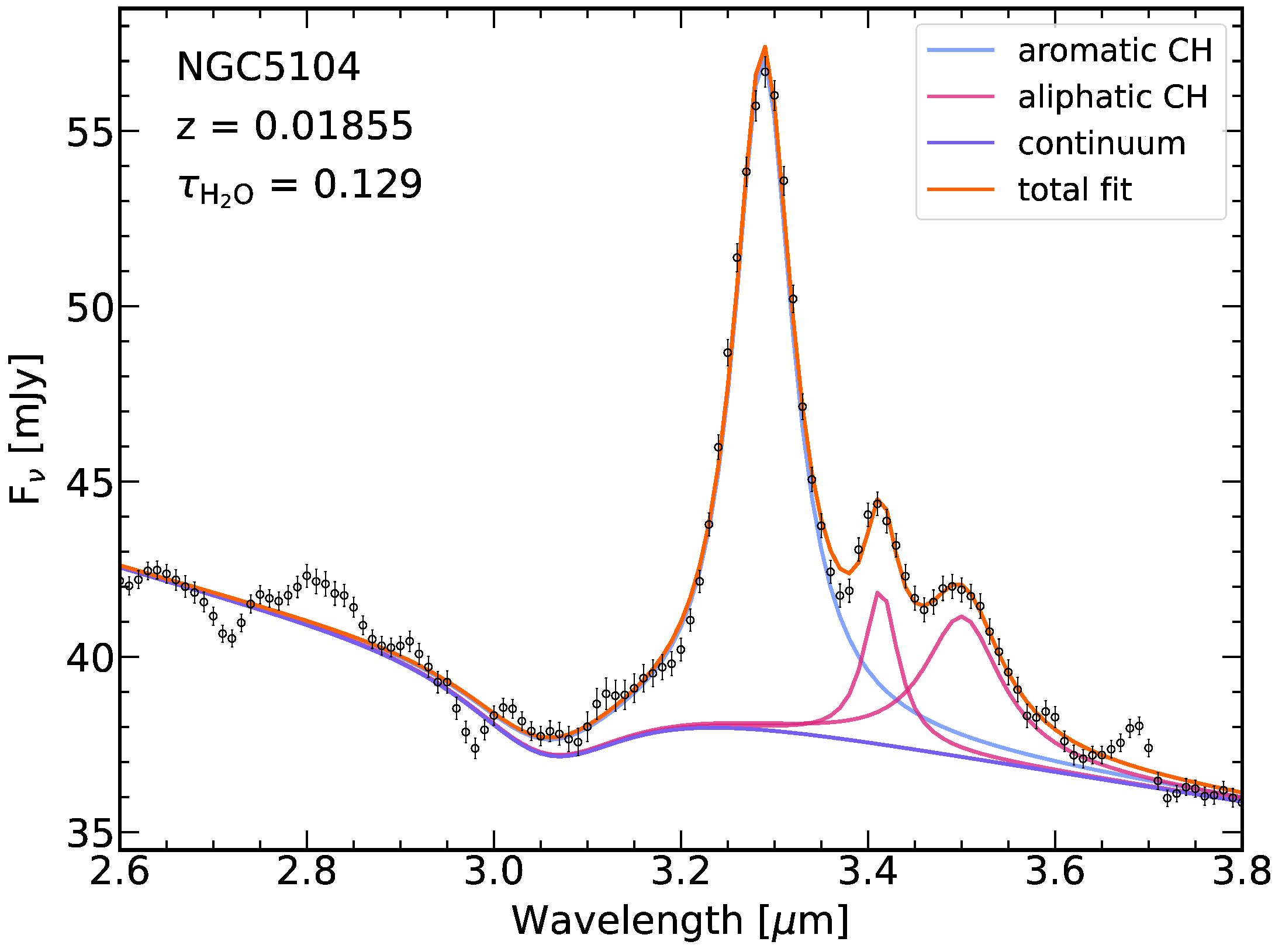}
\includegraphics[width=0.43\textwidth]{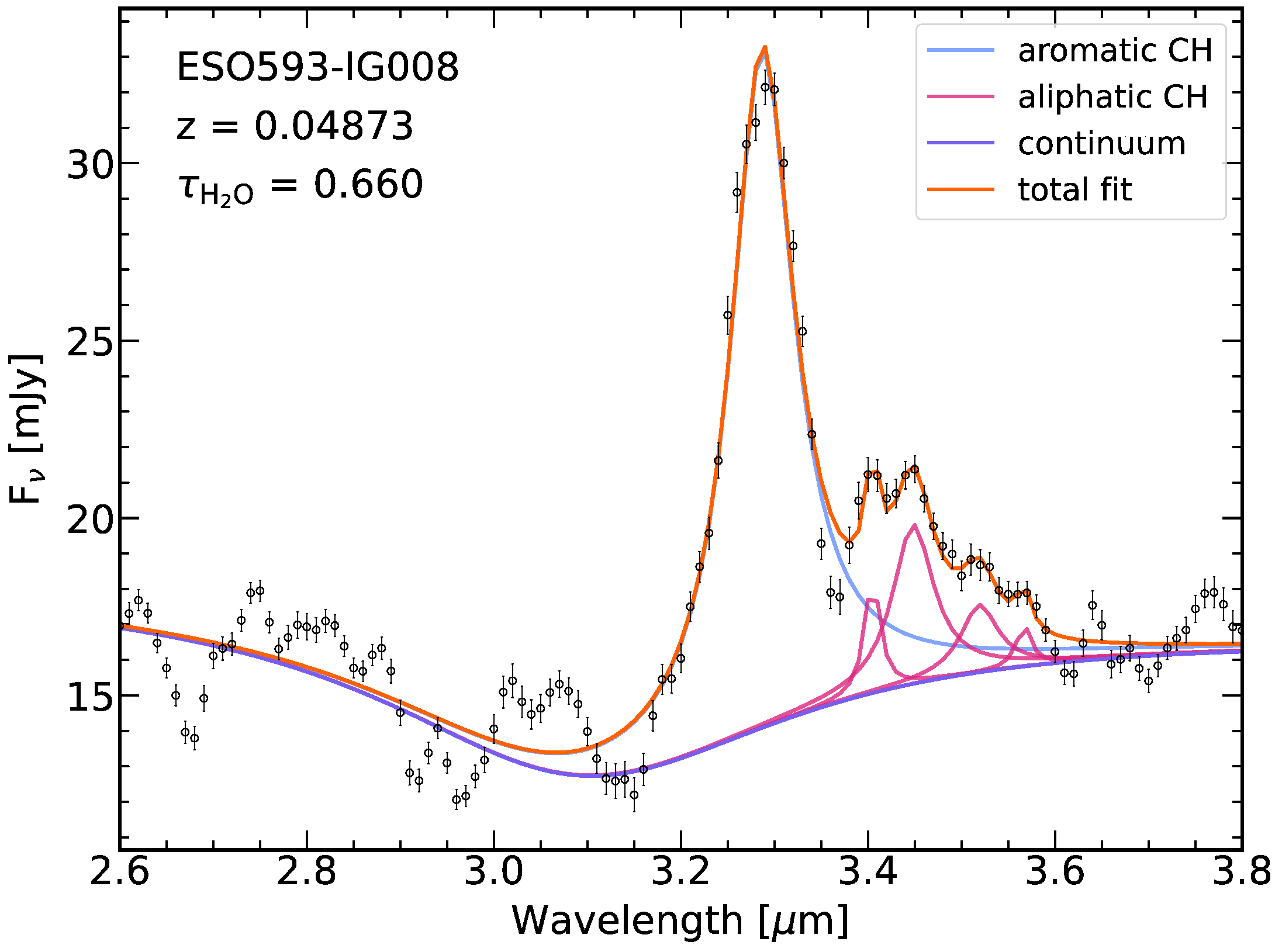}
\includegraphics[width=0.43\textwidth]{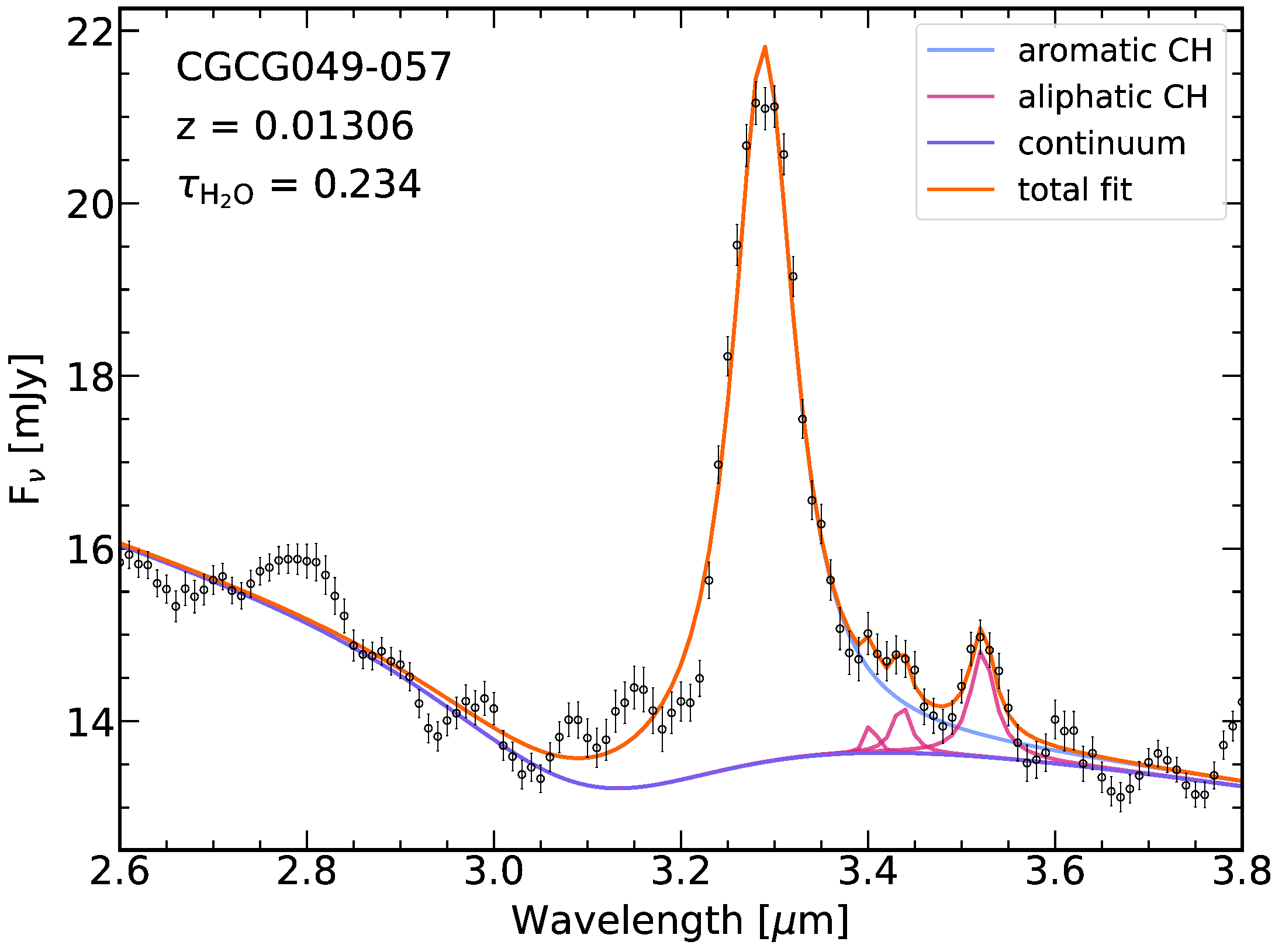}
\vspace{-0.2cm}
\caption{\label{fig:spec}
Examples of spectral fitting for the 3.3 and
3.4$\mum$ PAH features.
Galaxy name, redshift, and optical depth
of the 3.05$\mum$ H$_{2}$O ice absorption
feature are noted in the upper left corner.
The complete figure set (102 spectra) is
available in the online journal.
}
\end{figure*}

The spectral fitting was implemented
via the Levenberg-Marquardt least-square
minimization technique, with uncertainty
estimation for spectral feature intensities
derived from Monte Carlo simulations.
Figure~\ref{fig:spec} presents representative
examples of spectral fits for several sources
in our sample.

\section{Results} \label{sec:result}
\subsection{Galaxy Characteristics}

\begin{figure*}[h!]
\centering
\includegraphics[width=0.98\textwidth]{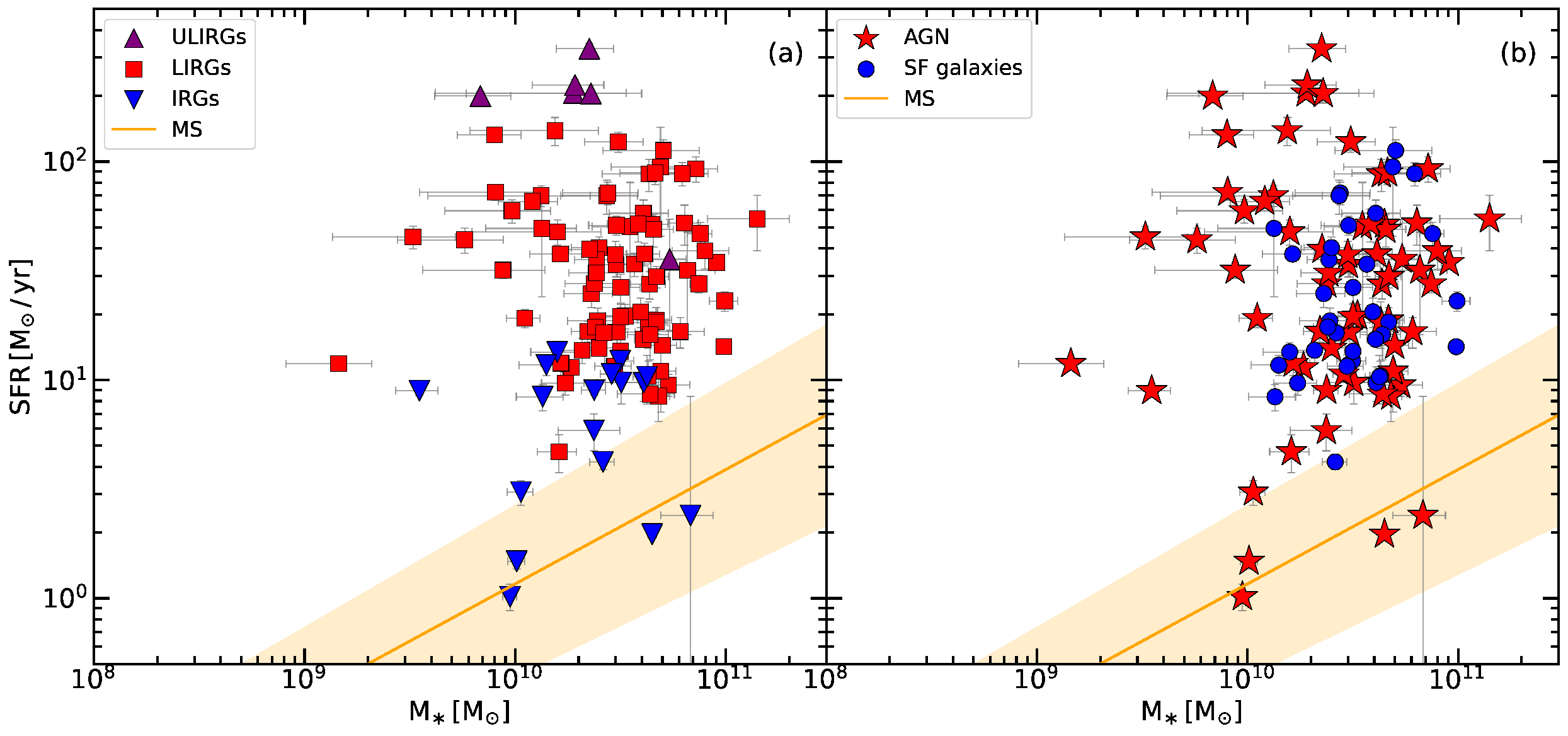}
\vspace{-0.2cm}
\caption{\label{fig:ms}
SFR versus stellar mass.
Based on the IR luminosity ($\LIR$)
obtained from SED analysis, the sample
was classified into three categories:
IR galaxies (IRGs, $\LIR < 10^{11}\Lsun$)
represented by blue downward triangles,
luminous IR galaxies (LIRGs,
$10^{11} < \LIR < 10^{12}\Lsun$)
represented by red squares,
and ultraluminous IR galaxies
(ULIRGs, $\LIR>10^{12}\Lsun$)
represented by purple upward triangles,
as shown in (a). In (b), galaxies with an active nucleus
(labelled ``Y'' in Table~3) are classified as AGN,
while the others are considered SF galaxies.
%***[AL: where is Y?]***
The orange-shaded region represents
the MS corresponding to the redshift
of our sample ($z$\,$\simali$0--0.17),
with the solid orange line in the center
indicating the MS at $z=0.1$,
as reported by \citet{Speagle2014}.
}
\end{figure*}

There exists a correlation between the SFR
and the stellar mass ($M_\bigstar$)
in star-forming galaxies, known as
the star-formation main sequence (MS),
which is an essential tool for studying
the characteristics of galaxies.
This strong correlation, which has been widely reported
\citep{Brinchmann2004, Elbaz2011, Elbaz2018, Speagle2014, Pearson2018},
can be used to distinguish between starburst galaxies
(above the MS), star-forming galaxies (on the MS),
and quenched galaxies (below the MS).
Figure~\ref{fig:ms} presents the SFR--$M_\bigstar$ relation
for the galaxy sample, with individual sources
color-coded by $\LIR$---the total luminosity
emitted by dust---and classified according to
AGN activity. Also shown is the MS corresponding
to redshift ($z$\,$\simali$ 0--0.17) of our sample
from \citet{Speagle2014}.

The majority of our sample lies above the MS,
indicating intense star formation activity,
which is consistent with our expectations
during the sample selection. Figure~\ref{fig:ms} (a)
demonstrates a positive correlation between $\LIR$
and SFR across the galaxy sample,
with all U/LIRGs positioned above the MS.
In Figure~\ref{fig:ms} (b), AGN and star-forming
galaxies (SF galaxies) exhibit significant overlap,
with no discernible features enabling clear separation
between these two populations.
This is consistent with the results
of \citet{Shangguan2019} and \citet{Yamada2023}.

\subsection{PAH Feature and Aliphatic Fraction}

%\begin{figure}[h!]
%\centering
%\includegraphics[width=0.6\textwidth]{averagespec.eps}
%\vspace{-0.2cm}
%\caption{Average spectrum of samples classified
%by $\LIR$, normalized at 3.65$\mum$.\label{fig:average}}
%\vspace{-0.3cm}
%\end{figure}

% Figure \ref{fig:average} shows the average spectra
%of different infrared luminosity levels.
%The spectrum displays prominent PAH
%emission features at 3.3 and 3.4$\mum$,
%accompanied by a 3.05$\mum$ ice absorption band.
%The intensity of aromatic features increases
%significantly with $\LIR$, while aliphatic features
%show the opposite trend, but their thresholds
%for large changes are different.
%In addition, spectra with elevated $\LIR$
%exhibit enhanced optical depths in
%the ice absorption bands, with a similar
%span of increase between each level.
%A strong emission feature is also detected
%at 4.05$\mum$, corresponding to
%the Br$_{\alpha}$ hydrogen recombination line.

\begin{figure}[h!]
\centering
\includegraphics[width=0.6\textwidth]{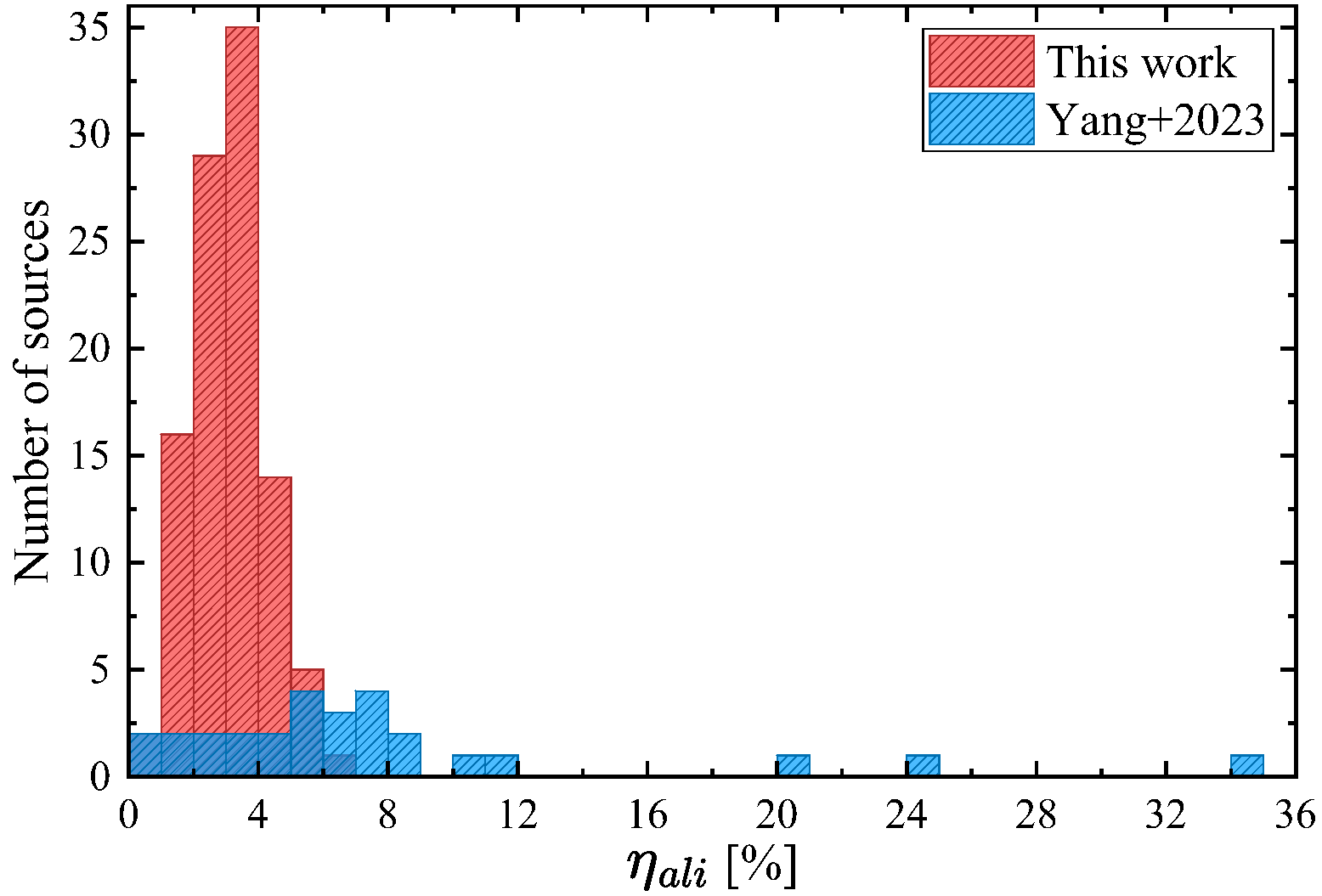}
\vspace{-0.1cm}
\caption{\label{fig:distri}
Comparison of the PAH aliphatic fraction distribution
between our sample and the Galactic sample
of \citet{Yang2023}.
}
\vspace{-0.3cm}
\end{figure}

With $P_{3.3}$ and $P_{3.4}$---the power
emitted from the 3.3 and 3.4$\mum$ bands,
respectively---obtained, we employed the formula
of \citet{Yang2023} to determine the aliphatic
fraction $\etaali$, defined as the fraction of
carbon atoms in aliphatic units:
\begin{gather}
\etaali = \frac{1}{1 + N_{\rm C,aro}/N_{\rm C,ali}} ~~,
~~ \frac{N_{\rm C,ali}}{N_{\rm C,aro}} \approx
\frac{1}{6.40}\ \frac{P_{3.4}}{P_{3.3}} ~~.
\end{gather}
Figure~\ref{fig:distri} presents the distribution
of the PAH aliphatic fractions for our sample,
compared with that of the Milky Way sources
reported in \citet{Yang2023}.
The majority of the galaxies in our sample
have $\etaali\simlt 6\%$,
ranging from 0.38\% to 6.8\%,
with a median of 3.1\%.
%, somewhat lower than
%those of Galactic sources.
% The physical drivers of this discrepancy
%will be explored in Section \ref{subsec:aliphatic}.
In Table~\ref{tabB:fitresult}, we list
the fitting results and $\etaali$.

\begin{figure}[ht]
 \vspace{-12mm}
  \begin{center}
\includegraphics[width=13.6cm,angle=0]{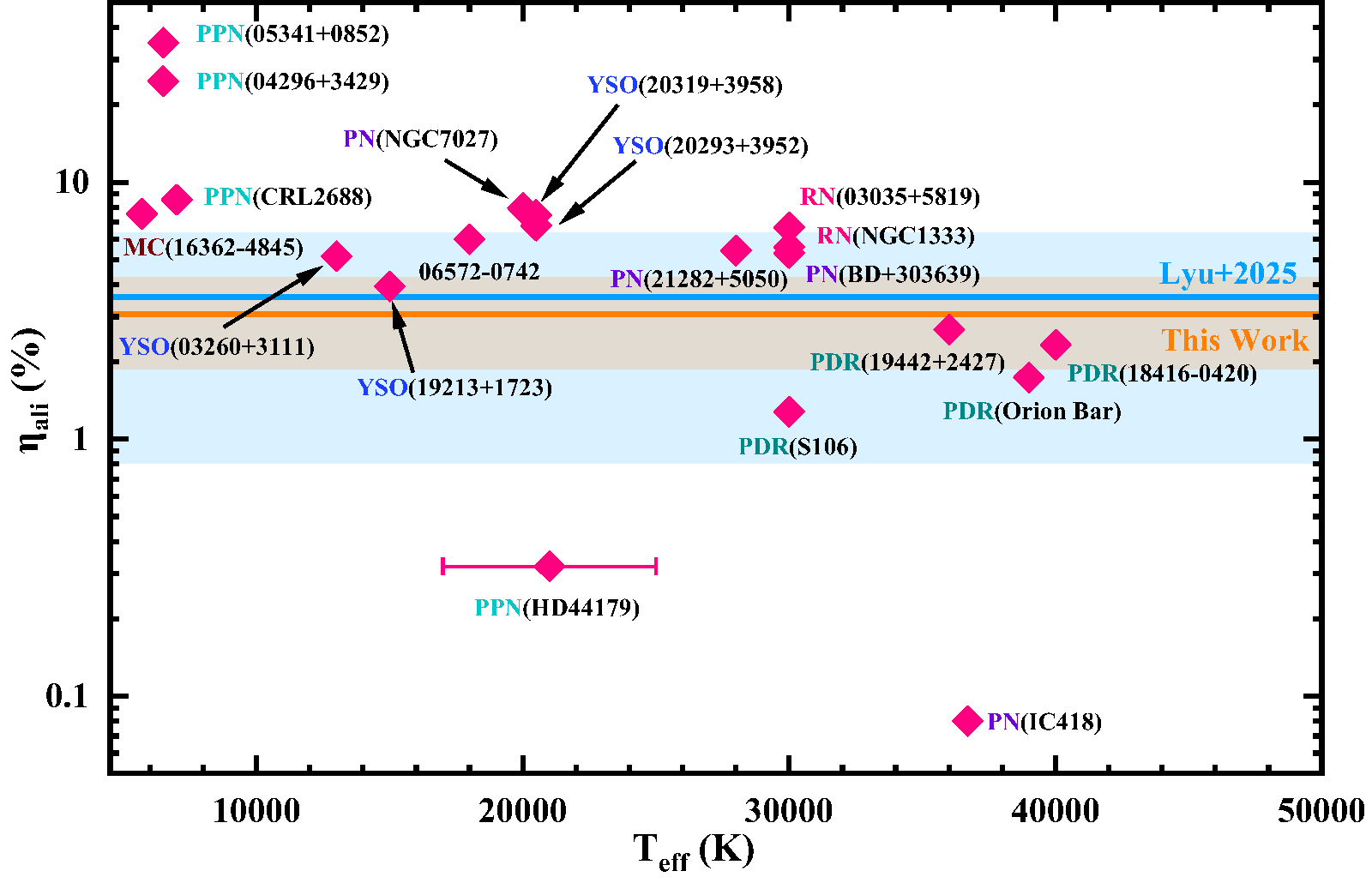}
  \end{center}
\vspace{-5mm}
\caption{\label{fig:Ali.vs.Teff} \footnotesize
               PAH aliphatic fraction ($\alifrac$)
               vs. stellar effective temperature ($\Teff$).
         PPN: protoplanetary nebula;
         PN: planetary nebula;
         RN: reflection nebula;
         MC: molecular cloud;
         PDR: photodissociated region;
         YSO: young stellar object.
The shaded orange horizontal belt
shows the mean aliphatic fraction
and its standard deviation
($\etaali\approx3.1\%\pm1.2\%$)
derived from our sample.
Also shown (as a shaded cyan horizontal belt)
is the mean aliphatic fraction
together with its standard deviation
($\etaali\approx3.6\%\pm2.8\%$)
derived by Lyu et al.\ (2025) for a sample
of 37 galaxies at redshifts $z$\,$\simali$0.2--0.5.
}
\vspace{-3mm}
\end{figure}

Previously, Yang \& Li (2023) had explored
whether (and how) the PAH aliphatic fraction
varies with astrophysical environments,
particularly with the hardness of the exciting
starlight photons as measured by
$\Teff$, the effective temperature
of the illuminating star.
As shown in Figure~\ref{fig:Ali.vs.Teff},
Yang \& Li (2023) found that
$\alifrac$ appears higher in regions
illuminated by stars with a lower $\Teff$,
indicating the survival of aliphatic sidegroups
attached to PAHs is more favorable
in regions lacking energetic photons.
Due to the lack of accurate knowledge
about the hardness of the starlight
photons of the galaxies studied here,
it is difficult to characterize the radiation
fields of these galaxies by a single stellar
temperature $\Teff$. Therefore, we plot
in Figure~\ref{fig:Ali.vs.Teff} the mean
aliphatic fraction and its standard deviation
($\etaali\approx3.1\%\pm1.2\%$)
derived from our sample
as a shaded horizontal belt.
Also shown is the mean aliphatic fraction
together with its standard deviation
($\etaali\approx3.6\%\pm2.8\%$)
derived by \citet{Lyu2025} for a sample
of 37 galaxies at redshifts $z$\,$\simali$0.2--0.5,
utilizing data from JWST's FRESCO program,
obtained with JWST's NIRCam/WFSS F444W
observations of two GOODS/CANDELS fields.
Figure~\ref{fig:Ali.vs.Teff} clearly shows that
the mean aliphatic fractions derived here
and derived by \citet{Lyu2025} are consistent
with those of \citet{Yang2023} for photodissociated
regions (PDRs), planetary nebulae
and reflection nebulae, but lower than that
for protoplanetary nebulae,
which are illuminated by UV-poor stars.

\section{Discussion} \label{sec:discussion}
\subsection{PAH and Infrared Luminosity}

\begin{figure*}[h!]
\centering
\includegraphics[width=0.48\textwidth]{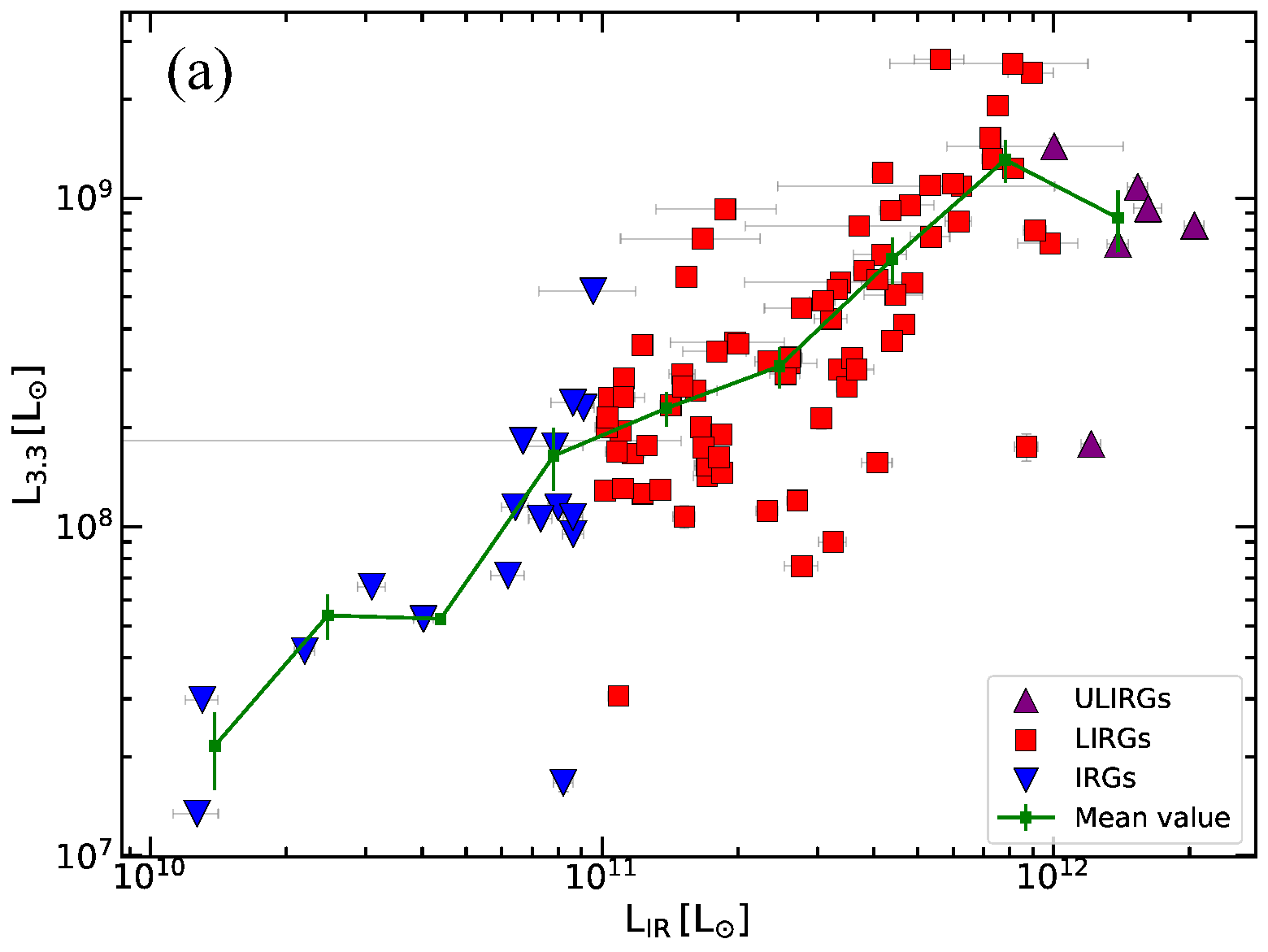}
\includegraphics[width=0.48\textwidth]{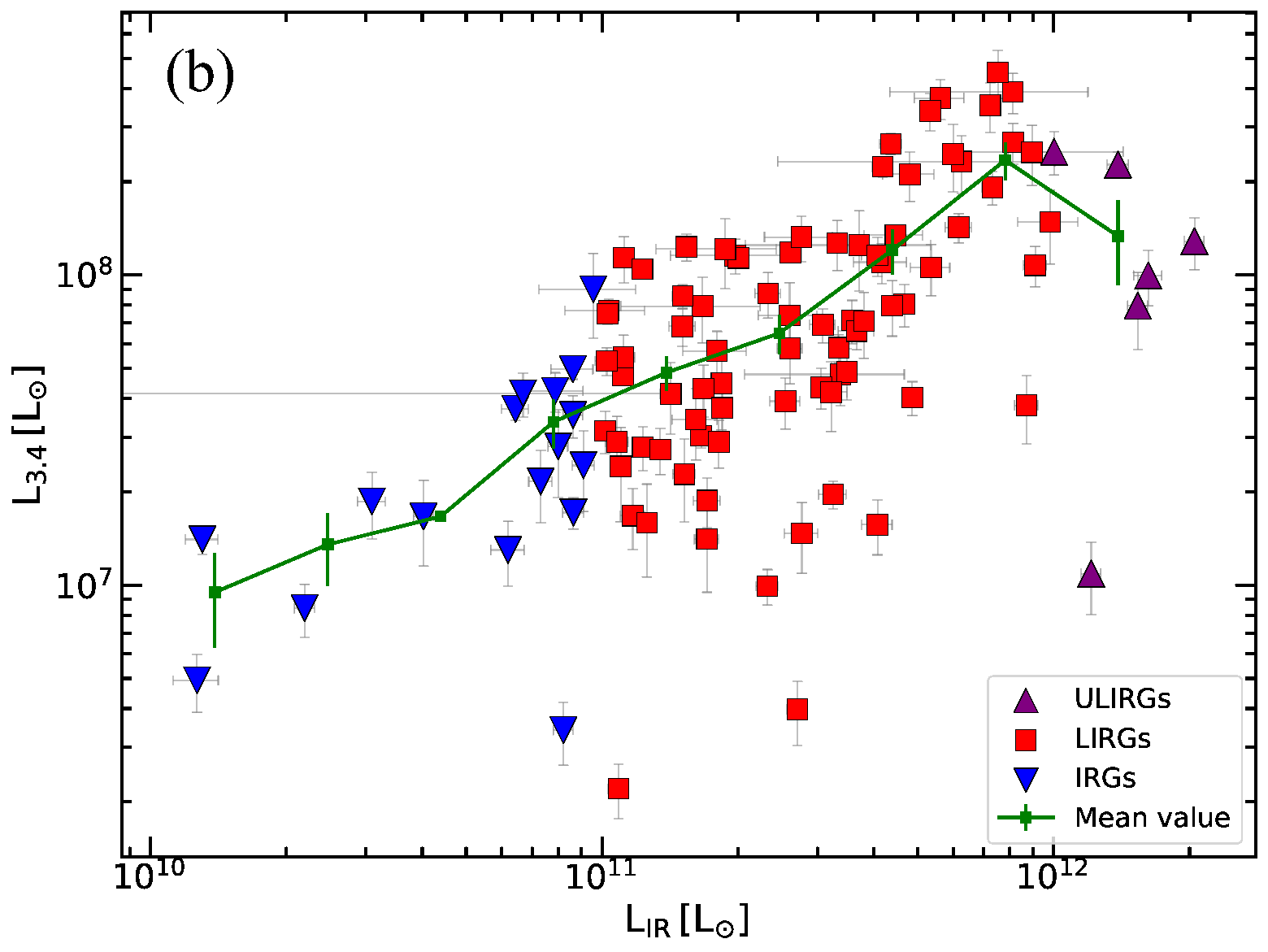}
\vspace{-0.2cm}
\caption{\label{fig:LIR}
Variations of $L_{3.3}$ and $L_{3.4}$
with $\LIR$. The data have been color-coded
by $\LIR$, with green squares representing
the average of the different $\LIR$ bins.
}
\end{figure*}

In the local universe, the luminosity of the 3.3$\mum$ PAH
emission $L_{3.3}$ has been observed to be strongly correlated
with the $\LIR$,  although $L_{3.3}/\LIR$ shows a declining
trend in ULIRGs, which may be due to the destruction of PAHs
by the intense UV radiation \citep{Yamada2013, Murata2017}.
As demonstrated in Figure~\ref{fig:LIR}a, our sample also
shows that $L_{3.3}$ and $\LIR$ correlate.
The apparent decrease in $L_{3.3}/\LIR$ at higher $\LIR$
in our sample is not pronounced; it is likely due to
the limited number of high $\LIR$ samples.

We further investigate the relation between $L_{3.4}$
and $\LIR$. As shown in Figure~\ref{fig:LIR}b,
$L_{3.4}$ also correlates with $\LIR$.
% The 3.4$\mum$ PAH follows a similar trend
% as the 3.3$\mum$ PAH, yet its {\Lali}/$\LIR$
% ratio is markedly lower, indicating enhanced
% destruction of aliphatic hydrocarbon carriers
%relative to aromatic structures in high-$\LIR$ environments.

\subsection{SFR Calibration with the 3.3$\mum$ PAH Emission}

\begin{figure}[h!]
\centering
\includegraphics[width=0.6\textwidth]{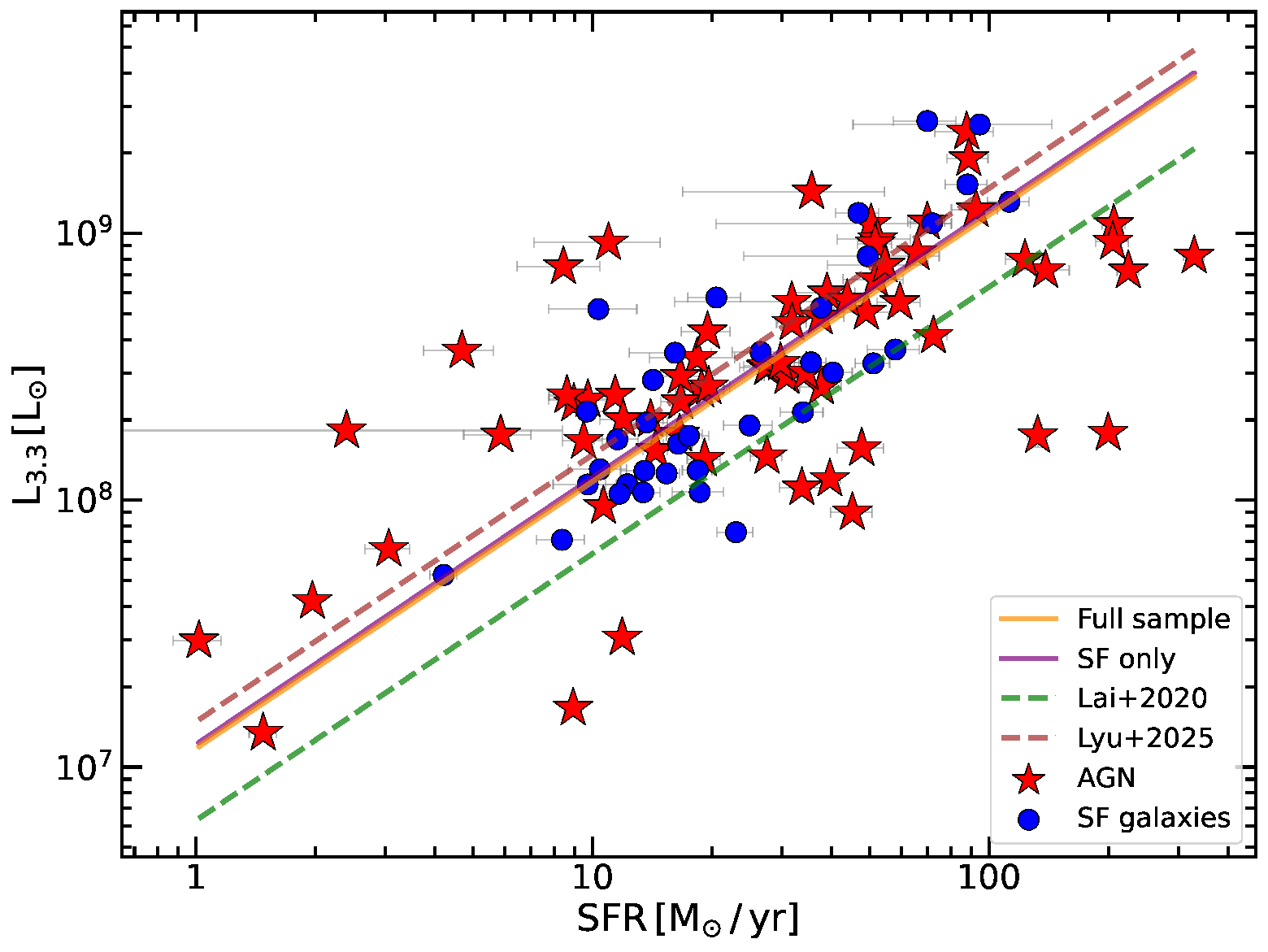}
\vspace{-0.2cm}
\caption{The correlation between {\Laro} and the SFR.
The orange solid line represents the best-fit calibration for the full sample, while the purple
solid line corresponds to the fit restricted to SF galaxies.
The green and brown dashed lines indicate the calibrations from \citet{Lai2020} and \citet{Lyu2025}, respectively, provided for comparison.
\label{fig:sfr}}
\vspace{-0.1cm}
\end{figure}

The 3.3$\mum$ PAH emission is a promising
star formation rate tracer \citep{Kim2012, Lai2020},
with its shorter wavelength providing a unique
advantage over other PAH spectral bands,
making it particularly valuable for studying
high-redshift galaxies.
Figure~\ref{fig:sfr} displays the relation
between $L_{3.3}$ and SFR in our sample.
While a significant positive correlation is
evident across the full dataset, AGN exhibit
markedly larger scatter compared to SF galaxies.
This discrepancy may arise from multiple factors.
In active galaxies, AGN activity can additionally
process PAHs, and the presence of an AGN
component may lead to over- or underestimation
of the SFR in SED fitting due to biases in
modeling the AGN contribution.

By fixing the slope to unity, we derived
the best-fit calibration between $L_{3.4}$ and SFR.
When considering the entire sample,
the relation can be expressed as:
\begin{equation}
\log\,\left(\frac{\rm SFR}{\Msun\yr^{-1}}\right)
= \log\,\left(\frac{L_{3.3}}{\Lsun}\right) - \left(7.07\pm0.36\right)~~,
\end{equation}
while for SF galaxies alone, it is represented as:
\begin{equation}
\log\,\left(\frac{\rm SFR}{\Msun\yr^{-1}}\right)
= \log\,\left(\frac{L_{3.3}}{\Lsun}\right) - \left(7.08\pm0.25\right)~~.
\end{equation}
Excluding AGN sources from our sample
has little impact on the fitting results

Compared to the calibration given by \citet{Lai2020},
our derived SFRs are systematically lower at a fixed $L_{3.3}$,
which we attribute to differences in the methodologies
used to estimate SFR. \citet{Lai2020} adopted the neon-based
SFR relation from \citet{Zhuang2019}, which relies on
[Ne\,\textsc{ii}] and [Ne\,\textsc{iii}] emission lines.
However, this method has been shown to systematically
yield slightly higher SFRs compared to those estimated
from SED fitting, especially when the metallicity is fixed
at the solar ($Z_{\odot}$), which is the same assumption
made by both \citet{Lai2020} and us.
For lower IR luminosity galaxies ($10^9<\LIR<10^{10}\Lsun$) at
redshifts $z$\,$\simali$0.2--0.5 observed with JWST/NIRCam,
\citet{Lyu2025} also calculated the $L_{3.3}$-SFR correlation
using the SFR determined from SED fitting.
Their calibration shows better agreement with ours,
differing only by $\simali$0.1 dex,
which is within the uncertainty.
Thus, the 3.3$\mum$ PAH emission
can effectively trace SFR, which is fully
demonstrated by the robustness of
the $L_{3.3}$--SFR correlation of
galaxies of different redshifts and types.

\subsection{PAH and Metallicity} \label{subsec:metallicity}

\begin{figure*}[h!]
\centering
\includegraphics[width=0.48\textwidth]{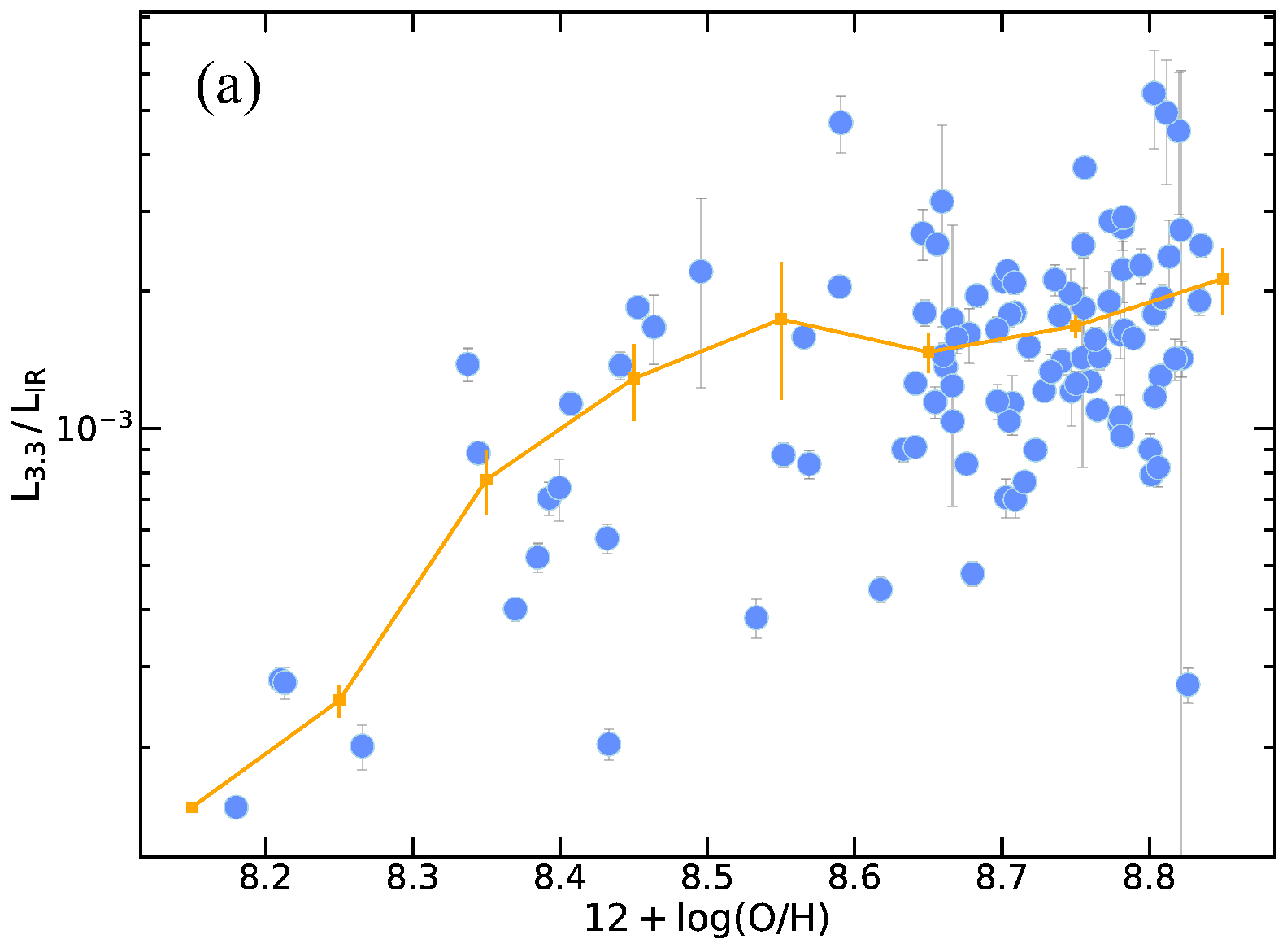}
\includegraphics[width=0.48\textwidth]{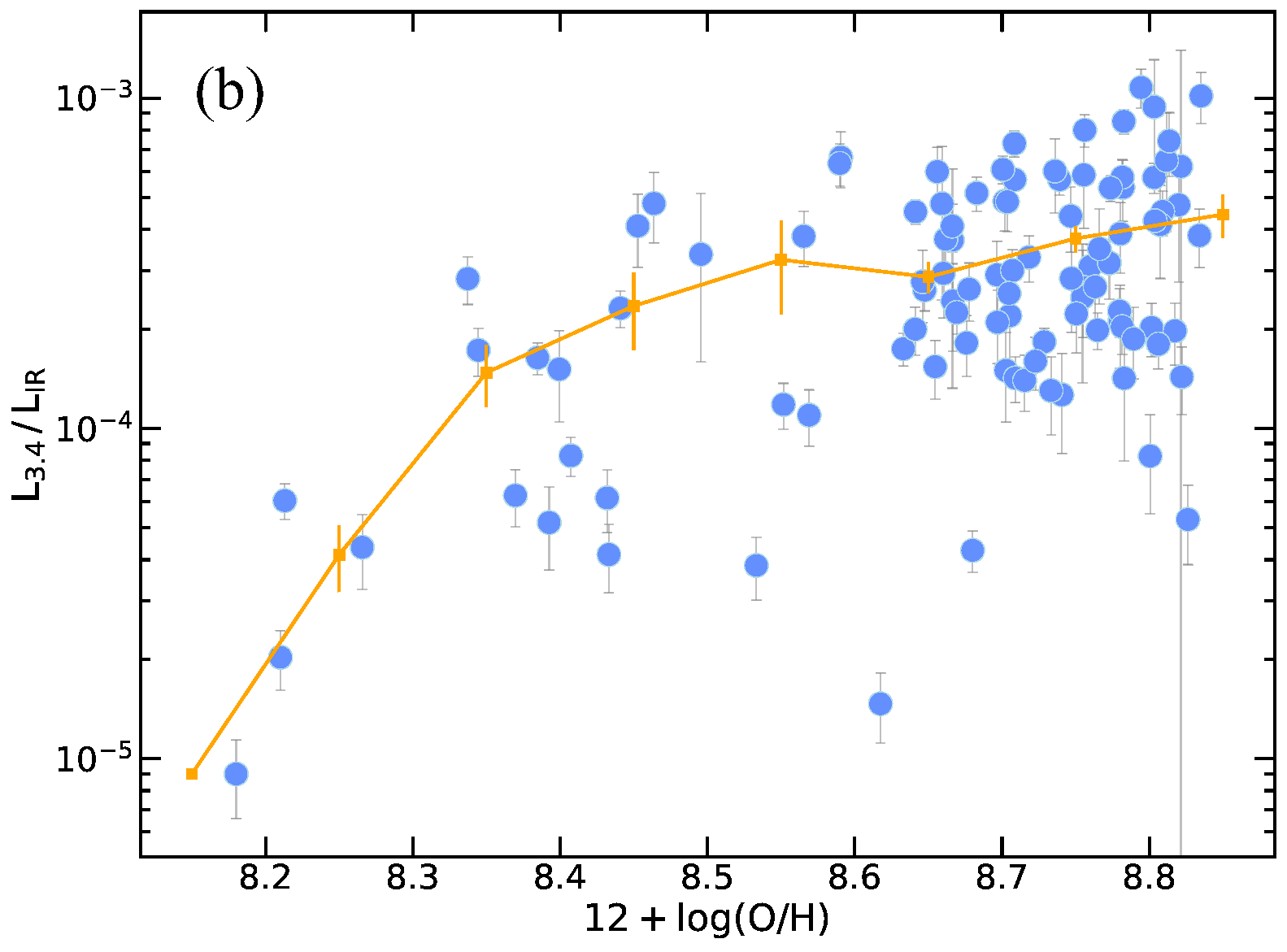}
\vspace{-0.2cm}
\caption{\label{fig:metallicity}
Left panel: Relation between $L_{3.3}/\LIR$
and metallicity $12\,+\,\log{\rm (O/H)}$.
Right panel: Relation between $L_{3.4}/\LIR$
and metallicity $12\,+\,\log{\rm (O/H)}$.
Orange squares show the mean
$L_{\rm PAH}/\LIR$ ratios
vs. $12\,+\,\log{\rm (O/H)}$.
}
\end{figure*}

There is a clear dependence between PAH emission
and metallicity, a relationship known as
the PZR (PAH-metallicity relation),
which has been extensively studied
\citep{Engelbracht2005, Smith2007, Chastenet2023, Shivaei2024}.
Previous studies have primarily focused on
the mid-IR emission of PAHs
at 6.2, 7.7, 8.6, and 11.3$\mum$.
In this work, we calculated the gas-phase
metallicities of our sample galaxies
using the FMR \citep{Sanders2021}
and investigated its relation with
$L_{3.3}/\LIR$ and $L_{3.4}/\LIR$,
as shown in Figure~\ref{fig:metallicity}.

Figure~\ref{fig:metallicity}a shows a clear
correlation between $L_{3.3}/\LIR$
and $12\,+\,\log{\rm (O/H)}$,
with a Pearson correlation coefficient
of $r\approx0.58$ and a statistical
significance of $p$\,$\simali$$10^{-10}$.
The binned mean values reveal a steep rise
in $L_{3.3}/\LIR$ at low metallicities
$12\,+\,\log{\rm (O/H)} < 8.5$,
while $L_{3.3}/\LIR$ flattens off
at $12\,+\,\log{\rm (O/H)} > 8.5$.
This behavior is similar to the trend
observed for PAHs in the mid-IR \citep{Whitcomb2024}.
Figure~\ref{fig:metallicity}b shows that
$L_{3.4}/\LIR$ exhibits a similar variation
with the metallicity
(Pearson $r\approx0.56$ and
$p$\,$\simali$$10^{-10}$),
suggesting that the aliphatic and aromatic units
are connected and influenced by metallicity
through the same mechanism.

There are various explanations for the PZR,
with one of the most common being photodestruction
\citep{Madden2006, Hunt2010, Egorov2023}.
In environments with lower metallicity,
dust is typically less abundant, and the extinction
is smaller so that UV photons from young stars
are less attenuated.
In such less shielded interstellar conditions,
PAHs are more likely to be destroyed by UV photons.
Recently, \citet{Whitcomb2024} systematically
assessed multiple proposed mechanisms,
demonstrating that the inhibited growth scenario
provides the most robust explanation.
In this scenario, individual carbon atoms
accumulate on the surface of larger grains for growth.
At lower metallicities, the availability of carbon atoms
is limited, which reduces the efficiency of growth
and consequently leads to a deficiency of PAHs.
More recent observational evidence supports this view.
\citet{Zhang2025} found that small PAHs,
which are more easily photodissociated, actually
dominate in the low-metallicity star-forming region
30 Doradus, suggesting that photodestruction
is not the primary mechanism driving the observed PZR.

\subsection{PAH Aliphatic Fraction}\label{subsec:aliphatic}

\begin{figure*}[p]
\centering
\includegraphics[width=0.48\textwidth]{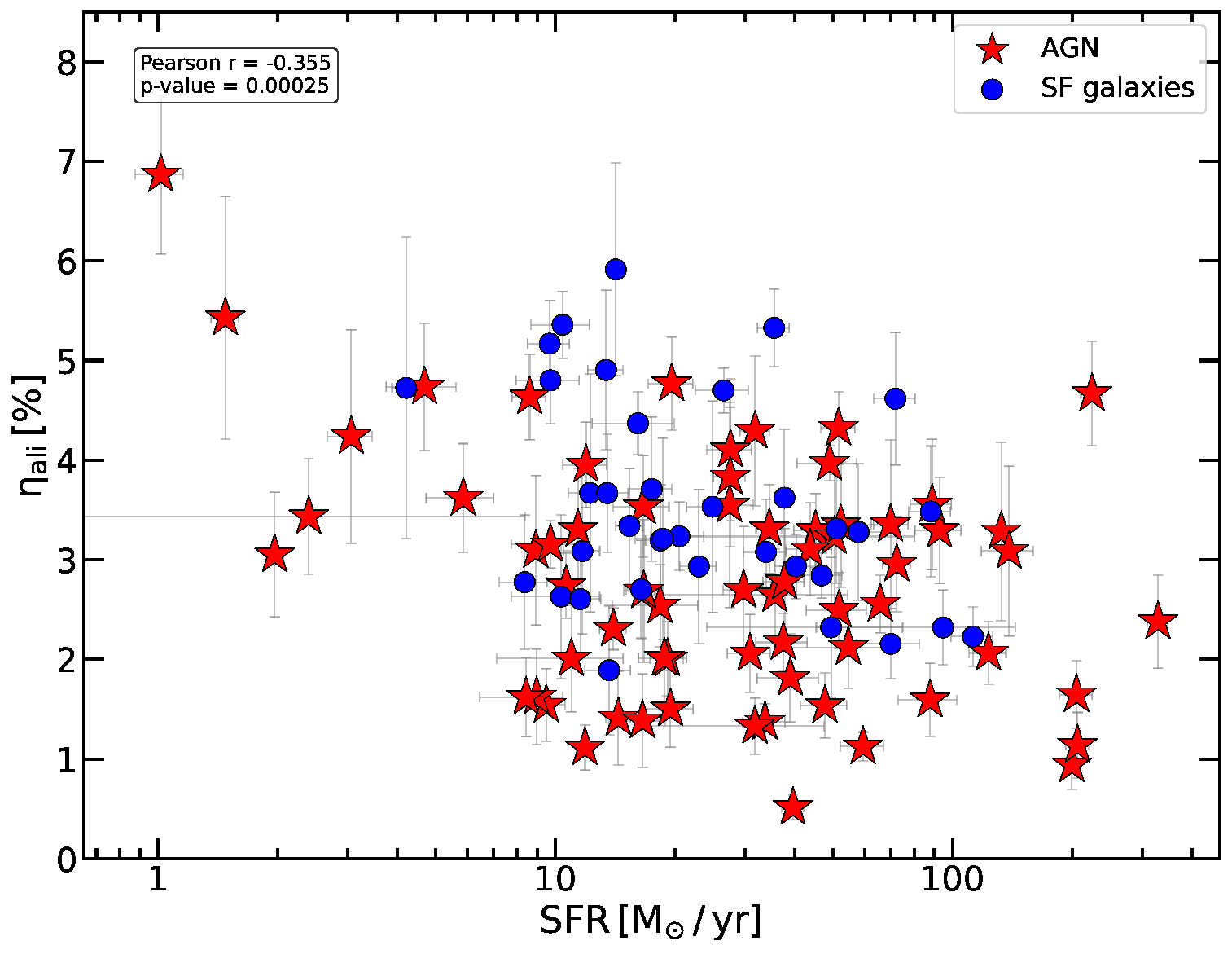}
\includegraphics[width=0.48\textwidth]{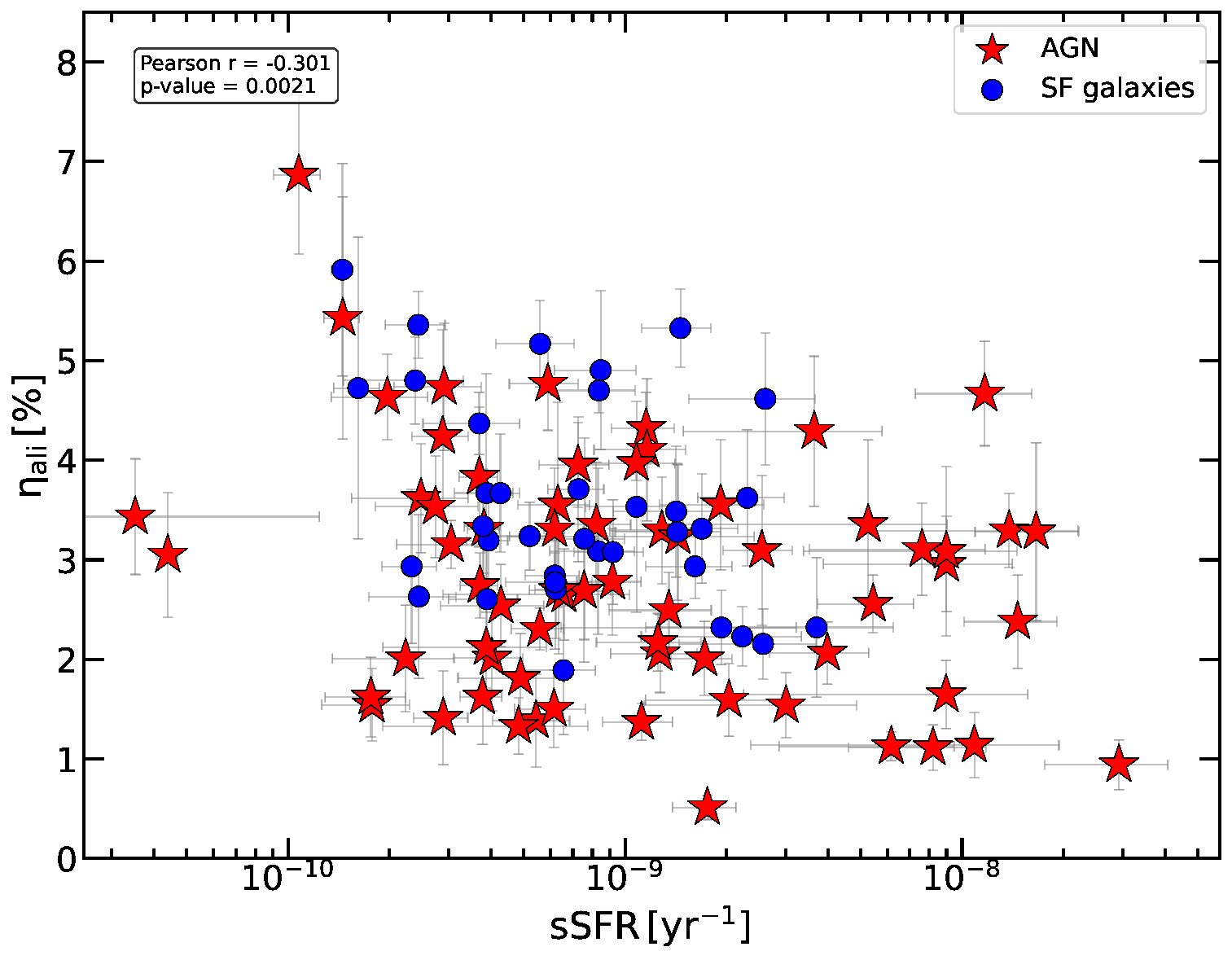}
\includegraphics[width=0.48\textwidth]{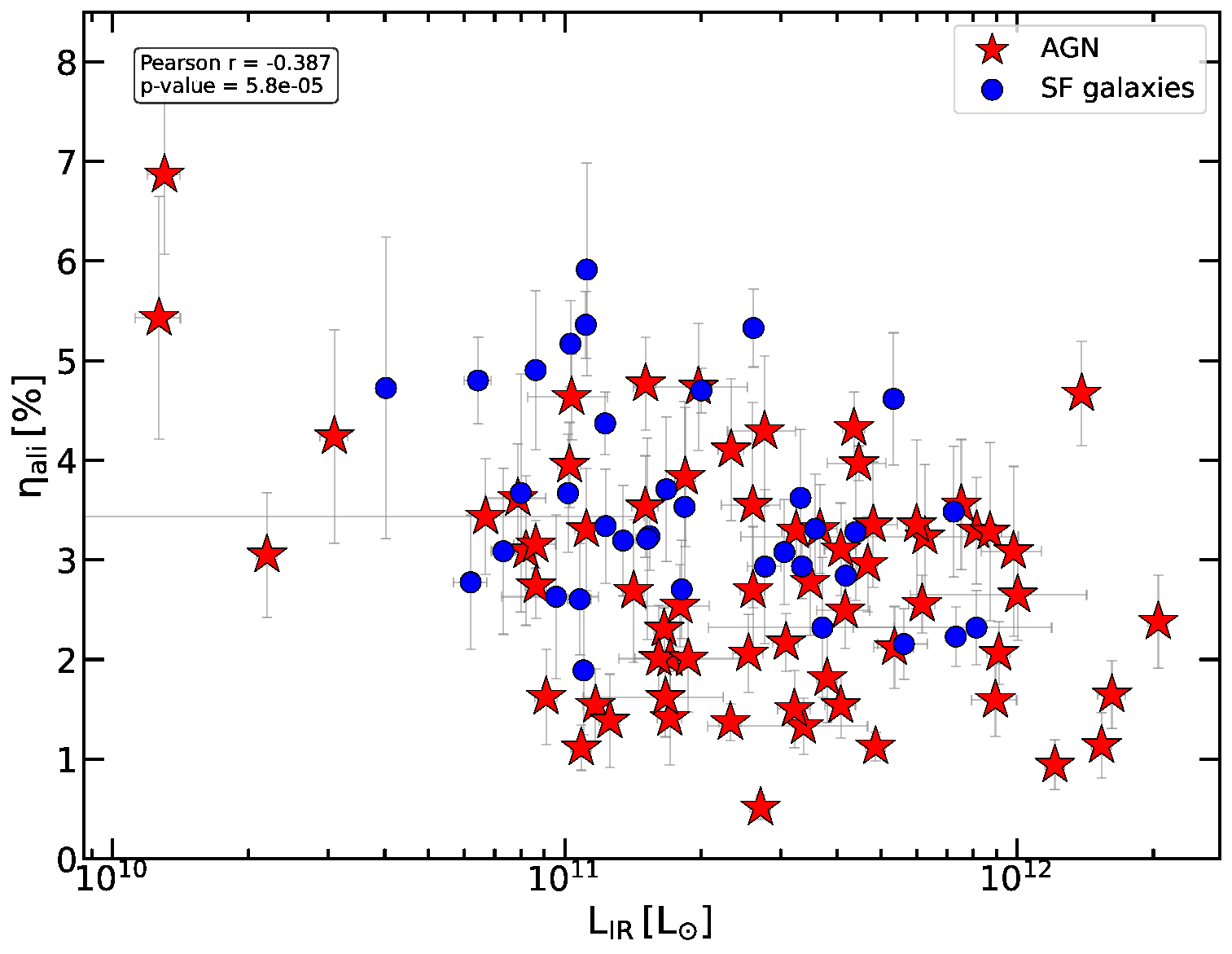}
\includegraphics[width=0.48\textwidth]{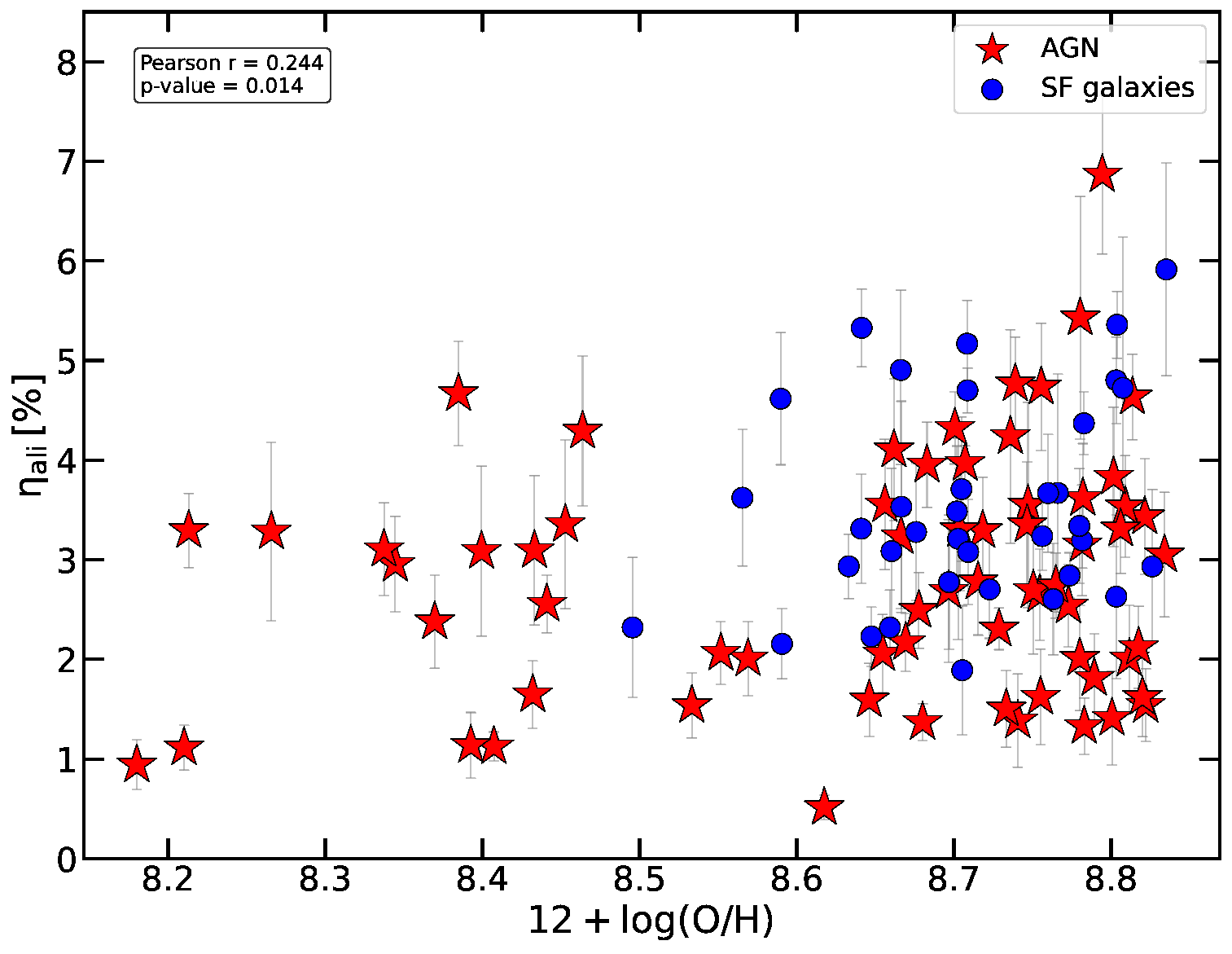}
\includegraphics[width=0.48\textwidth]{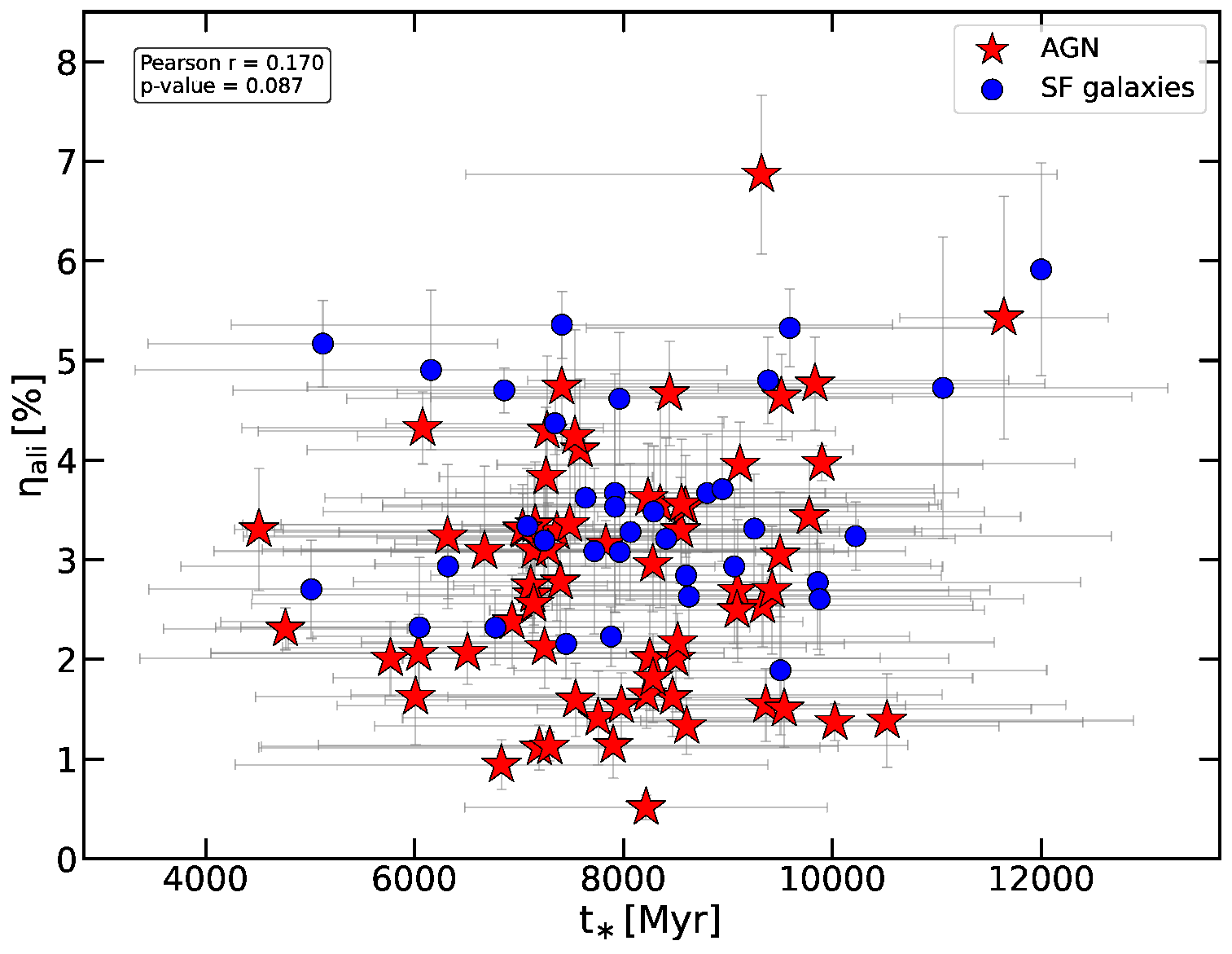}
\includegraphics[width=0.48\textwidth]{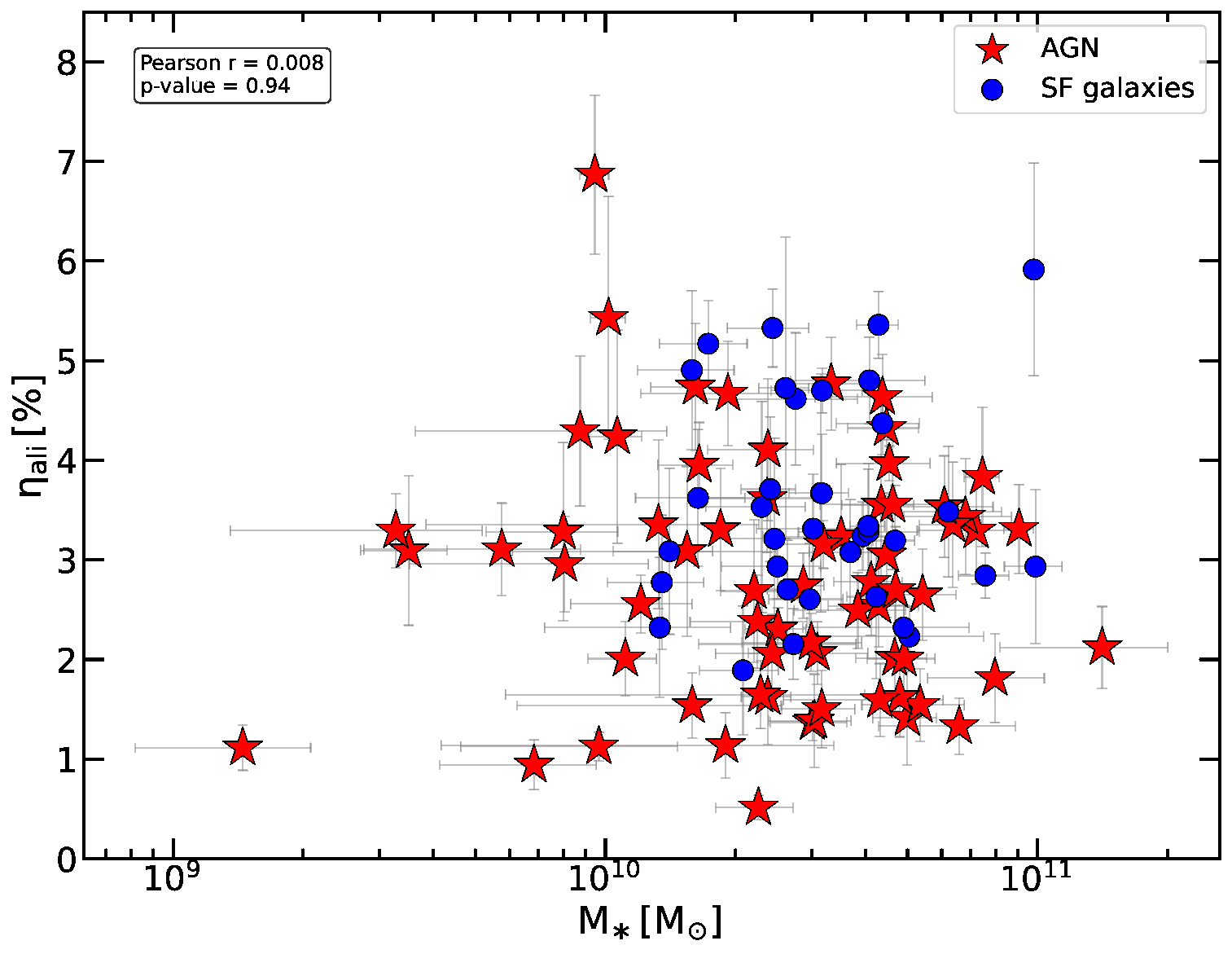}
\caption{\label{fig:fracali}
Comparison of the aliphatic fractions of PAHs
with various galaxy properties.
Pearson correlation coefficients $r$
and statistical significance $p$
are annotated in the top-left corner of each panel.
}
\end{figure*}

The PAH aliphatic fraction $\etaali$ provides
a quantitative diagnostic of hydrocarbon processing.
In Figure~\ref{fig:fracali}, we compare $\etaali$
against various galaxy properties (such as SFR,
stellar mass, and metallicity) and apply Pearson
correlation analysis to uncover their physical connections
in galaxy evolution. Below, we discuss these results.

\paragraph{Star Formation Rate:}
The top-left panel of Figure \ref{fig:fracali}
shows the relation between the SFR and
the aliphatic fraction.
We observed a weak negative correlation,
which was statistically supported by
a  Pearson correlation coefficient of
$r\approx-0.355$ and $p<0.01$.
This trend was also reported by \citet{Lyu2025}
in lower IR luminosity galaxies
at redshifts of $z$\,$\simali$0.2--0.5.
In our exploration of two other galaxy properties
related to SFR, specific SFR
(sSFR, defined as SFR/$M_\bigstar$)
and $\LIR$, similar weak correlations were observed,
both of which passed the Pearson
correlation test ($p<0.01$).

These results suggest that in environments
with a higher SFR, aliphatic features are more
likely suppressed relative to aromatic features.
The birth of young stars is always accompanied
by an increase in UV photons, and in such UV-rich
environments, aliphatic sidegroups will be preferentially
destroyed by high-energy photons.
Laboratory experiments have demonstrated that
aliphatic C--H bonds are more susceptible to UV
photodissociation than their aromatic counterparts
\citep{Marciniak2021}, which corroborates our interpretation.

This negative correlation explains
the elevated aliphatic fraction observed
in Galactic environments compared to
local star-forming galaxies (see Figure~\ref{fig:distri}).
The weaker star formation activity in the Milky Way
results in a weaker UV radiation field,
which allows for the preservation of aliphatic structures.
In contrast, starburst galaxies with intense SFRs
exhibit a higher proportion of aromatic structures,
as the aliphatic components are more easily
destroyed by the intense UV radiation.

\paragraph{Metallicity:}
We examined the relation between the aliphatic
fraction and metallicity and found no statistically
significant correlation, as shown in the middle-right
panel of Figure~\ref{fig:fracali}.
This implies that the PZR trend seen in
\S\ref{subsec:metallicity} is not primarily driven
by UV photodestruction, aligning with the interpretation
proposed by \citet{Whitcomb2024}.
If photodissociation is the primary mechanism
responsible for the depletion of PAHs in
low-metallicity environments, the more UV-sensitive
aliphatic sidegroups should be preferentially destroyed,
leading to a decrease in the aliphatic fraction.
However, if the PZR effect primarily suppresses
PAH growth, both aliphatic and aromatic components
would be affected simultaneously,
which is more consistent with our findings.

\paragraph{Galaxy Age and Stellar Mass:}
Galaxy age ($t_\bigstar$) and stellar mass
($M_\bigstar$) are fundamental properties
directly linked to galaxy evolution and are
therefore likely to be associated with PAH emission.
However, as shown in the bottom panels of
Figure~\ref{fig:fracali}, our analysis reveals no
statistically significant correlation between these
parameters and the aliphatic fraction.
Nevertheless, it should be noted that the uncertainty
in $t_\bigstar$ is large, so it needs to be treated with caution.

\begin{figure}[h!]
\centering
\includegraphics[width=0.6\textwidth]{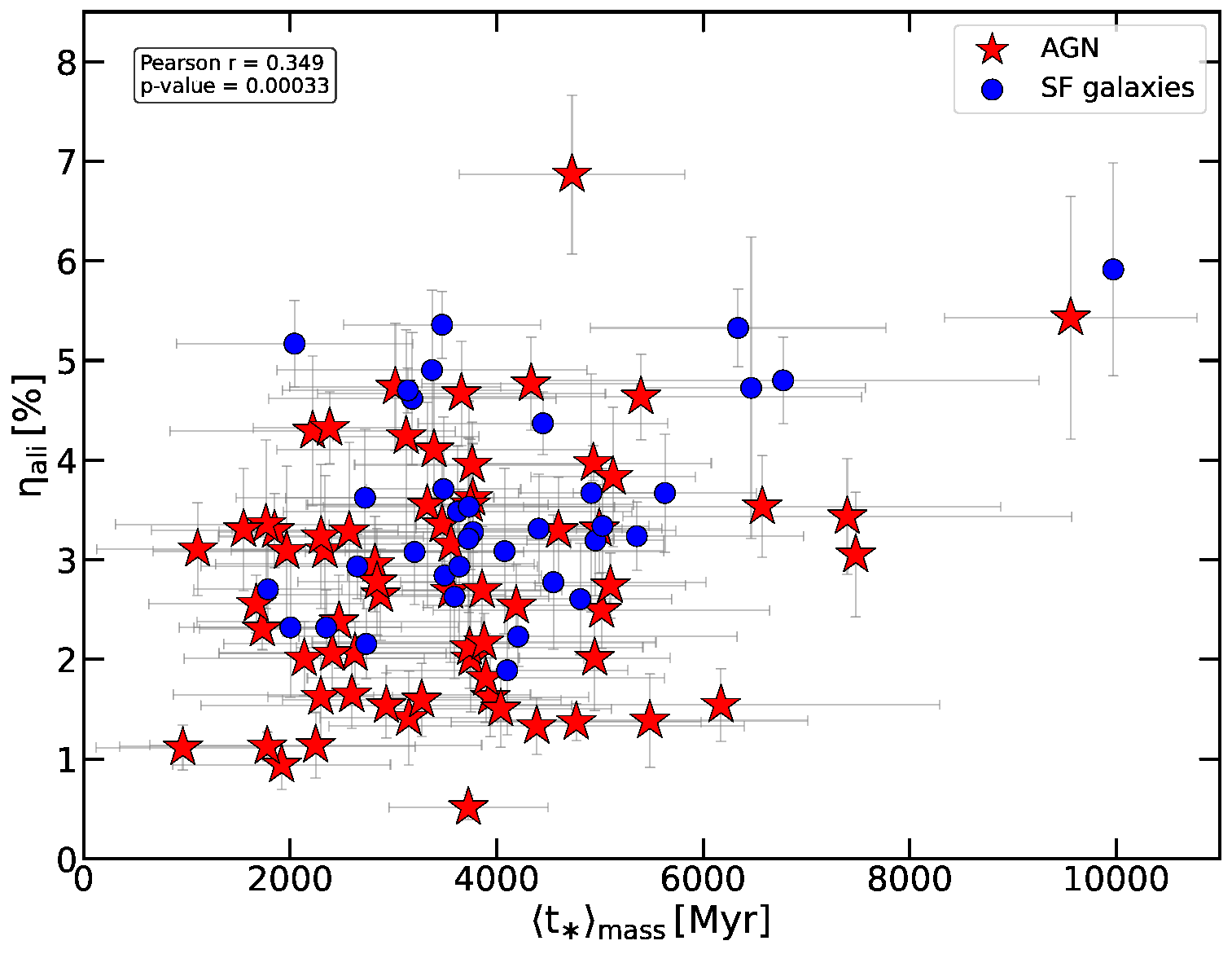}
\vspace{-0.1cm}
\caption{\label{fig:MWage}
The relation between the aliphatic fraction
and the mass-weighted age.
The upper left corner shows
the Pearson correlation test result.
}
\vspace{-0.3cm}
\end{figure}

The mass-weighted galaxy age
($\langle t_\bigstar\rangle_{\mathrm{mass}}$)
is a parameter that provides a comprehensive
representation of star formation history \citep{Conroy2013}.
It is defined as:
\begin{eqnarray}
  \langle t_\bigstar \rangle_{\mathrm{mass}}
= \frac{\sum_{i} t_{i} \times M_{i}}{\sum_{i} M_{i}} ~~,
\end{eqnarray}
where $t_{i}$ and $M_{i}$ are the age and stellar mass
of the $i$-th stellar subpopulation, respectively.
In Figure~\ref{fig:MWage}, we compared the mass-weighted
age with the aliphatic fraction and found a weak correlation,
with a Pearson correlation coefficient of
$r\approx0.349$ and $p<0.01$.
A higher mass-weighted age indicates
that a greater proportion of the galaxy's stellar mass
was formed at an early age, while a lower mass-weighted
age implies more recent active star formation.
Therefore, this correlation provides additional
support for our earlier argument,
suggesting that intense recent star formation
generates abundant UV photons, which in turn
lead to the destruction of aliphatic structures.

\paragraph{AGN Activity:}
In Figures~\ref{fig:fracali} and \ref{fig:MWage},
we distinguish AGN from SF galaxies to investigate
whether nuclear activity influences the correlations
between the aliphatic fraction and galaxy properties.
Clearly, the presence of an AGN does not alter any
observed trends related to the aliphatic fraction.
However, it is noteworthy that galaxies with lower
aliphatic fractions are predominantly  AGN hosts,
which may suggest that high-energy photons
produced by AGN activity contribute to the destruction
of aliphatic structures.

Finally, we note that, while this study
suggests that the aliphatic fraction of PAHs
(or any other UIE carriers) is the result of
hydrocarbon processing, planetary nebulae
actually provide the best sources to study this
because the $P_{3.4}/P_{3.3}$ ratio can be observed
over thousand-year time intervals. Also, the aliphatic
fraction is strongest in protoplanetary nebulae
and young planetary nebulae
(see Figure~\ref{fig:Ali.vs.Teff}),
suggesting that radiative processing
decreases the $P_{3.4}/P_{3.3}$ ratio as
the nebulae evolve.

\section{Summary} \label{sec:summary}
We have presented a comprehensive investigation
of the 3.3$\mum$ aromatic and 3.4$\mum$
aliphatic PAH emission in a sample of 102 local
star-forming galaxies. The PAH spectral characteristics
were quantified through spectral fitting of
the AKARI/IRC near-IR spectroscopy.
We compiled multi-wavelength photometric data
from cross-matched sky surveys and performed
UV-to-IR SED analysis using the CIGALE SED fitting code
to derive fundamental galaxy properties.
Our principal findings are as follows:
\begin{enumerate}
\item Both the 3.3 ($L_{3.3}$) and 3.4$\mum$ ($L_{3.4}$)
PAH emission exhibit a strong correlation with $\LIR$.
\item The 3.3$\mum$ PAH emission was shown to
be a reasonably accurate calibrator of the SFR,
in agreement with earlier studies.
This agreement highlights the robustness
of the 3.3$\mum$ PAH emission as a SFR tracer.
\item The PAH aliphatic fraction ($\etaali$) ranges
from 0.38\% to 6.8\% (median 3.1\%), consistent
with that of Galactic PDRs, reflection nebulae,
and planetary nebulae, appreciably lower than that
of protoplanetary nebulae excited by UV-poor stars.
The aliphatic fraction exhibits a weak negative
correlation with SFR, sSFR, and $\LIR$,
suggesting that the UV photons produced by
the intense star formation activity lead to
the destruction of aliphatic structures.
\item $L_{3.3}/\LIR$ and $L_{3.4}/\LIR$ exhibit
a dependence on metallicity, remaining constant
at ${\rm 12\,+\,\log(O/H) \simgt 8.5}$.
In contrast, the aliphatic fraction shows no correlation
with metallicity, suggesting that photodestruction
is unlikely to be the primary driver of this metallicity dependence.
\item While the aliphatic fraction shows no correlation
with galaxy age or stellar mass, it correlates with
mass-weighted galaxy age. This may arise from recent
star formation activity, which suppresses aliphatic PAHs,
highlighting its potential
as a tracer of galactic evolutionary processes.
\item Galaxies with low aliphatic fractions
are predominantly AGN hosts, likely due to AGN-driven
high-energy radiation, which dissociates the aliphatic
units in PAHs. However, AGN's presence does not
systematically alter the observed correlations
between the aliphatic fraction and other galactic properties.
\end{enumerate}

Our study unveils the hidden physical insights
of the shortest-wavelength PAH features,
particularly the 3.4$\mum$ aliphatic feature.
As JWST opens a new window into the early universe,
these findings will offer valuable perspectives
for future exploration.

\acknowledgments
We thank the anonymous referee,
J.W.~Lyu,  B.~Yang, and C.C.~Zhang
for helpful comments and suggestions.
JHP and XJY are supported in part by NSFC\,12333005
and 12122302, CMS-CSST-2021-A09,
and the Innovative Research Group Project
of Natural Science Foundation of Hunan Province
of China No. 2024JJ1008.
This research has made use of the SIMBAD database,
operated at CDS, Strasbourg, France \citep{Wenger2000}.
This research makes use of data from the NASA/IPAC
Extragalactic Database (NED) and NASA/IPAC
Infrared Science Archive (IRSA),
operated by JPL/California Institute of Technology
under contract with the National Aeronautics and Space Administration.

This research is based on observations made with the Galaxy Evolution Explorer, obtained from the
MAST data archive at the Space Telescope Science Institute, which is operated by the Association of
Universities for Research in Astronomy, Inc., under NASA contract NAS 5–26555.
Swift UVOT was designed and built in collaboration between MSSL, PSU, SwRI, Swales Aerospace, and
GSFC, and was launched by NASA.

The Pan-STARRS1 Surveys (PS1) and the PS1 public science archive have been made possible through
contributions by the Institute for Astronomy, the University of Hawaii, the Pan-STARRS Project
Office, the Max-Planck Society and its participating institutes, the Max Planck Institute for
Astronomy, Heidelberg and the Max Planck Institute for Extraterrestrial Physics, Garching, The
Johns Hopkins University, Durham University, the University of Edinburgh, the Queen's University
Belfast, the Harvard-Smithsonian Center for Astrophysics, the Las Cumbres Observatory Global
Telescope Network Incorporated, the National Central University of Taiwan, the Space Telescope
Science Institute, the National Aeronautics and Space Administration under Grant No. NNX08AR22G
issued through the Planetary Science Division of the NASA Science Mission Directorate, the National
Science Foundation Grant No. AST-1238877, the University of Maryland, Eotvos Lorand University (ELTE),
the Los Alamos National Laboratory, and the Gordon and Betty Moore Foundation.

The national facility capability for SkyMapper has been funded through ARC LIEF grant
LE130100104 from the Australian Research Council, awarded to the University of Sydney, the
Australian National University, Swinburne University of Technology, the University of Queensland,
the University of Western Australia, the University of Melbourne, Curtin University of
Technology, Monash University and the Australian Astronomical Observatory. SkyMapper is owned
and operated by The Australian National University's Research School of Astronomy and
Astrophysics. The survey data were processed and provided by the SkyMapper Team at ANU.
The SkyMapper node of the All-Sky Virtual Observatory (ASVO) is hosted at the National
Computational Infrastructure (NCI). Development and support of the SkyMapper node of the ASVO
has been funded in part by Astronomy Australia Limited (AAL) and the Australian Government
through the Commonwealth's Education Investment Fund (EIF) and National Collaborative Research
Infrastructure Strategy (NCRIS), particularly the National eResearch Collaboration Tools and
Resources (NeCTAR) and the Australian National Data Service Projects (ANDS).

Funding for the Sloan Digital Sky Survey V has been provided by the Alfred P. Sloan Foundation,
the Heising-Simons Foundation, the National Science Foundation, and the Participating Institutions.
SDSS acknowledges support and resources from the Center for High-Performance Computing at the University
of Utah. SDSS telescopes are located at Apache Point Observatory, funded by the Astrophysical Research
Consortium and operated by New Mexico State University, and at Las Campanas Observatory, operated by
the Carnegie Institution for Science. The SDSS web site is \url{www.sdss.org}.
SDSS is managed by the Astrophysical Research Consortium for the Participating Institutions of the SDSS
Collaboration, including Caltech, the Carnegie Institution for Science, Chilean National Time Allocation
Committee (CNTAC) ratified researchers, The Flatiron Institute, the Gotham Participation Group, Harvard
University, Heidelberg University, The Johns Hopkins University, L’Ecole polytechnique fédérale de
Lausanne (EPFL), Leibniz-Institut für Astrophysik Potsdam (AIP), Max-Planck-Institut für Astronomie
(MPIA Heidelberg), Max-Planck-Institut für Extraterrestrische Physik (MPE), Nanjing University, National
Astronomical Observatories of China (NAOC), New Mexico State University, The Ohio State University,
Pennsylvania State University, Smithsonian Astrophysical Observatory, Space Telescope Science Institute
(STScI), the Stellar Astrophysics Participation Group, Universidad Nacional Autónoma de México,
University of Arizona, University of Colorado Boulder, University of Illinois at Urbana-Champaign,
University of Toronto, University of Utah, University of Virginia, Yale University, and Yunnan
University.

This publication makes use of data products from the Two Micron All Sky Survey, which is a joint
project of the University of Massachusetts and the Infrared Processing and Analysis
Center/California Institute of Technology, funded by the National Aeronautics and Space
Administration and the National Science Foundation.
This publication makes use of data products from the Wide-field Infrared Survey Explorer, which
is a joint project of the University of California, Los Angeles, and the Jet Propulsion
Laboratory/California Institute of Technology, funded by the National Aeronautics and Space
Administration.
This research is based on observations with AKARI, a JAXA project with the participation of ESA.

\appendix

\section{SED-derived Physical Properties}

After constructing the model grid,
CIGALE performs $\chi^2$ minimization
against the observational data, computes
the Bayesian posterior probability for each model,
and derives the final parameter estimates
through Bayesian-weighted averaging.
Table~\ref{tabA:SEDresult} shows the physical
properties of our samples derived from the SED decomposition.

\section{Spectral Fitting Results}

We present the results of the spectral fitting in Table \ref{tabB:fitresult}.
The uncertainty of the PAH feature intensities is estimated based on Monte Carlo simulations.

\section{Mock Analysis}

CIGALE provides the ability to evaluate constraints on physical properties by analyzing mock catalogues.
Based on the best-fit parameters for each object, the program perturbs the values by adding
random noise drawn from a Gaussian distribution with a standard deviation equal to the
observational uncertainty.
This mock catalogue is a valuable reference for evaluating the reliability of inferred physical properties.

\begin{figure*}[h!]
\centering
\includegraphics[width=0.48\textwidth]{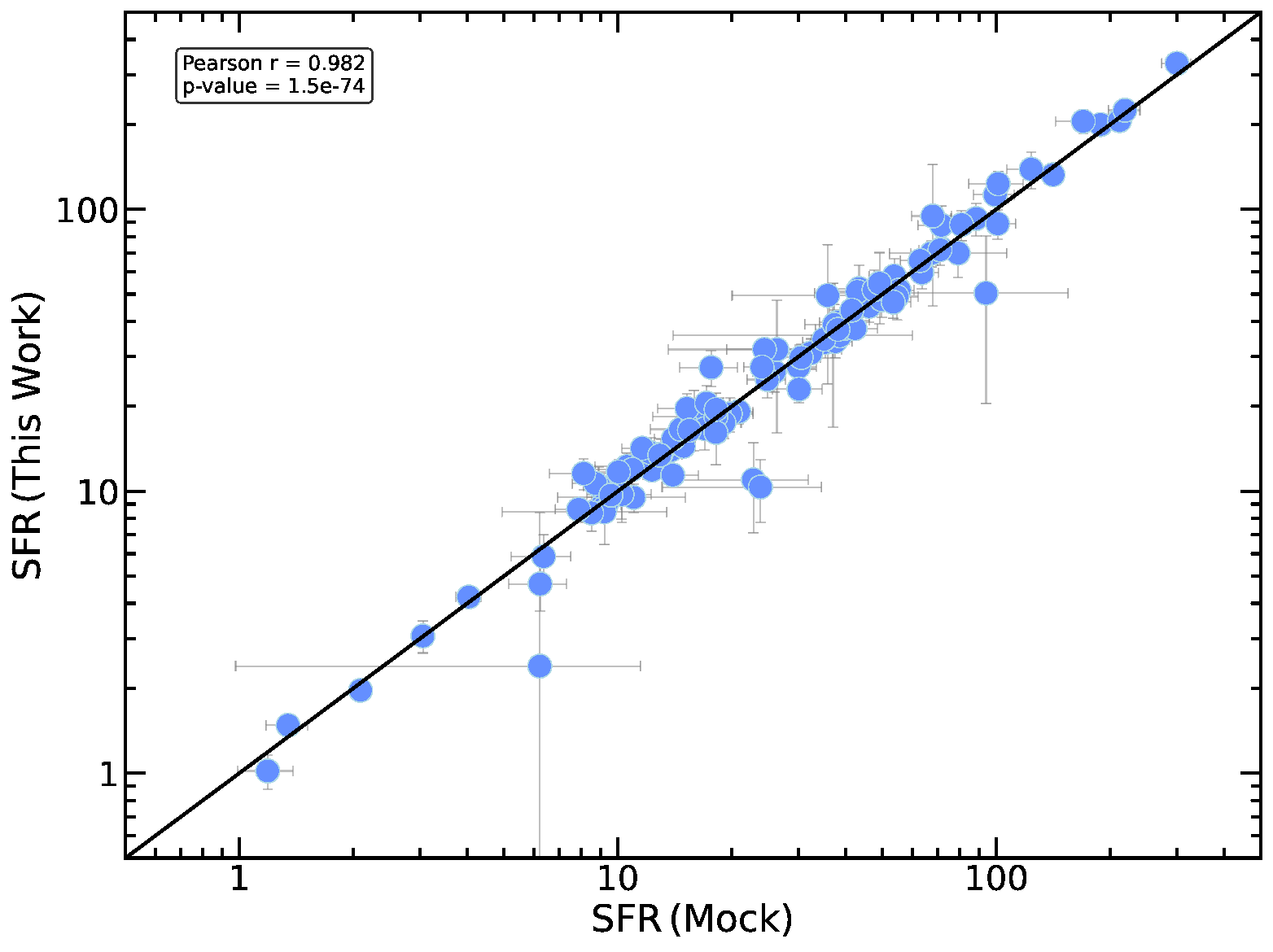}
\includegraphics[width=0.48\textwidth]{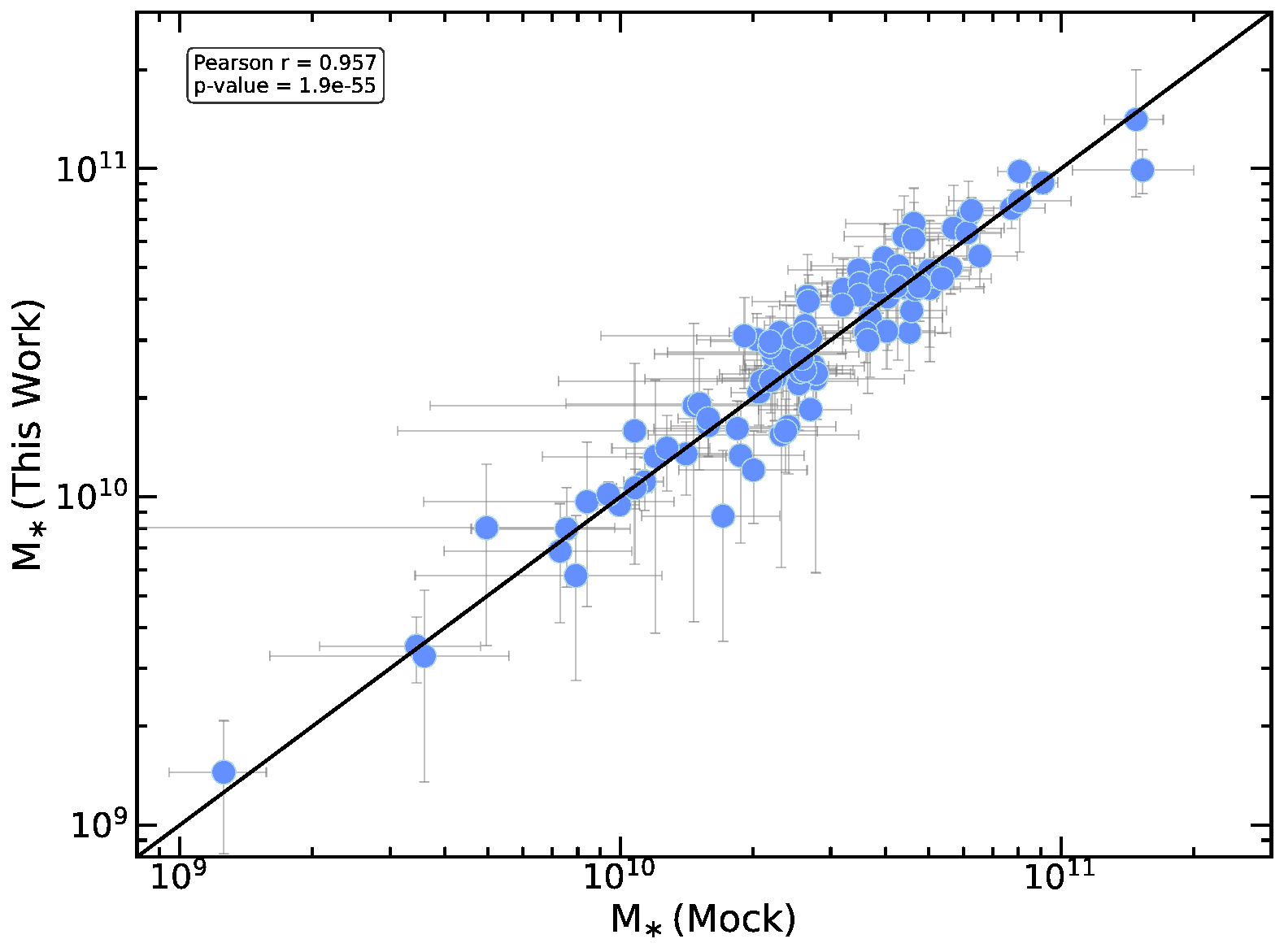}
\includegraphics[width=0.48\textwidth]{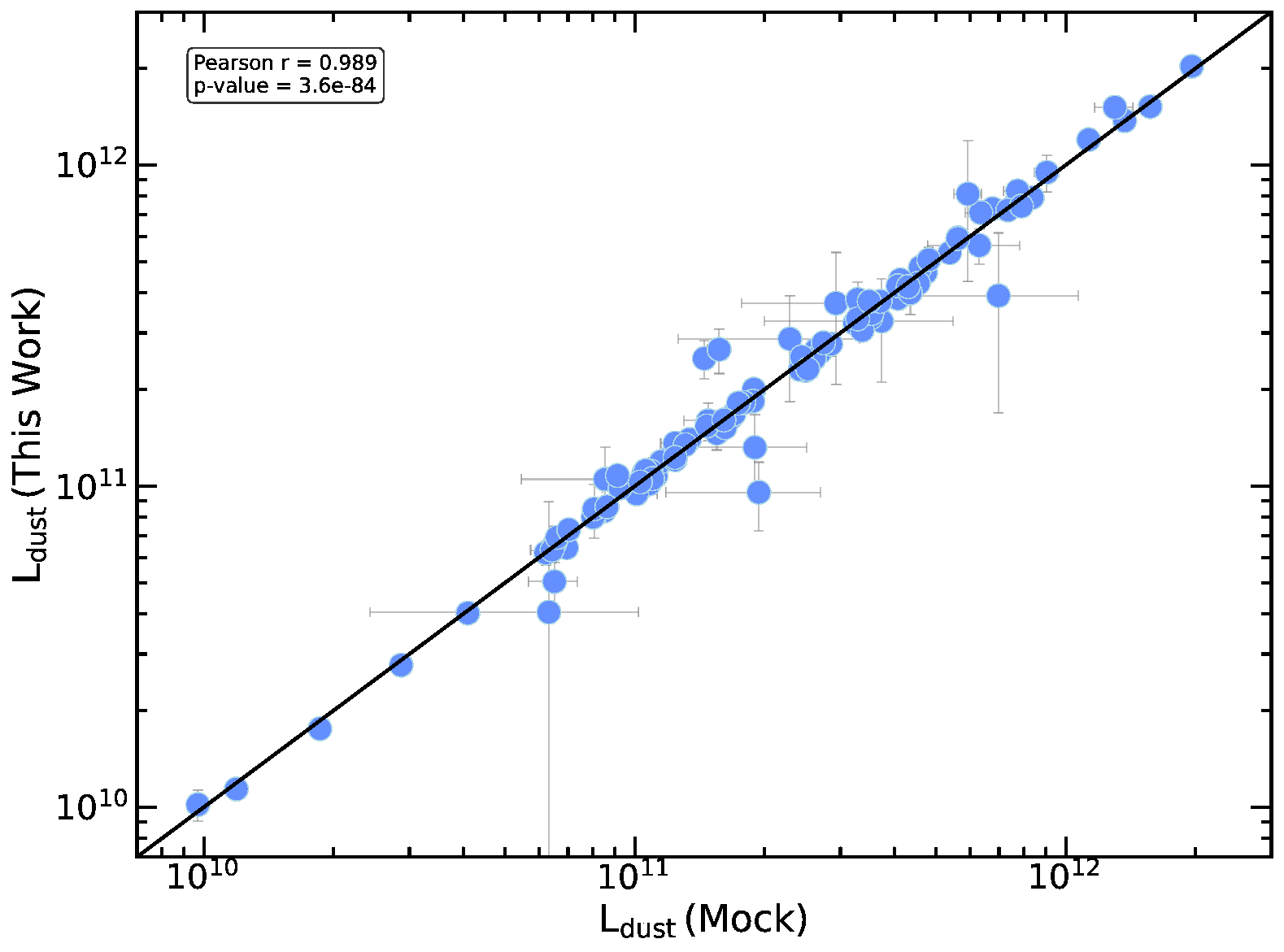}
\hspace*{0.2cm}
\includegraphics[width=0.48\textwidth]{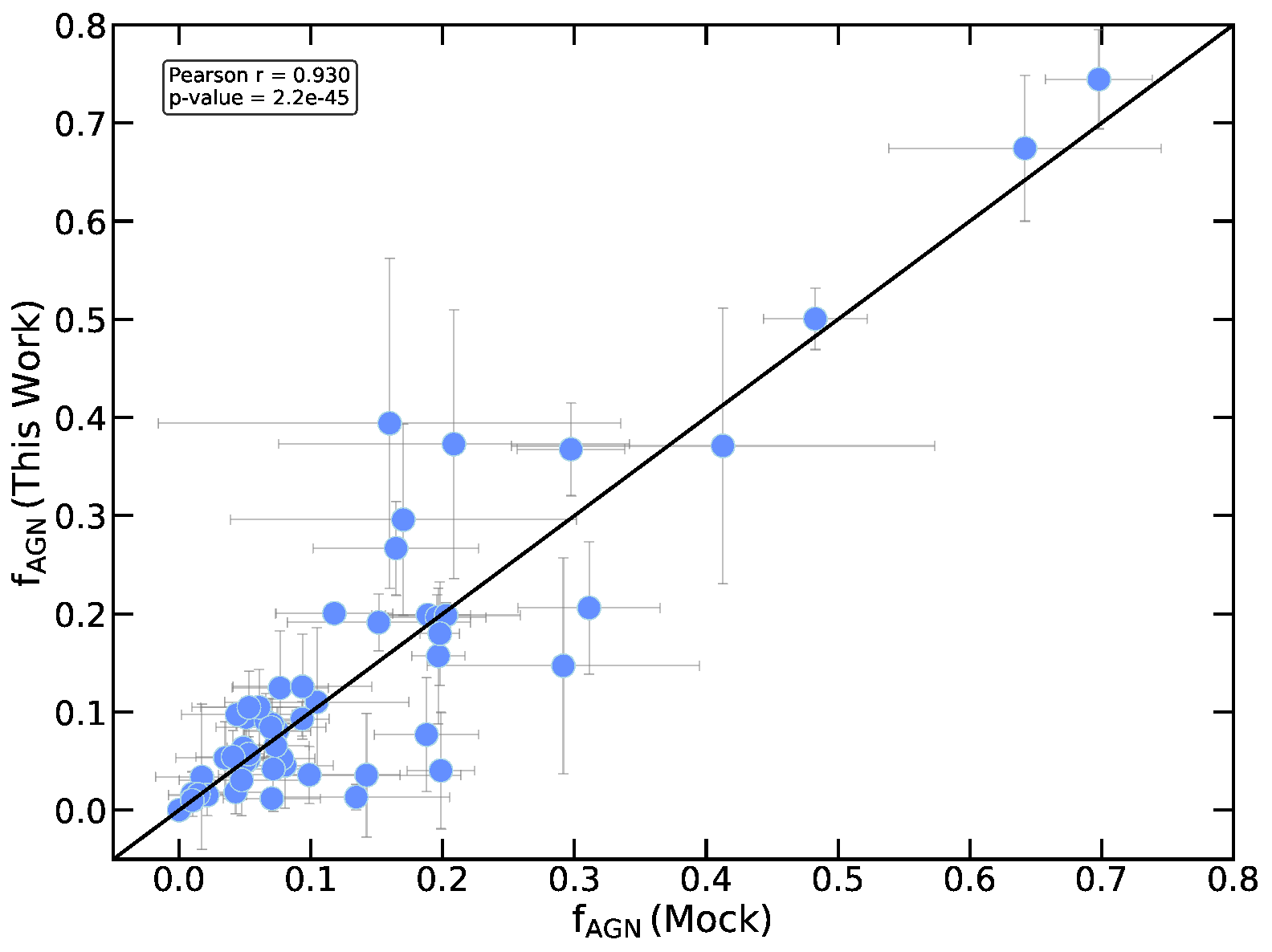}
\includegraphics[width=0.48\textwidth]{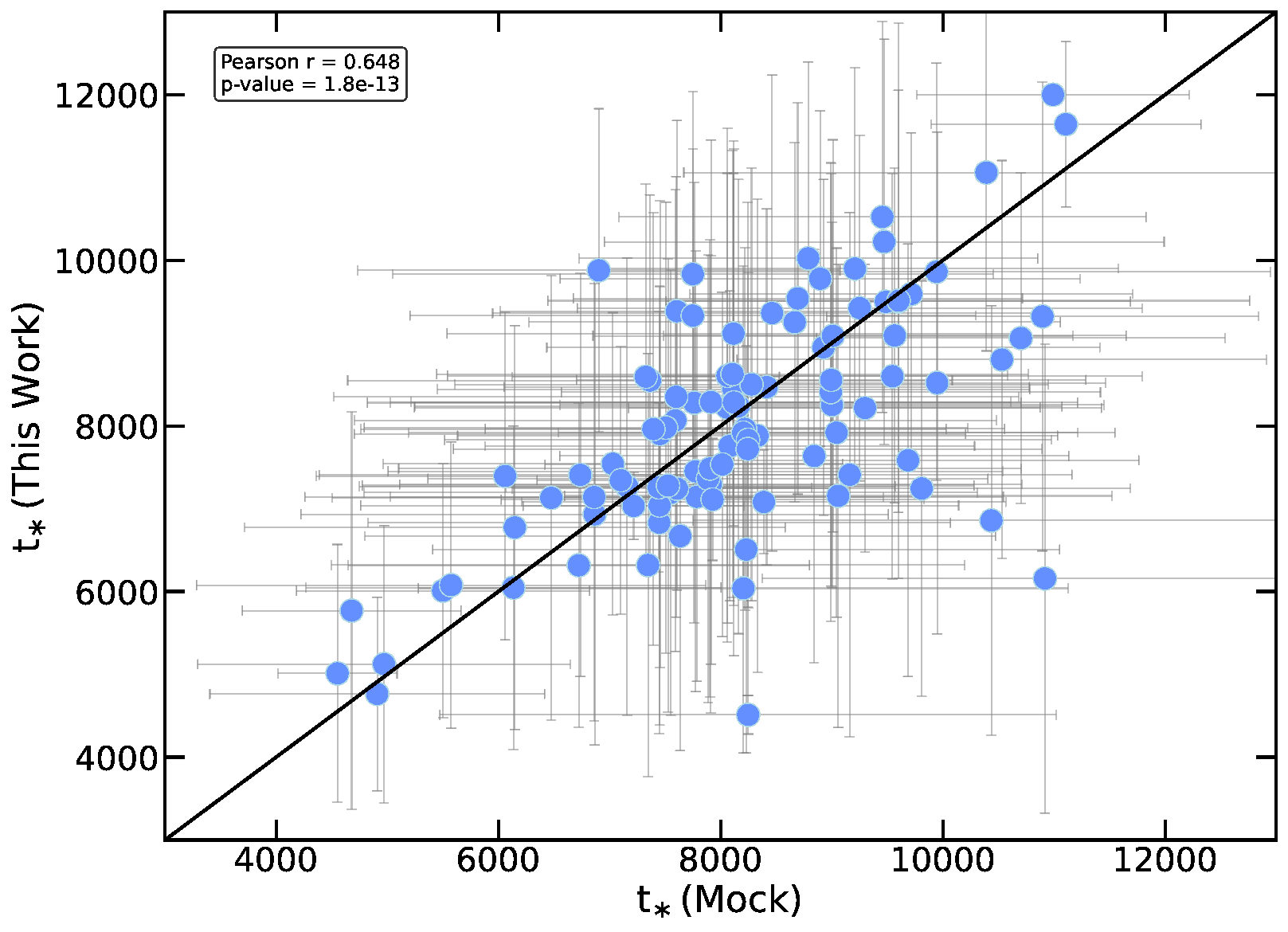}
\includegraphics[width=0.48\textwidth]{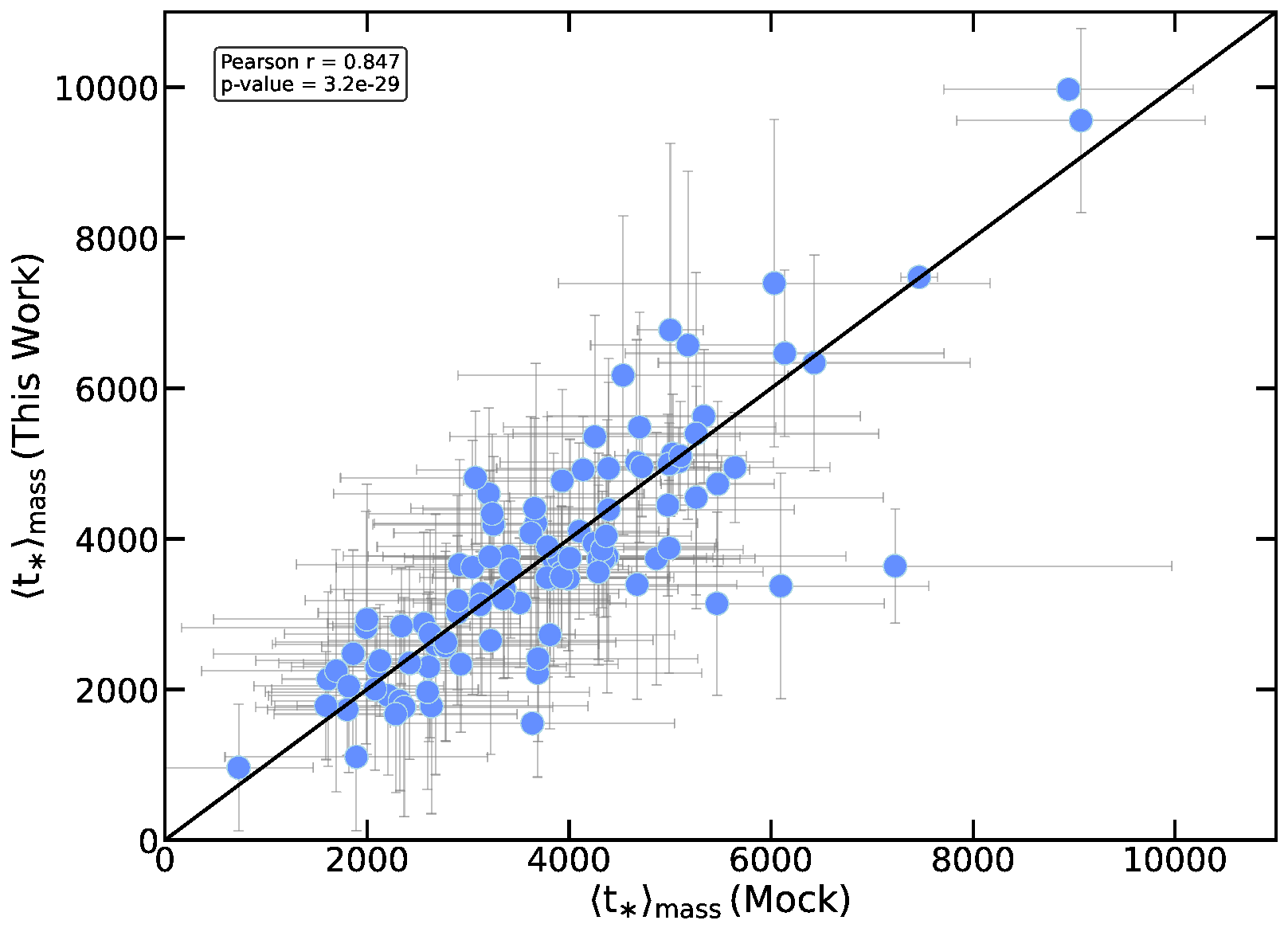}
\hspace*{0.2cm}
\caption{Comparison between SED-derived parameters ($y$-axis) and those obtained from the mock analysis ($x$-axis). The black solid line represents the $y = x$ relation, and the Pearson correlation coefficient is displayed in the upper left corner of each panel.\label{figC:mock}}
\end{figure*}

Figure \ref{figC:mock} compares parameter estimates derived from mock analysis with those obtained
through Bayesian inference.
All parameters except $t_{\bigstar}$ exhibit strong correlations (Pearson r $>$ 0.8), demonstrating the
reliability of SED fitting methodology.
The weaker correlation and larger uncertainties in $t_{\bigstar}$ indicate lower accuracy in its estimation.
However, since we do not find any significant correlations involving $t_{\bigstar}$, we conclude that this
does not impact our results.

\section{Comparison with Other SED Studies}

The local LIRG sample used in this study has been previously analyzed by other researchers
using different software or SED fitting parameters \citep{Yamada2023, Shangguan2019}.
Here, we compare our SED fitting results with those from previous studies to assess any
potential discrepancies.

\begin{figure*}[h!]
\centering
\includegraphics[width=0.48\textwidth]{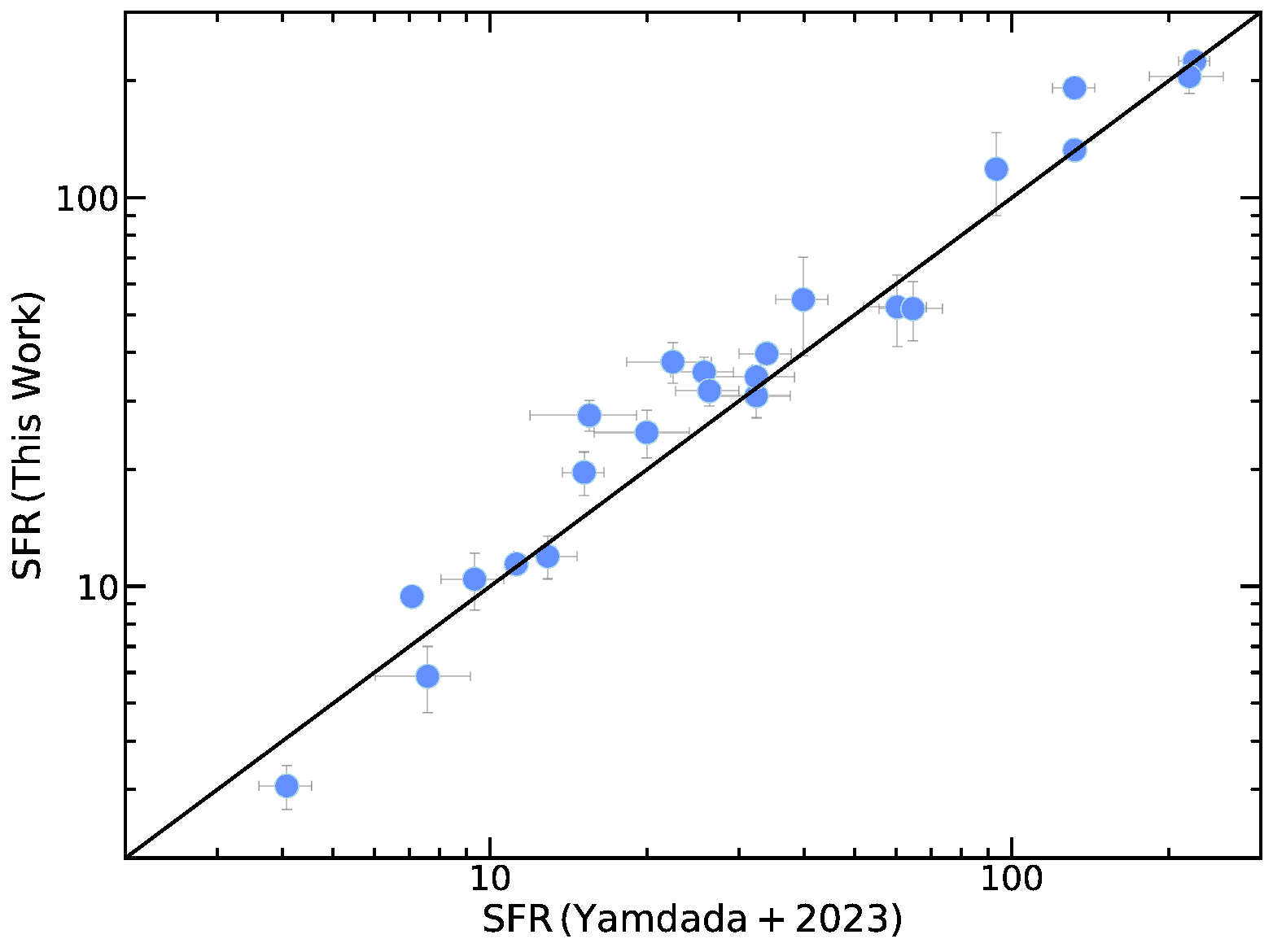}
\includegraphics[width=0.48\textwidth]{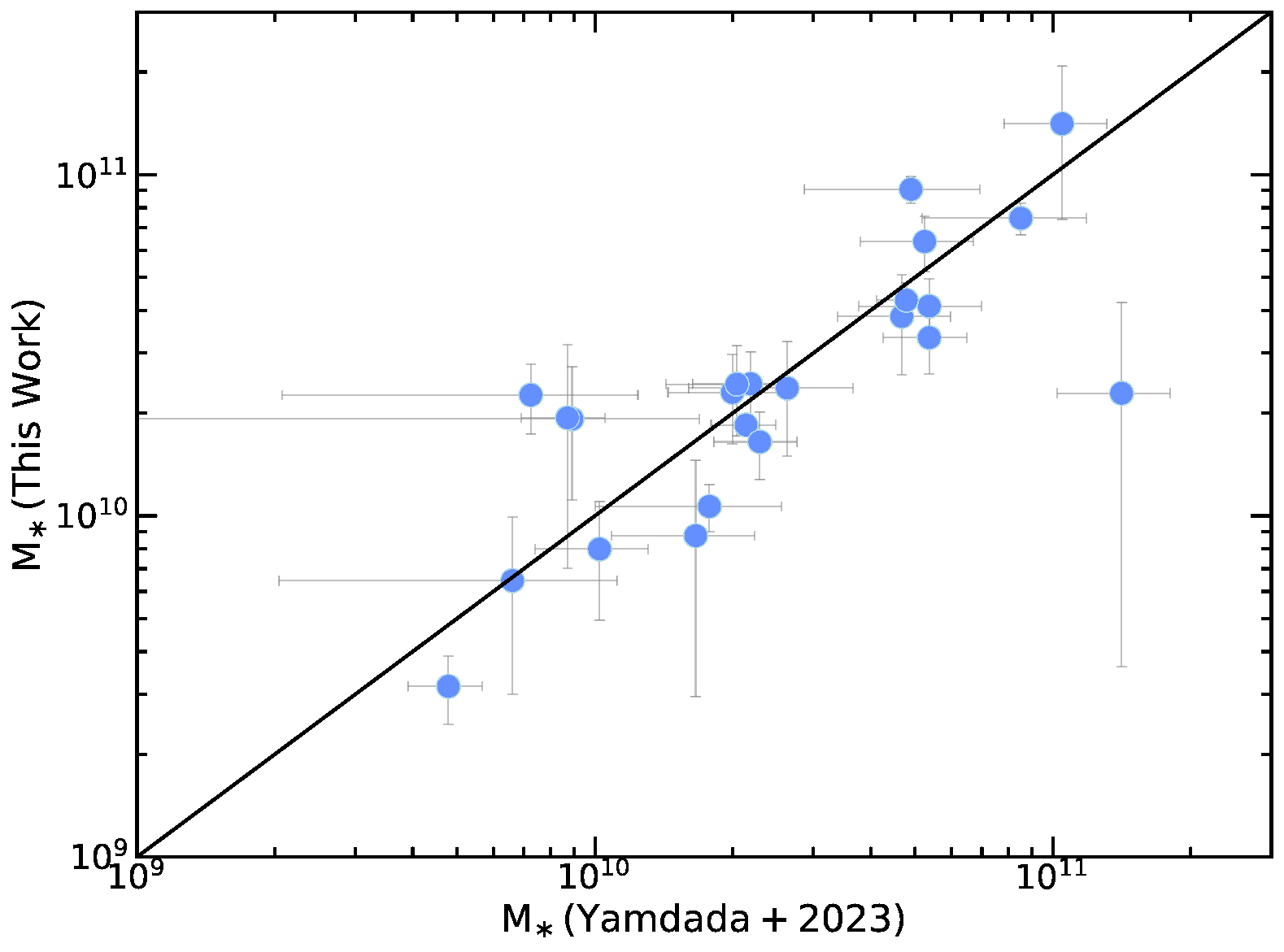}
\includegraphics[width=0.48\textwidth]{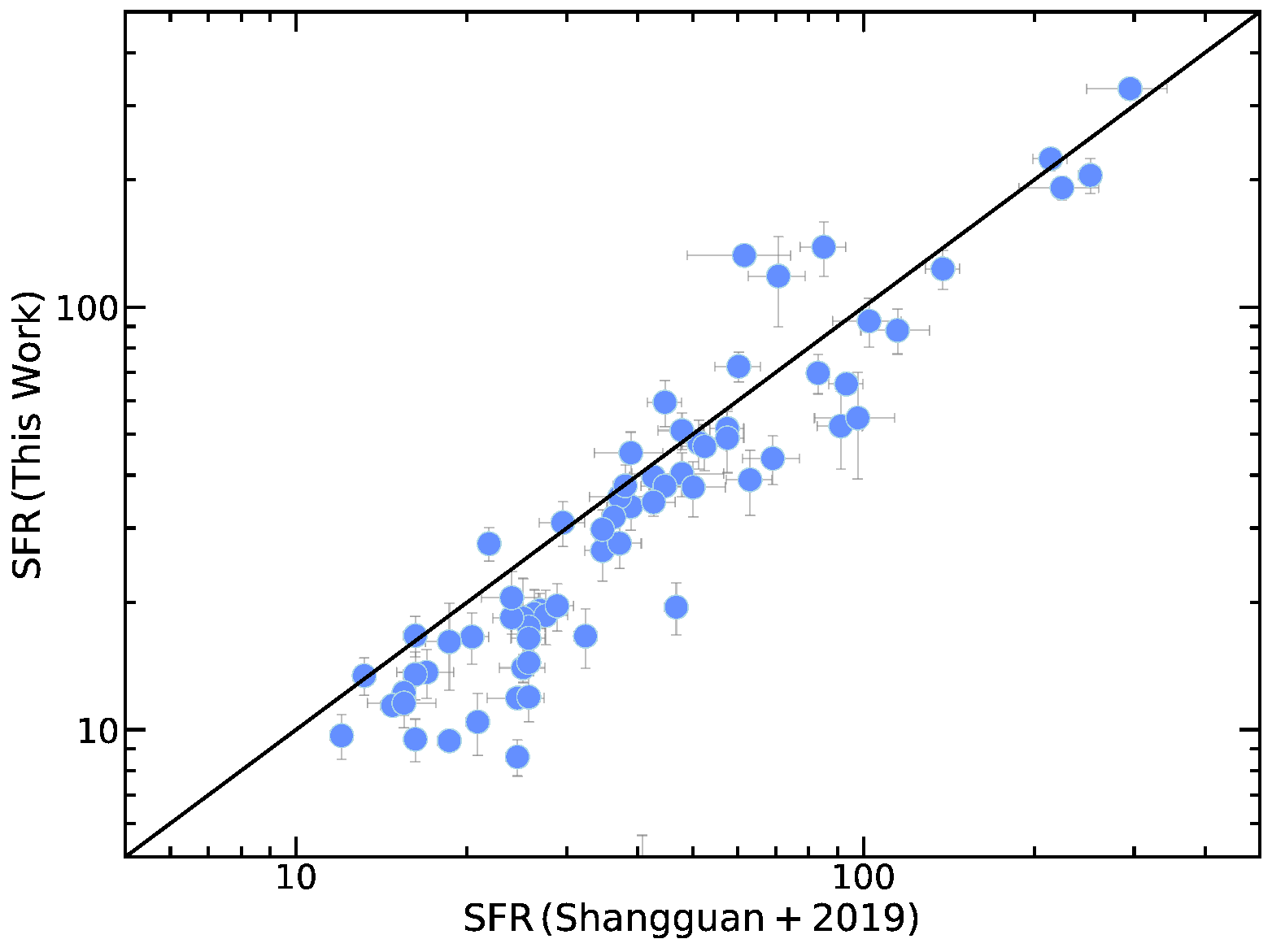}
\includegraphics[width=0.48\textwidth]{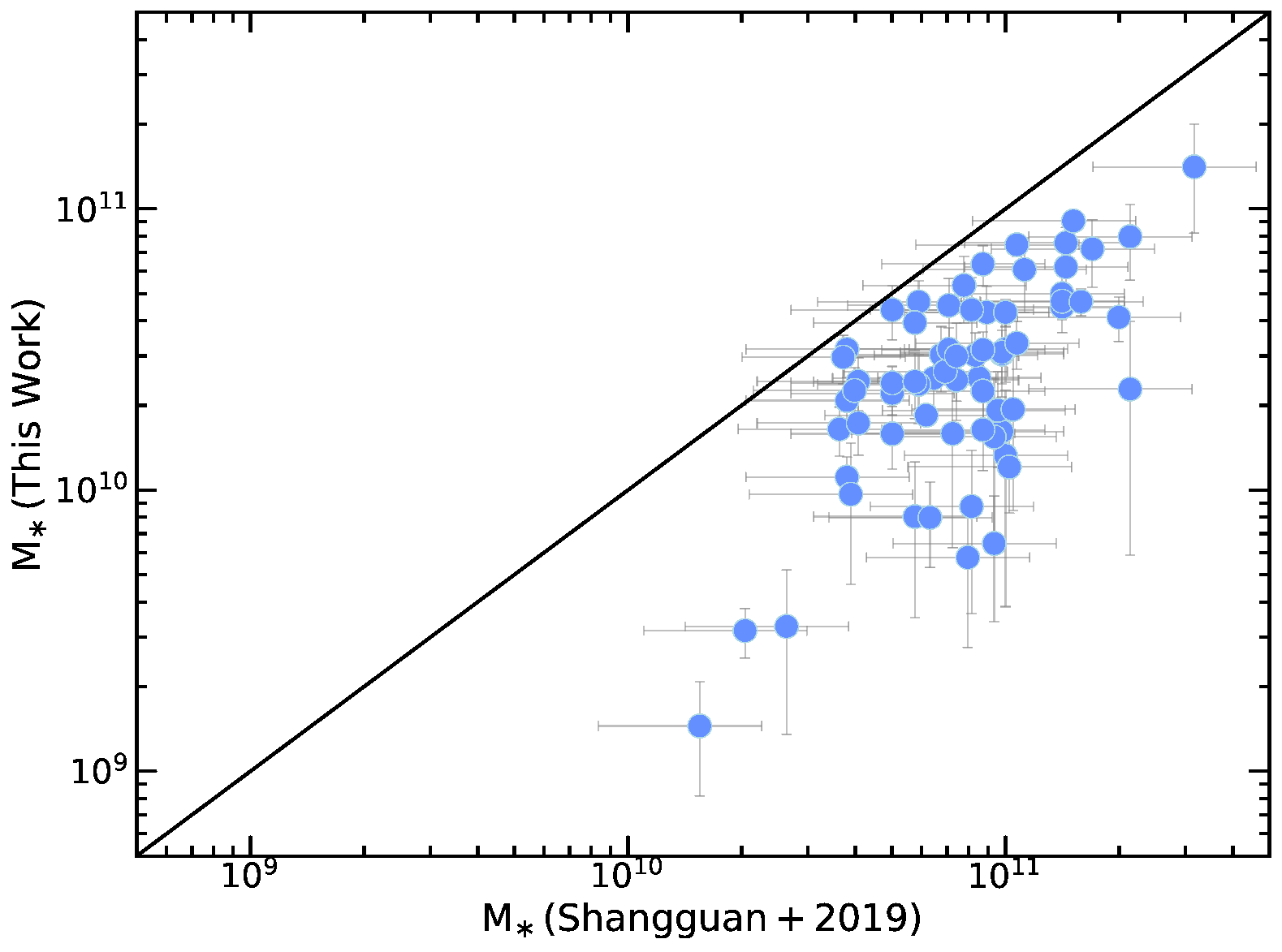}
\caption{Comparison with other SED studies. \label{figD:check}}
\end{figure*}

The top panel of Figure \ref{figD:check} demonstrates excellent consistency between our SFR and $M_{\bigstar}$
with those reported in \citet{Yamada2023}.
In contrast, the bottom panel reveals a systematic offset in $M_{\bigstar}$ estimates when
compared to \citet{Shangguan2019}, although SFR measurements are basically consistent with ours.
This $M_{\bigstar}$ discrepancy mirrors findings in \citet{Yamada2023}, which they explained
as being caused by the lack of optical photometric data.
This absence leads to an overestimation of the fraction of old stellar populations, reducing
the average stellar luminosity and consequently resulting in an overestimated total
stellar mass.

\section{Multiwavelength Photometry}

Table \ref{tabE:photometry1} and \ref{tabE:photometry2} present compiled photometric data from the
ultraviolet to the infrared, converted into flux density units in mJy.
The tabulated fluxes remain uncorrected for Galactic extinction, while $E(B-V)$ values from \citetalias{Schlegel1998}.
are provided for reference purposes.

\clearpage

%%% Figure 1 %%%
% This image is the sample figure of the figure set below.
%%% Figure 1 %%%

%%% Figure 2 %%%
%%% Figure 2 %%%

%%% Fugure 3 %%%
%%% Fugure 3 %%%

%%% Figure 4 %%%
%%% Figure 4 %%%

%%% Figure 5 *new* %%%
%%% Figure 5 *new* %%%

%%% Figure 6 %%%
%%% Figure 6 %%%

%%% Figure 7 %%%
%%% Figure 7 %%%

%%% Figure 8 %%%
%%% Figure 8 %%%

%%% Figure 9 %%%
%%% Figure 9 %%%

%%% Figure 10 %%%
%%% Figure 10 %%%

%%% Figure 11 %%%
%%% Figure 11 %%%

%%% Figure 12 %%%
%%% Figure 12 %%%

\clearpage

%%% Table 1 %%%
\begin{table}[h!]
\begin{center}
\footnotesize
\caption{Observation Programs\label{tab:programs}}
\begin{tabular}{cccc}
\hline
Program & Phase & Number of galaxies & Targets \\
\hline
AGNUL & 1-2 & 67 & LIRG \& ULIRG \\
AGNUL & 3 & 265 & LIRG \& ULIRG \\
AMUSE & 1-2 & 28 & $F_{24\mum} >5\mJy$ \\
BRSFR & 3 & 32 & $L_{\rm IR} < 10^{11}\Lsun$ \\
COABS & 3 & 1 & Bright Seyfert 2 \\
DTIRC & 3 & 83 & Data check \\
GOALS & 3 & 99 & LIRG \\
ISBEG & 3 & 37 & Blue early type \\
MSAGN & 3 & 141 & AKARI/mid-IR AGN \\
NULIZ & 1-2 & 26 & ULIRG \\
QSONG & 3 & 242 & QSO \\
SYDUS & 3 & 14 & Seyfert 1, 2 \\
\hline
\end{tabular}
\smallskip
\\
{\footnotesize
The target type comes from \citet{Murata2017}.
}
\vspace{-0.6cm}
\end{center}
\end{table}
%%% Table 1 %%%

%%% Table 2 %%%
\begin{landscape}
\begin{table*}[h!]
\centering
\scriptsize
\caption{Model Parameters for CIGALE SED Fitting\label{tab:parameters}}
\begin{tabular}{llc}
\hline
Model & Parameter & Values \\
\hline
Star-formation History ($\mathtt{sfhdelayedbq}$\textsuperscript{[i]}) & $\mathtt{tau\_main}$ (e-folding time) [Myr] & 1000, 3000, 5000, 10000\\
 & $\mathtt{age\_main}$ (Galaxy age) [Myr] & 4500, 7000, 9500, 12000\\
 & $\mathtt{age\_bq}$ (Age of the burst/quench episode) [Myr] & 10, 20, 30, 100\\
 & $\mathtt{r\_sfr}$ (Ratio of the SFR after/before $\mathtt{age\_bq}$) & 1, 3.16, 10, 31.6, 100, 1000\\
\hline
Stellar Population ($\mathtt{bc03}$\textsuperscript{[ii]}) & $\mathtt{imf}$ & 1 (Chabrier\textsuperscript{[iii]})\\
  & $\mathtt{metallicity}$ & 0.02 (solar metallicity)\\
\hline
Nebular Emission ($\mathtt{nebular}$\textsuperscript{[iv]}) & $\mathtt{logU}$ (Ionisation parameter) & -3.0\\
  & $\mathtt{zgas}$ (Gas metallicity) & 0.02 (solar metallicity)\\
  & $\mathtt{f\_esc}$ (Lyman continuum photon escape fraction) & 0.0\\
  & $\mathtt{f\_dust}$ (Lyman continuum photon absorb fraction) & 0.0\\
\hline
Attenuation Law ($\mathtt{dustatt\_modified\_starburst}$\textsuperscript{[v]}) & $\mathtt{E\_BV\_lines}$ (colour excess of the nebular lines light) & 0.0, 0.1, 0.2, 0.3, 0.4, 0.5\\
  & $\mathtt{E\_BV\_factor}$ (Reduction factor to apply on $\mathtt{E\_BV\_lines}$) & 0.44\\
  & $\mathtt{uv\_bump\_wavelength}$ [nm] & 217.5\\
  & $\mathtt{uv\_bump\_width}$ [nm] & 35.0\\
  & $\mathtt{uv\_bump\_amplitude}$ & 0.0\\
  & $\mathtt{powerlaw\_slope}$ & -0.8, -0.4, 0.0\\
  & $\mathtt{Ext\_law\_emission\_lines}$ & 1 (Milky Way\textsuperscript{[vi]})\\
  & $\mathtt{Rv}$ & 3.1\\
\hline
Dust Emission ($\mathtt{dl2014}$\textsuperscript{[vii]}) &
                                                           $q_{\rm pah}$ (Mass fraction of PAH) & 0.47, 1.12, 2.50, 4.60\\
  & $\mathtt{umin}$ (Minimum radiation field) & 0.1-50.0 (all possible values)\\
  & $\mathtt{alpha}$ (Power law index of the radiation field) & 2.0, 2.5\\
  & $\mathtt{gamma}$ (Fraction illuminated from $U_{\rm min}$ to
    $U_{\rm max}$) & 0.02, 0.15\\
\hline
Active Galactic Nucleus ($\mathtt{skirtor2016}$\textsuperscript{[viii]}) & $\mathtt{t}$ (Average edge-on optical depth at 9.7$\mum$) & 5, 9, 11\\
  & $\mathtt{pl}$ (Torus density radial parameter) & 1.0\\
  & $\mathtt{q}$ (Torus density angular parameter) & 1.0\\
  & $\mathtt{oa}$ (Angle between the equatorial plane and torus edge) & 40\\
  & $\mathtt{R}$ (Ratio of outer to inner radius) & 20\\
  & $\mathtt{Mcl}$ (Fraction of total dust mass inside clumps) & 0.97\\
  & $\mathtt{i}$ (Viewing angle) & 30 (type 1), 70 (type 2)\\
  & $\mathtt{fracAGN}$ & 0.0 (for SF galaxies)\\
  & & or 0.01, 0.05, 0.1, 0.2, 0.3, 0.4,\\
  & & 0.5, 0.6, 0.7, 0.8 (for AGNs)\\
  & $\mathtt{law}$ (Extinction law of the polar dust) & 0 (SMC)\\
  & $\mathtt{EBV}$ & 0.05, 0.3, 0.8\\
  & $\mathtt{temperature}$ (Temperature of the polar dust) [K] & 100, 150, 200, 250\\
  & $\mathtt{emissivity}$ (Emissivity index of the polar dust) & 1.6\\
\hline
\end{tabular}
\\
\smallskip
{\footnotesize
The brackets in the first column indicate the name of each module.
}
\\
\smallskip
{\footnotesize [i] \citet{Ciesla2017}; [ii] \citet{Bruzual2003}; [iii] \citet{Chabrier2003};
[iv] \citet{Inoue2011}; [v] \citet{Calzetti2000}; [vi] \citet{Cardelli1989}, \citet{Odonnell1994};
[vii] \citet{Draine2014}; [viii] \citet{Stalevski2012, Stalevski2016}.}
\vspace{-0.8cm}
\end{table*}
\end{landscape}
%%% Table 2 %%%

%%% Table 3 %%%
\begin{landscape}
{\scriptsize
\setlength{\LTpre}{0pt}%
\setlength{\LTpost}{0pt}%
\begin{longtable}{ccccccccccccc}
\caption{SED Fitting Results\label{tabA:SEDresult}}\\
\hline
Name & R.A. & Decl. & Redshift & $\log\,M_\bigstar$ &
                                                      $\log\,t_{\bigstar}$
  & $\log\,\langle t_{\bigstar} \rangle_{\rm mass}$ & SFR & $\log\LIR$
  & 12+$\log$\,(O/H) & $\chi^2_{\rm red}$ & AGN & Reference(s)\\
 &  &  &  & ($M_{\odot}$) & (Myr) & (Myr) & ($M_{\odot}\yr^{-1}$) & ($L_{\odot}$) &  &  &  & \\
(1) & (2) & (3) & (4) & (5) & (6) & (7) & (8) & (9) & (10) & (11) & (12) & (13)\\
\hline
\endfirsthead

\hline
Name & R.A. & Decl. & Redshift & $\log\,M_{\bigstar}$ &
                                                        $\log\,t_{\bigstar}$
  & $\log\,\langle t_{\bigstar} \rangle_{\rm mass}$ & SFR & $\log\LIR$
  & 12+$\log$\,(O/H) & $\chi^2_{\rm red}$ & AGN & Reference(s)\\
 &  &  &  & ($M_{\odot}$) & (Myr) & (Myr) & ($M_{\odot}/{\rm yr}$) & ($L_{\odot}$) &  &  &  & \\
 (1) & (2) & (3) & (4) & (5) & (6) & (7) & (8) & (9) & (10) & (11) & (12) & (13)\\
\hline
\endhead

\hline
\multicolumn{13}{r}{(\textit{Continued on next page})}\\
\hline
\endfoot

\hline
\endlastfoot
AM0702-601 & 105.8510 & -60.2562 & 0.03132 & 10.21 $\pm$ 0.09 & 3.87 $\pm$ 0.14 & 3.48 $\pm$ 0.15 & 4.69 $\pm$ 0.93 & 11.30 $\pm$ 0.12 & 8.76 & 0.69 & Y & 17, 21 \\
Arp193 & 200.1472 & 34.1395 & 0.02306 & 9.91 $\pm$ 0.24 & 3.92 $\pm$ 0.14 & 3.45 $\pm$ 0.24 & 72.36 $\pm$ 5.81 & 11.67 $\pm$ 0.02 & 8.34 & 1.44 & Y & 17, 21 \\
Arp220 & 233.7384 & 23.5037 & 0.01840 & 9.83 $\pm$ 0.17 & 3.83 $\pm$ 0.16 & 3.28 $\pm$ 0.24 & 199.66 $\pm$ 9.98 & 12.08 $\pm$ 0.02 & 8.18 & 6.24 & Y & 21 \\
CGCG011-076 & 170.3010 & -2.9841 & 0.02429 & 10.40 $\pm$ 0.07 & 3.68 $\pm$ 0.11 & 3.24 $\pm$ 0.15 & 14.00 $\pm$ 1.08 & 11.22 $\pm$ 0.02 & 8.73 & 0.88 & Y & 11 \\
CGCG049-057 & 228.3046 & 7.2255 & 0.01306 & 9.16 $\pm$ 0.19 & 3.86 $\pm$ 0.16 & 2.98 $\pm$ 0.38 & 11.88 $\pm$ 0.59 & 11.04 $\pm$ 0.02 & 8.21 & 4.25 & Y & 3 \\
CGCG052-037 & 247.7356 & 4.0829 & 0.02449 & 10.50 $\pm$ 0.10 & 3.84 $\pm$ 0.16 & 3.50 $\pm$ 0.17 & 26.53 $\pm$ 4.05 & 11.30 $\pm$ 0.02 & 8.71 & 2.03 & $\cdots$ & $\cdots$ \\
CGCG247-020 & 214.9302 & 49.2366 & 0.02553 & 10.05 $\pm$ 0.08 & 3.76 $\pm$ 0.18 & 3.33 $\pm$ 0.24 & 19.16 $\pm$ 1.83 & 11.23 $\pm$ 0.03 & 8.57 & 0.58 & Y & 20 \\
CGCG436-030 & 20.0110 & 14.3618 & 0.03152 & 9.98 $\pm$ 0.23 & 3.86 $\pm$ 0.16 & 3.25 $\pm$ 0.35 & 59.55 $\pm$ 7.37 & 11.69 $\pm$ 0.02 & 8.41 & 1.33 & Y & 14 \\
CGCG453-062 & 346.2356 & 19.5523 & 0.02510 & 10.67 $\pm$ 0.08 & 3.92 $\pm$ 0.12 & 3.69 $\pm$ 0.06 & 18.82 $\pm$ 2.61 & 11.21 $\pm$ 0.05 & 8.78 & 2.21 & Y & 17 \\
ESO099-G004 & 231.2425 & -63.1264 & 0.02928 & 10.61 $\pm$ 0.14 & 3.91 $\pm$ 0.15 & 3.58 $\pm$ 0.19 & 57.99 $\pm$ 8.67 & 11.64 $\pm$ 0.02 & 8.68 & 0.43 & $\cdots$ & $\cdots$ \\
ESO264-G036 & 160.7820 & -46.2124 & 0.02107 & 10.70 $\pm$ 0.07 & 3.89 $\pm$ 0.10 & 3.50 $\pm$ 0.12 & 14.41 $\pm$ 1.01 & 11.23 $\pm$ 0.03 & 8.80 & 3.22 & Y & 6 \\
ESO286-G035 & 316.0464 & -43.5930 & 0.01736 & 10.32 $\pm$ 0.09 & 3.98 $\pm$ 0.12 & 3.61 $\pm$ 0.12 & 13.67 $\pm$ 1.80 & 11.04 $\pm$ 0.02 & 8.71 & 1.55 & N & 8 \\
ESO297-G011 & 24.0976 & -37.3216 & 0.01731 & 10.61 $\pm$ 0.15 & 3.97 $\pm$ 0.11 & 3.83 $\pm$ 0.16 & 9.72 $\pm$ 1.77 & 10.81 $\pm$ 0.03 & 8.80 & 3.36 & $\cdots$ & $\cdots$ \\
ESO320-G030 & 178.2988 & -39.1303 & 0.01078 & 10.50 $\pm$ 0.02 & 3.94 $\pm$ 0.12 & 3.75 $\pm$ 0.07 & 13.51 $\pm$ 1.75 & 11.01 $\pm$ 0.02 & 8.76 & 4.19 & $\cdots$ & $\cdots$ \\
ESO339-G011 & 299.4066 & -37.9357 & 0.01920 & 10.73 $\pm$ 0.12 & 3.97 $\pm$ 0.13 & 3.79 $\pm$ 0.15 & 9.49 $\pm$ 1.10 & 11.07 $\pm$ 0.03 & 8.82 & 2.15 & Y & 20 \\
ESO353-G020 & 23.7137 & -36.1372 & 0.01592 & 10.61 $\pm$ 0.04 & 3.85 $\pm$ 0.04 & 3.70 $\pm$ 0.04 & 15.37 $\pm$ 1.18 & 11.09 $\pm$ 0.02 & 8.78 & 3.74 & $\cdots$ & $\cdots$ \\
ESO507-G070 & 195.7182 & -23.9216 & 0.02170 & 10.48 $\pm$ 0.09 & 4.00 $\pm$ 0.10 & 3.68 $\pm$ 0.11 & 33.71 $\pm$ 4.08 & 11.37 $\pm$ 0.02 & 8.68 & 2.61 & Y & 20 \\
ESO593-IG008 & 288.6291 & -21.3186 & 0.04873 & 10.86 $\pm$ 0.12 & 3.93 $\pm$ 0.11 & 3.66 $\pm$ 0.11 & 92.67 $\pm$ 12.34 & 11.91 $\pm$ 0.02 & 8.72 & 3.80 & Y & 21 \\
ESO602-G025 & 337.8561 & -19.0333 & 0.02504 & 10.63 $\pm$ 0.10 & 3.97 $\pm$ 0.09 & 3.62 $\pm$ 0.09 & 18.36 $\pm$ 4.44 & 11.25 $\pm$ 0.07 & 8.77 & 2.14 & Y & 6 \\
IC4280 & 203.2225 & -24.2070 & 0.01631 & 10.38 $\pm$ 0.06 & 3.78 $\pm$ 0.11 & 3.36 $\pm$ 0.10 & 8.97 $\pm$ 0.47 & 10.96 $\pm$ 0.02 & 8.76 & 4.50 & Y & 6 \\
IC5179 & 334.0380 & -36.8437 & 0.01141 & 10.48 $\pm$ 0.09 & 4.02 $\pm$ 0.10 & 3.74 $\pm$ 0.12 & 16.58 $\pm$ 2.31 & 11.10 $\pm$ 0.02 & 8.74 & 10.31 & Y & 6 \\
IC860 & 198.7646 & 24.6188 & 0.01291 & 9.55 $\pm$ 0.10 & 3.85 $\pm$ 0.14 & 3.37 $\pm$ 0.17 & 8.93 $\pm$ 0.45 & 10.91 $\pm$ 0.02 & 8.43 & 5.41 & Y & 19 \\
IIIZw035 & 26.1272 & 17.1022 & 0.02744 & 9.52 $\pm$ 0.25 & 3.85 $\pm$ 0.16 & 3.27 $\pm$ 0.28 & 45.19 $\pm$ 5.41 & 11.51 $\pm$ 0.03 & 8.21 & 2.15 & Y & 21 \\
IRAS00456-2904 & 12.0282 & -28.8051 & 0.10989 & 10.70 $\pm$ 0.21 & 3.90 $\pm$ 0.16 & 3.62 $\pm$ 0.22 & 112.49 $\pm$ 13.52 & 11.86 $\pm$ 0.02 & 8.65 & 1.22 & $\cdots$ & $\cdots$ \\
IRAS03209-0806 & 50.8453 & -7.9376 & 0.16641 & 10.55 $\pm$ 0.15 & 3.80 $\pm$ 0.13 & 3.36 $\pm$ 0.19 & 50.39 $\pm$ 29.90 & 11.80 $\pm$ 0.26 & 8.67 & 0.82 & Y & 20 \\
IRAS04103-2838 & 63.0810 & -28.5069 & 0.11788 & 10.73 $\pm$ 0.08 & 3.85 $\pm$ 0.16 & 3.46 $\pm$ 0.18 & 35.69 $\pm$ 18.84 & 12.00 $\pm$ 0.18 & 8.75 & 0.94 & Y & 20 \\
IRAS09111-1007E & 138.4119 & -10.3222 & 0.05414 & 10.64 $\pm$ 0.08 & 3.92 $\pm$ 0.14 & 3.52 $\pm$ 0.15 & 27.47 $\pm$ 3.98 & 11.42 $\pm$ 0.06 & 8.75 & 1.60 & Y & 17 \\
IRAS09111-1007W & 138.4019 & -10.3250 & 0.07199 & 10.28 $\pm$ 0.34 & 3.90 $\pm$ 0.15 & 3.35 $\pm$ 0.31 & 206.19 $\pm$ 13.78 & 12.19 $\pm$ 0.02 & 8.39 & 1.32 & Y & 17 \\
IRAS10494+4424 & 163.0982 & 44.1464 & 0.09208 & 10.68 $\pm$ 0.06 & 3.93 $\pm$ 0.11 & 3.60 $\pm$ 0.10 & 8.44 $\pm$ 1.99 & 11.22 $\pm$ 0.15 & 8.82 & 2.03 & Y & 15 \\
IRAS12112+0305 & 183.4419 & 2.8115 & 0.07332 & 10.28 $\pm$ 0.16 & 3.93 $\pm$ 0.13 & 3.56 $\pm$ 0.17 & 224.32 $\pm$ 13.98 & 12.14 $\pm$ 0.02 & 8.38 & 3.95 & Y & 21 \\
IRAS12116-5615 & 183.5921 & -56.5425 & 0.02710 & 10.40 $\pm$ 0.11 & 3.80 $\pm$ 0.18 & 3.42 $\pm$ 0.25 & 40.35 $\pm$ 4.81 & 11.52 $\pm$ 0.02 & 8.63 & 0.57 & $\cdots$ & $\cdots$ \\
IRAS13052-5711 & 197.0779 & -57.4584 & 0.02123 & 10.39 $\pm$ 0.12 & 3.92 $\pm$ 0.14 & 3.57 $\pm$ 0.16 & 18.64 $\pm$ 2.74 & 11.18 $\pm$ 0.02 & 8.70 & 2.73 & $\cdots$ & $\cdots$ \\
IRAS13120-5453 & 198.7764 & -55.1563 & 0.03076 & 10.36 $\pm$ 0.32 & 3.91 $\pm$ 0.15 & 3.41 $\pm$ 0.29 & 205.02 $\pm$ 19.49 & 12.21 $\pm$ 0.03 & 8.43 & 0.41 & Y & 20 \\
IRAS13539+2920 & 209.0417 & 29.0931 & 0.10876 & 10.82 $\pm$ 0.15 & 3.93 $\pm$ 0.15 & 3.64 $\pm$ 0.20 & 31.81 $\pm$ 15.71 & 11.53 $\pm$ 0.17 & 8.78 & 0.54 & Y & 10 \\
IRAS14060+2919 & 212.0792 & 29.0797 & 0.11675 & 10.43 $\pm$ 0.17 & 3.87 $\pm$ 0.16 & 3.44 $\pm$ 0.22 & 69.85 $\pm$ 12.59 & 11.75 $\pm$ 0.05 & 8.59 & 1.76 & $\cdots$ & $\cdots$ \\
IRAS14121-0126 & 213.6896 & -1.6822 & 0.15020 & 10.64 $\pm$ 0.17 & 3.88 $\pm$ 0.11 & 3.52 $\pm$ 0.18 & 87.76 $\pm$ 14.76 & 11.95 $\pm$ 0.05 & 8.65 & 0.74 & Y & 18 \\
IRAS15043+5754 & 226.4147 & 57.7187 & 0.15058 & 10.13 $\pm$ 0.20 & 3.78 $\pm$ 0.14 & 3.30 $\pm$ 0.23 & 49.45 $\pm$ 25.39 & 11.57 $\pm$ 0.19 & 8.50 & 0.63 & $\cdots$ & $\cdots$ \\
IRAS15250+3609 & 231.7476 & 35.9770 & 0.05520 & 9.90 $\pm$ 0.15 & 3.87 $\pm$ 0.16 & 3.41 $\pm$ 0.21 & 132.53 $\pm$ 7.56 & 11.94 $\pm$ 0.03 & 8.27 & 1.80 & Y & 18 \\
IRAS15335-0513 & 234.0487 & -5.3979 & 0.02700 & 10.83 $\pm$ 0.12 & 3.99 $\pm$ 0.09 & 3.87 $\pm$ 0.13 & 2.40 $\pm$ 6.00 & 10.83 $\pm$ 0.54 & 8.82 & 0.99 & Y & 18 \\
IRAS16474+3430 & 252.3091 & 34.4194 & 0.11147 & 10.69 $\pm$ 0.18 & 3.83 $\pm$ 0.16 & 3.37 $\pm$ 0.24 & 94.61 $\pm$ 49.22 & 11.91 $\pm$ 0.20 & 8.66 & 0.82 & $\cdots$ & $\cdots$ \\
IRAS17028+5817 & 255.9248 & 58.2291 & 0.10609 & 10.69 $\pm$ 0.08 & 3.93 $\pm$ 0.13 & 3.57 $\pm$ 0.13 & 10.97 $\pm$ 3.85 & 11.27 $\pm$ 0.13 & 8.81 & 1.55 & Y & 12 \\
IRAS17132+5313 & 258.5825 & 53.1747 & 0.05094 & 10.12 $\pm$ 0.31 & 3.87 $\pm$ 0.16 & 3.25 $\pm$ 0.36 & 69.81 $\pm$ 7.49 & 11.78 $\pm$ 0.02 & 8.45 & 0.51 & Y & 9 \\
IRAS23128-5919 & 348.9448 & -59.0544 & 0.04460 & 10.19 $\pm$ 0.26 & 3.82 $\pm$ 0.17 & 3.29 $\pm$ 0.29 & 138.69 $\pm$ 20.52 & 11.99 $\pm$ 0.07 & 8.40 & 1.83 & Y & 17 \\
IRAS23436+5257 & 356.5231 & 53.2334 & 0.03413 & 10.57 $\pm$ 0.09 & 3.90 $\pm$ 0.12 & 3.51 $\pm$ 0.14 & 33.92 $\pm$ 4.20 & 11.49 $\pm$ 0.02 & 8.71 & 0.54 & $\cdots$ & $\cdots$ \\
IRASF06076-2139 & 92.4408 & -21.6733 & 0.03745 & 10.20 $\pm$ 0.26 & 3.90 $\pm$ 0.15 & 3.47 $\pm$ 0.27 & 47.72 $\pm$ 6.35 & 11.61 $\pm$ 0.03 & 8.53 & 0.32 & Y & 6 \\
IRASF10565+2448 & 164.8256 & 24.5429 & 0.04310 & 10.49 $\pm$ 0.13 & 3.81 $\pm$ 0.16 & 3.42 $\pm$ 0.22 & 123.15 $\pm$ 13.12 & 11.96 $\pm$ 0.03 & 8.55 & 1.98 & Y & 18 \\
IRASF16330-6820 & 249.5498 & -68.4358 & 0.04697 & 10.79 $\pm$ 0.14 & 3.92 $\pm$ 0.17 & 3.56 $\pm$ 0.20 & 88.12 $\pm$ 10.69 & 11.86 $\pm$ 0.02 & 8.70 & 1.53 & $\cdots$ & $\cdots$ \\
IRASF17207-0014 & 260.8415 & -0.2835 & 0.04281 & 10.35 $\pm$ 0.13 & 3.84 $\pm$ 0.17 & 3.39 $\pm$ 0.24 & 328.86 $\pm$ 19.36 & 12.31 $\pm$ 0.02 & 8.37 & 0.59 & Y & 13 \\
IRASF18293-3413 & 278.1713 & -34.1909 & 0.01818 & 10.80 $\pm$ 0.07 & 3.85 $\pm$ 0.17 & 3.54 $\pm$ 0.14 & 52.29 $\pm$ 10.93 & 11.68 $\pm$ 0.06 & 8.75 & 1.99 & Y & 6 \\
LEDA19076 & 96.8403 & -47.1767 & 0.03940 & 10.44 $\pm$ 0.17 & 3.90 $\pm$ 0.14 & 3.50 $\pm$ 0.19 & 71.80 $\pm$ 8.55 & 11.73 $\pm$ 0.02 & 8.59 & 1.66 & N & 21 \\
LEDA59176 & 253.1539 & 81.0047 & 0.04915 & 10.63 $\pm$ 0.06 & 3.94 $\pm$ 0.14 & 3.56 $\pm$ 0.11 & 10.34 $\pm$ 2.59 & 10.98 $\pm$ 0.10 & 8.80 & 1.12 & $\cdots$ & $\cdots$ \\
MCG+04-48-002 & 307.1461 & 25.7334 & 0.01390 & 10.37 $\pm$ 0.14 & 3.92 $\pm$ 0.14 & 3.58 $\pm$ 0.21 & 5.87 $\pm$ 1.13 & 10.90 $\pm$ 0.07 & 8.78 & 0.75 & Y & 20 \\
MCG+07-23-019 & 165.9748 & 40.8500 & 0.03452 & 10.48 $\pm$ 0.11 & 3.97 $\pm$ 0.10 & 3.64 $\pm$ 0.12 & 51.06 $\pm$ 5.06 & 11.55 $\pm$ 0.02 & 8.64 & 1.68 & $\cdots$ & $\cdots$ \\
MCG+08-18-013 & 144.1550 & 48.4744 & 0.02596 & 10.36 $\pm$ 0.11 & 3.90 $\pm$ 0.12 & 3.57 $\pm$ 0.16 & 24.87 $\pm$ 3.47 & 11.26 $\pm$ 0.02 & 8.67 & 1.75 & $\cdots$ & $\cdots$ \\
MCG+12-02-001 & 13.5171 & 73.0854 & 0.01570 & 10.39 $\pm$ 0.09 & 3.98 $\pm$ 0.09 & 3.80 $\pm$ 0.10 & 35.55 $\pm$ 3.26 & 11.42 $\pm$ 0.02 & 8.64 & 1.39 & N & 1 \\
MCG-01-60-022 & 355.5035 & -3.6152 & 0.02324 & 10.34 $\pm$ 0.07 & 3.96 $\pm$ 0.10 & 3.55 $\pm$ 0.11 & 16.69 $\pm$ 1.84 & 11.15 $\pm$ 0.02 & 8.70 & 1.18 & Y & 6, 8 \\
MCG-02-01-051 & 4.7121 & -10.3769 & 0.02710 & 9.94 $\pm$ 0.25 & 3.86 $\pm$ 0.17 & 3.35 $\pm$ 0.27 & 31.83 $\pm$ 2.71 & 11.44 $\pm$ 0.07 & 8.46 & 1.82 & Y & 6 \\
MCG-03-04-014 & 17.5373 & -16.8528 & 0.03514 & 10.65 $\pm$ 0.08 & 3.78 $\pm$ 0.12 & 3.38 $\pm$ 0.14 & 51.63 $\pm$ 5.06 & 11.64 $\pm$ 0.02 & 8.70 & 0.63 & Y & 6 \\
MCG-07-03-014 & 19.5347 & -44.4619 & 0.02093 & 10.78 $\pm$ 0.13 & 3.93 $\pm$ 0.12 & 3.82 $\pm$ 0.15 & 16.65 $\pm$ 2.66 & 11.18 $\pm$ 0.03 & 8.81 & 5.13 & Y & 6 \\
Mrk18 & 135.4934 & 60.1517 & 0.01113 & 9.98 $\pm$ 0.03 & 3.97 $\pm$ 0.13 & 3.67 $\pm$ 0.10 & 1.02 $\pm$ 0.14 & 10.12 $\pm$ 0.04 & 8.79 & 0.42 & Y & 4 \\
Mrk331 & 357.8615 & 20.5862 & 0.01814 & 10.38 $\pm$ 0.12 & 3.88 $\pm$ 0.15 & 3.53 $\pm$ 0.19 & 27.60 $\pm$ 3.55 & 11.37 $\pm$ 0.03 & 8.66 & 1.21 & Y & 13 \\
NGC1614 & 68.5001 & -8.5792 & 0.01594 & 10.66 $\pm$ 0.11 & 4.00 $\pm$ 0.11 & 3.69 $\pm$ 0.10 & 49.00 $\pm$ 8.45 & 11.65 $\pm$ 0.06 & 8.71 & 2.45 & Y & 6, 20 \\
NGC232 & 10.6909 & -23.5614 & 0.02264 & 10.61 $\pm$ 0.08 & 3.87 $\pm$ 0.12 & 3.45 $\pm$ 0.12 & 37.78 $\pm$ 4.51 & 11.54 $\pm$ 0.02 & 8.72 & 2.48 & Y & 6 \\
NGC2388 & 112.2228 & 33.8191 & 0.01379 & 10.38 $\pm$ 0.06 & 3.95 $\pm$ 0.10 & 3.54 $\pm$ 0.09 & 17.49 $\pm$ 2.15 & 11.22 $\pm$ 0.02 & 8.70 & 3.87 & $\cdots$ & $\cdots$ \\
NGC2623 & 129.6003 & 25.7547 & 0.01851 & 10.35 $\pm$ 0.09 & 3.91 $\pm$ 0.09 & 3.57 $\pm$ 0.09 & 39.67 $\pm$ 2.50 & 11.43 $\pm$ 0.02 & 8.62 & 4.02 & Y & 17, 19 \\
NGC3110 & 151.0087 & -6.4749 & 0.01686 & 10.52 $\pm$ 0.08 & 3.99 $\pm$ 0.10 & 3.64 $\pm$ 0.11 & 19.63 $\pm$ 2.52 & 11.18 $\pm$ 0.02 & 8.74 & 2.79 & Y & 6 \\
NGC3256 & 156.9626 & -43.9051 & 0.00935 & 10.58 $\pm$ 0.12 & 3.96 $\pm$ 0.11 & 3.70 $\pm$ 0.14 & 51.78 $\pm$ 8.93 & 11.62 $\pm$ 0.06 & 8.68 & 3.04 & Y & 6 \\
NGC34 & 2.7773 & -12.1077 & 0.01962 & 10.39 $\pm$ 0.11 & 3.78 $\pm$ 0.14 & 3.38 $\pm$ 0.20 & 30.90 $\pm$ 3.78 & 11.41 $\pm$ 0.03 & 8.65 & 2.23 & Y & 6, 21 \\
NGC4102 & 181.5960 & 52.7110 & 0.00282 & 10.01 $\pm$ 0.04 & 4.07 $\pm$ 0.04 & 3.98 $\pm$ 0.06 & 1.48 $\pm$ 0.12 & 10.10 $\pm$ 0.05 & 8.78 & 6.97 & Y & 16 \\
NGC5010 & 198.1100 & -15.7979 & 0.00992 & 10.42 $\pm$ 0.06 & 4.04 $\pm$ 0.08 & 3.81 $\pm$ 0.07 & 4.21 $\pm$ 0.33 & 10.60 $\pm$ 0.02 & 8.81 & 2.06 & $\cdots$ & $\cdots$ \\
NGC5104 & 200.3463 & 0.3426 & 0.01855 & 10.63 $\pm$ 0.05 & 3.87 $\pm$ 0.19 & 3.54 $\pm$ 0.12 & 10.43 $\pm$ 1.74 & 11.05 $\pm$ 0.02 & 8.80 & 2.57 & N & 21 \\
NGC5135 & 201.4335 & -29.8334 & 0.01369 & 10.87 $\pm$ 0.04 & 3.86 $\pm$ 0.06 & 3.71 $\pm$ 0.07 & 27.53 $\pm$ 2.51 & 11.27 $\pm$ 0.02 & 8.80 & 2.17 & Y & 17, 21 \\
NGC520 & 21.1466 & 3.7920 & 0.00761 & 10.13 $\pm$ 0.11 & 3.99 $\pm$ 0.11 & 3.66 $\pm$ 0.14 & 8.37 $\pm$ 1.15 & 10.79 $\pm$ 0.04 & 8.70 & 9.01 & $\cdots$ & $\cdots$ \\
NGC5256 & 204.5741 & 48.2764 & 0.02786 & 10.67 $\pm$ 0.06 & 3.97 $\pm$ 0.10 & 3.59 $\pm$ 0.09 & 29.78 $\pm$ 3.32 & 11.42 $\pm$ 0.03 & 8.75 & 2.48 & Y & 17, 21 \\
NGC5728 & 220.5995 & -17.2530 & 0.00932 & 10.65 $\pm$ 0.02 & 3.98 $\pm$ 0.00 & 3.87 $\pm$ 0.00 & 1.97 $\pm$ 0.10 & 10.34 $\pm$ 0.02 & 8.83 & 8.20 & Y & 16 \\
NGC6090 & 242.9192 & 52.4569 & 0.02930 & 10.21 $\pm$ 0.12 & 3.88 $\pm$ 0.14 & 3.44 $\pm$ 0.20 & 37.73 $\pm$ 1.89 & 11.52 $\pm$ 0.02 & 8.57 & 1.97 & N & 21 \\
NGC6240 & 253.2453 & 2.4009 & 0.02431 & 11.10 $\pm$ 0.18 & 3.86 $\pm$ 0.15 & 3.57 $\pm$ 0.19 & 54.64 $\pm$ 15.49 & 11.73 $\pm$ 0.04 & 8.82 & 3.75 & Y & 17 \\
NGC6285 & 254.5999 & 58.9561 & 0.01898 & 10.03 $\pm$ 0.06 & 3.88 $\pm$ 0.12 & 3.49 $\pm$ 0.10 & 3.06 $\pm$ 0.40 & 10.49 $\pm$ 0.03 & 8.74 & 1.64 & Y & 21 \\
NGC6621 & 273.2306 & 68.3633 & 0.02065 & 10.67 $\pm$ 0.05 & 3.86 $\pm$ 0.07 & 3.70 $\pm$ 0.06 & 18.41 $\pm$ 1.56 & 11.13 $\pm$ 0.02 & 8.78 & 2.87 & N & 21 \\
NGC6670b & 278.3925 & 59.8883 & 0.02811 & 10.99 $\pm$ 0.02 & 4.08 $\pm$ 0.00 & 4.00 $\pm$ 0.00 & 14.20 $\pm$ 0.71 & 11.05 $\pm$ 0.02 & 8.84 & 4.95 & N & 21 \\
NGC6701 & 280.8022 & 60.6533 & 0.01323 & 10.50 $\pm$ 0.07 & 3.90 $\pm$ 0.11 & 3.69 $\pm$ 0.06 & 12.23 $\pm$ 1.44 & 10.90 $\pm$ 0.02 & 8.77 & 3.69 & N & 21 \\
NGC6786 & 287.7245 & 73.4102 & 0.02524 & 10.50 $\pm$ 0.10 & 3.89 $\pm$ 0.11 & 3.55 $\pm$ 0.14 & 9.73 $\pm$ 1.98 & 10.94 $\pm$ 0.05 & 8.78 & 5.48 & Y & 20 \\
NGC695 & 27.8096 & 22.5822 & 0.03247 & 10.88 $\pm$ 0.06 & 3.93 $\pm$ 0.12 & 3.54 $\pm$ 0.12 & 46.81 $\pm$ 5.84 & 11.62 $\pm$ 0.02 & 8.77 & 1.53 & N & 21 \\
NGC7469 & 345.8151 & 8.8739 & 0.01627 & 10.96 $\pm$ 0.03 & 3.85 $\pm$ 0.03 & 3.70 $\pm$ 0.02 & 34.55 $\pm$ 2.55 & 11.56 $\pm$ 0.04 & 8.81 & 0.89 & Y & 13, 17, 21 \\
NGC7552 & 349.0447 & -42.5847 & 0.00537 & 10.46 $\pm$ 0.05 & 3.85 $\pm$ 0.04 & 3.71 $\pm$ 0.06 & 10.65 $\pm$ 0.53 & 10.94 $\pm$ 0.02 & 8.76 & 7.96 & Y & 6 \\
NGC7592W & 349.5909 & -4.4159 & 0.02444 & 10.64 $\pm$ 0.13 & 3.98 $\pm$ 0.15 & 3.73 $\pm$ 0.17 & 8.62 $\pm$ 0.85 & 11.02 $\pm$ 0.09 & 8.81 & 4.51 & Y & 17, 21 \\
NGC7679 & 352.1944 & 3.5114 & 0.01715 & 10.27 $\pm$ 0.02 & 3.65 $\pm$ 0.02 & 3.19 $\pm$ 0.03 & 11.41 $\pm$ 0.57 & 11.05 $\pm$ 0.02 & 8.70 & 0.96 & Y & 17, 21 \\
NGC7771 & 357.8536 & 20.1118 & 0.01446 & 10.99 $\pm$ 0.07 & 3.96 $\pm$ 0.10 & 3.56 $\pm$ 0.09 & 22.99 $\pm$ 2.39 & 11.44 $\pm$ 0.04 & 8.83 & 1.43 & $\cdots$ & $\cdots$ \\
NGC838 & 32.4105 & -10.1467 & 0.01284 & 10.22 $\pm$ 0.09 & 3.96 $\pm$ 0.11 & 3.58 $\pm$ 0.13 & 11.95 $\pm$ 1.50 & 11.01 $\pm$ 0.03 & 8.68 & 0.83 & Y & 2 \\
NGC992 & 39.3563 & 21.1009 & 0.01378 & 10.24 $\pm$ 0.10 & 3.71 $\pm$ 0.14 & 3.31 $\pm$ 0.24 & 9.68 $\pm$ 1.18 & 11.01 $\pm$ 0.02 & 8.71 & 2.47 & $\cdots$ & $\cdots$ \\
UGC12150 & 340.3010 & 34.2491 & 0.02139 & 10.42 $\pm$ 0.04 & 3.70 $\pm$ 0.13 & 3.25 $\pm$ 0.17 & 16.44 $\pm$ 1.38 & 11.26 $\pm$ 0.02 & 8.72 & 2.76 & N & 21 \\
UGC1385 & 28.7241 & 36.9179 & 0.01875 & 10.20 $\pm$ 0.11 & 3.79 $\pm$ 0.20 & 3.53 $\pm$ 0.19 & 13.42 $\pm$ 1.36 & 10.94 $\pm$ 0.02 & 8.67 & 2.50 & $\cdots$ & $\cdots$ \\
UGC1845 & 36.0333 & 47.9697 & 0.01514 & 10.47 $\pm$ 0.08 & 3.99 $\pm$ 0.09 & 3.68 $\pm$ 0.08 & 11.55 $\pm$ 1.46 & 11.03 $\pm$ 0.03 & 8.76 & 1.23 & $\cdots$ & $\cdots$ \\
UGC2238 & 41.5730 & 13.0957 & 0.02151 & 10.60 $\pm$ 0.09 & 4.01 $\pm$ 0.10 & 3.73 $\pm$ 0.13 & 20.52 $\pm$ 3.13 & 11.19 $\pm$ 0.02 & 8.76 & 2.98 & N & 21 \\
UGC2982 & 63.0941 & 5.5472 & 0.01770 & 10.64 $\pm$ 0.09 & 3.87 $\pm$ 0.10 & 3.65 $\pm$ 0.12 & 16.14 $\pm$ 3.76 & 11.09 $\pm$ 0.02 & 8.78 & 2.13 & N & 21 \\
UGC8335W & 198.8781 & 62.1292 & 0.03134 & 10.15 $\pm$ 0.11 & 3.89 $\pm$ 0.17 & 3.61 $\pm$ 0.16 & 11.70 $\pm$ 1.24 & 10.86 $\pm$ 0.03 & 8.66 & 0.97 & $\cdots$ & $\cdots$ \\
UGC9618N & 224.2529 & 24.6171 & 0.03367 & 10.90 $\pm$ 0.13 & 3.92 $\pm$ 0.16 & 3.59 $\pm$ 0.19 & 39.02 $\pm$ 6.84 & 11.58 $\pm$ 0.02 & 8.79 & 1.54 & Y & 17 \\
VIIZw031 & 79.1935 & 79.6701 & 0.05367 & 10.66 $\pm$ 0.14 & 3.93 $\pm$ 0.13 & 3.57 $\pm$ 0.18 & 88.81 $\pm$ 10.50 & 11.88 $\pm$ 0.02 & 8.66 & 1.00 & Y & 17 \\
VV114b & 16.9479 & -17.5070 & 0.02007 & 10.50 $\pm$ 0.08 & 3.98 $\pm$ 0.11 & 3.61 $\pm$ 0.12 & 19.50 $\pm$ 2.74 & 11.51 $\pm$ 0.04 & 8.73 & 4.74 & Y & 6 \\
VV250a & 198.8867 & 62.1269 & 0.03079 & 9.76 $\pm$ 0.23 & 3.86 $\pm$ 0.16 & 3.04 $\pm$ 0.39 & 43.83 $\pm$ 5.79 & 11.61 $\pm$ 0.03 & 8.34 & 0.93 & Y & 14 \\
VV283 & 195.4617 & 4.3333 & 0.03748 & 10.48 $\pm$ 0.13 & 3.93 $\pm$ 0.15 & 3.59 $\pm$ 0.19 & 37.48 $\pm$ 5.58 & 11.49 $\pm$ 0.03 & 8.67 & 1.15 & Y & 5 \\
VV705 & 229.5263 & 42.7436 & 0.04019 & 10.08 $\pm$ 0.14 & 3.85 $\pm$ 0.16 & 3.22 $\pm$ 0.27 & 65.76 $\pm$ 3.46 & 11.79 $\pm$ 0.03 & 8.44 & 1.54 & Y & 7 \\
\end{longtable}

\begin{minipage}{\linewidth}\footnotesize
Columns: (1) Object name; (2) Redshift from NASA/IPAC Extragalactic Database (NED);
(3) Right Ascension in decimal degrees (J2000); (4) Declination in
decimal degrees (J2000);  (5) Logarithmic stellar mass; (6)
Logarithmic galaxy age; (7) Logarithmic mass-weighted galaxy age; (8)
Star formation rate; (9) Logarithmic IR luminosity. The AGN $\LIR$ is
defined as the dust luminosity plus the AGN luminosity contribution in
the IR band, and the non-AGN $\LIR$ is defined as the dust luminosity alone;
(10) Gas-phase metallicity 12+log\,(O/H), which is calculated based on the FMR of \citet{Sanders2021}; (11) Reduced $\chi^2$ of the best-fit model;
(12) Type of nuclear activity; (13) Reference for nuclear activity.\\
References: (1) \citet{Alonso-herrero2009}; (2) \citet{Asmus2020}; (3) \citet{Baan1998};
(4) \citet{Baumgartner2013}; (5) \citet{Best2012}; (6) \citet{Chen2022}; (7) \citet{Comerford2020};
(8) \citet{Corbett2003}; (9) \citet{Farrah2003}; (10) \citet{Ge2012}; (11) \citet{Hernan-caballero2011};
(12) \citet{Kim1998}; (13) \citet{Koss2013}; (14) \citet{Monroe2016}; (15) \citet{Nisbet2016};
(16) \citet{Pena-herazo2022}; (17) \citet{Petric2011}; (18) \citet{Lambrides2019};
(19) \citet{Toba2014}; (20) \citet{Veron-cetty2010}; (21) \citet{Yuan2010}.\\
(This table is available in its entirety in machine-readable form.)
\end{minipage}
}
\end{landscape}
%%% Table 3 %%%

%%% Table 4 %%%
{\scriptsize
\setlength{\LTpre}{0pt}%
\setlength{\LTpost}{0pt}%
\begin{longtable}{cccccccc}
\caption{Spectral Fitting Results\label{tabB:fitresult}}\\
\hline
Name & ObsID & $\tau({\rm H_2O})$ & $f_{\mathrm{aro}}$ & $f_{\mathrm{ali}}$ & $f_{\mathrm{ali}}/f_{\mathrm{aro}}$ & $\etaali$ & $\sigma$\\
 &  & & ($10^{-16}$\,W\,m$^{-2}$) & ($10^{-16}$\,W\,m$^{-2}$) &    & (\%) & \\
(1) & (2) & (3) & (4) & (5) & (6) & (7) & (8)\\
\hline
\endfirsthead

\hline
Name & ObsID & $\tau({\rm H_2O})$ & $f_{\mathrm{aro}}$ & $f_{\mathrm{ali}}$ & $f_{\mathrm{ali}}/f_{\mathrm{aro}}$ & $\etaali$ & $\sigma$\\
 &  & & ($10^{-16}$\,W\,m$^{-2}$) & ($10^{-16}$\,W\,m$^{-2}$) &  & (\%) & \\
\hline
\endhead

\hline
\multicolumn{8}{r}{(\textit{Continued on next page})}\\
\hline
\endfoot

\hline
\endlastfoot
AM0702-601 & 3370033 & 0.33 $\pm$ 0.024 & 6.16 $\pm$ 0.139 & 1.96 $\pm$ 0.250 & 0.32 $\pm$ 0.041 & 4.74 $\pm$ 0.639 & 0.02 \\
Arp193 & 1120151 & 0.96 $\pm$ 0.055 & 13.08 $\pm$ 0.253 & 2.55 $\pm$ 0.400 & 0.20 $\pm$ 0.031 & 2.96 $\pm$ 0.480 & 0.02 \\
Arp220 & 1120017 & 0.71 $\pm$ 0.038 & 8.96 $\pm$ 0.198 & 0.55 $\pm$ 0.143 & 0.06 $\pm$ 0.016 & 0.94 $\pm$ 0.249 & 0.02 \\
CGCG011-076 & 1120123 & 0.24 $\pm$ 0.029 & 5.72 $\pm$ 0.117 & 0.86 $\pm$ 0.075 & 0.15 $\pm$ 0.013 & 2.31 $\pm$ 0.210 & 0.02 \\
CGCG049-057 & 1120137 & 0.23 $\pm$ 0.028 & 3.07 $\pm$ 0.111 & 0.22 $\pm$ 0.044 & 0.07 $\pm$ 0.015 & 1.11 $\pm$ 0.227 & 0.02 \\
CGCG052-037 & 1120141 & 0.31 $\pm$ 0.034 & 10.08 $\pm$ 0.188 & 3.18 $\pm$ 0.132 & 0.32 $\pm$ 0.014 & 4.70 $\pm$ 0.224 & 0.01 \\
CGCG247-020 & 1120135 & 0.19 $\pm$ 0.024 & 3.68 $\pm$ 0.073 & 0.48 $\pm$ 0.087 & 0.13 $\pm$ 0.024 & 2.01 $\pm$ 0.372 & 0.02 \\
CGCG436-030 & 1120105 & 0.91 $\pm$ 0.070 & 9.24 $\pm$ 0.210 & 0.67 $\pm$ 0.085 & 0.07 $\pm$ 0.009 & 1.13 $\pm$ 0.145 & 0.02 \\
CGCG453-062 & 1120145 & 0.19 $\pm$ 0.024 & 6.93 $\pm$ 0.134 & 0.91 $\pm$ 0.235 & 0.13 $\pm$ 0.034 & 2.01 $\pm$ 0.529 & 0.02 \\
ESO099-G004 & 3370023 & 0.21 $\pm$ 0.023 & 7.14 $\pm$ 0.142 & 1.55 $\pm$ 0.314 & 0.22 $\pm$ 0.044 & 3.28 $\pm$ 0.686 & 0.02 \\
ESO264-G036 & 1122231 & 0.38 $\pm$ 0.054 & 5.85 $\pm$ 0.322 & 0.54 $\pm$ 0.175 & 0.09 $\pm$ 0.030 & 1.41 $\pm$ 0.472 & 0.03 \\
ESO286-G035 & 3371046 & 0.42 $\pm$ 0.022 & 11.03 $\pm$ 0.359 & 1.36 $\pm$ 0.458 & 0.12 $\pm$ 0.042 & 1.89 $\pm$ 0.647 & 0.01 \\
ESO297-G011 & 1122273 & 0.19 $\pm$ 0.019 & 6.50 $\pm$ 0.227 & 2.10 $\pm$ 0.167 & 0.32 $\pm$ 0.028 & 4.80 $\pm$ 0.436 & 0.02 \\
ESO320-G030 & 5200050 & 0.27 $\pm$ 0.044 & 19.04 $\pm$ 0.458 & 4.64 $\pm$ 0.717 & 0.24 $\pm$ 0.038 & 3.67 $\pm$ 0.592 & 0.01 \\
ESO339-G011 & 1122264 & 0.28 $\pm$ 0.114 & 7.67 $\pm$ 0.508 & 0.77 $\pm$ 0.172 & 0.10 $\pm$ 0.023 & 1.54 $\pm$ 0.364 & 0.03 \\
ESO353-G020 & 1122308 & 0.52 $\pm$ 0.029 & 8.47 $\pm$ 0.396 & 1.87 $\pm$ 0.300 & 0.22 $\pm$ 0.037 & 3.34 $\pm$ 0.574 & 0.01 \\
ESO507-G070 & 1120128 & 0.31 $\pm$ 0.021 & 4.00 $\pm$ 0.097 & 0.36 $\pm$ 0.047 & 0.09 $\pm$ 0.012 & 1.37 $\pm$ 0.185 & 0.02 \\
ESO593-IG008 & 3370013 & 0.66 $\pm$ 0.144 & 8.40 $\pm$ 0.384 & 1.83 $\pm$ 0.277 & 0.22 $\pm$ 0.034 & 3.29 $\pm$ 0.535 & 0.05 \\
ESO602-G025 & 1120144 & 0.42 $\pm$ 0.026 & 9.15 $\pm$ 0.269 & 1.52 $\pm$ 0.238 & 0.17 $\pm$ 0.027 & 2.54 $\pm$ 0.413 & 0.02 \\
IC4280 & 1122276 & 0.19 $\pm$ 0.042 & 14.73 $\pm$ 0.421 & 1.56 $\pm$ 0.452 & 0.11 $\pm$ 0.031 & 1.62 $\pm$ 0.480 & 0.02 \\
IC5179 & 1122259 & 0.27 $\pm$ 0.079 & 23.30 $\pm$ 0.963 & 2.10 $\pm$ 0.695 & 0.09 $\pm$ 0.030 & 1.39 $\pm$ 0.467 & 0.02 \\
IC860 & 1120040 & 0.10 $\pm$ 0.027 & 1.71 $\pm$ 0.103 & 0.35 $\pm$ 0.080 & 0.20 $\pm$ 0.048 & 3.09 $\pm$ 0.751 & 0.01 \\
IIIZw035 & 3370035 & 0.45 $\pm$ 0.067 & 2.00 $\pm$ 0.079 & 0.44 $\pm$ 0.045 & 0.22 $\pm$ 0.024 & 3.29 $\pm$ 0.372 & 0.03 \\
IRAS00456-2904 & 1100221 & 1.27 $\pm$ 0.109 & 1.62 $\pm$ 0.067 & 0.24 $\pm$ 0.030 & 0.15 $\pm$ 0.019 & 2.23 $\pm$ 0.299 & 0.04 \\
IRAS03209-0806 & 1100210 & 0.82 $\pm$ 0.081 & 0.54 $\pm$ 0.030 & 0.12 $\pm$ 0.024 & 0.21 $\pm$ 0.047 & 3.23 $\pm$ 0.721 & 0.04 \\
IRAS04103-2838 & 1120092 & 0.65 $\pm$ 0.130 & 1.52 $\pm$ 0.088 & 0.27 $\pm$ 0.042 & 0.17 $\pm$ 0.029 & 2.65 $\pm$ 0.458 & 0.05 \\
IRAS09111-1007E & 3371018 & 0.45 $\pm$ 0.111 & 1.73 $\pm$ 0.111 & 0.41 $\pm$ 0.112 & 0.24 $\pm$ 0.067 & 3.55 $\pm$ 1.030 & 0.06 \\
IRAS09111-1007W & 3371018 & 1.16 $\pm$ 0.149 & 3.27 $\pm$ 0.219 & 0.24 $\pm$ 0.067 & 0.07 $\pm$ 0.021 & 1.14 $\pm$ 0.328 & 0.07 \\
IRAS10494+4424 & 1100266 & 1.91 $\pm$ 0.128 & 1.35 $\pm$ 0.062 & 0.14 $\pm$ 0.034 & 0.11 $\pm$ 0.026 & 1.62 $\pm$ 0.398 & 0.05 \\
IRAS12112+0305 & 1120031 & 0.94 $\pm$ 0.130 & 2.11 $\pm$ 0.101 & 0.66 $\pm$ 0.064 & 0.31 $\pm$ 0.034 & 4.67 $\pm$ 0.524 & 0.05 \\
IRAS12116-5615 & 3370030 & 0.61 $\pm$ 0.047 & 6.87 $\pm$ 0.216 & 1.33 $\pm$ 0.136 & 0.19 $\pm$ 0.021 & 2.93 $\pm$ 0.323 & 0.02 \\
IRAS13052-5711 & 5200055 & 0.19 $\pm$ 0.052 & 4.02 $\pm$ 0.316 & 0.85 $\pm$ 0.255 & 0.21 $\pm$ 0.065 & 3.21 $\pm$ 1.012 & 0.04 \\
IRAS13120-5453 & 3370004 & 0.96 $\pm$ 0.044 & 16.37 $\pm$ 0.304 & 1.75 $\pm$ 0.356 & 0.11 $\pm$ 0.022 & 1.65 $\pm$ 0.340 & 0.02 \\
IRAS13539+2920 & 1100235 & 0.47 $\pm$ 0.061 & 0.70 $\pm$ 0.023 & 0.06 $\pm$ 0.013 & 0.09 $\pm$ 0.018 & 1.33 $\pm$ 0.283 & 0.03 \\
IRAS14060+2919 & 1120169 & 1.41 $\pm$ 0.222 & 2.86 $\pm$ 0.191 & 0.40 $\pm$ 0.059 & 0.14 $\pm$ 0.023 & 2.16 $\pm$ 0.353 & 0.07 \\
IRAS14121-0126 & 1100011 & 2.50 $\pm$ 0.517 & 1.51 $\pm$ 0.088 & 0.16 $\pm$ 0.034 & 0.10 $\pm$ 0.024 & 1.59 $\pm$ 0.366 & 0.08 \\
IRAS15043+5754 & 1100213 & 0.42 $\pm$ 0.129 & 0.51 $\pm$ 0.034 & 0.08 $\pm$ 0.023 & 0.15 $\pm$ 0.045 & 2.32 $\pm$ 0.702 & 0.06 \\
IRAS15250+3609 & 1122003 & 0.48 $\pm$ 0.227 & 0.92 $\pm$ 0.089 & 0.20 $\pm$ 0.050 & 0.22 $\pm$ 0.058 & 3.28 $\pm$ 0.895 & 0.11 \\
IRAS15335-0513 & 1120139 & 0.35 $\pm$ 0.031 & 4.20 $\pm$ 0.127 & 0.95 $\pm$ 0.154 & 0.23 $\pm$ 0.037 & 3.43 $\pm$ 0.581 & 0.02 \\
IRAS16474+3430 & 1120171 & 1.48 $\pm$ 0.210 & 3.07 $\pm$ 0.174 & 0.47 $\pm$ 0.069 & 0.15 $\pm$ 0.024 & 2.32 $\pm$ 0.374 & 0.07 \\
IRAS17028+5817 & 1100248 & 1.57 $\pm$ 0.095 & 1.23 $\pm$ 0.080 & 0.16 $\pm$ 0.041 & 0.13 $\pm$ 0.034 & 2.01 $\pm$ 0.535 & 0.05 \\
IRAS17132+5313 & 1120143 & 0.67 $\pm$ 0.060 & 6.89 $\pm$ 0.202 & 1.53 $\pm$ 0.375 & 0.22 $\pm$ 0.055 & 3.36 $\pm$ 0.848 & 0.02 \\
IRAS23128-5919 & 1122118 & 0.58 $\pm$ 0.040 & 5.97 $\pm$ 0.158 & 1.22 $\pm$ 0.327 & 0.20 $\pm$ 0.055 & 3.09 $\pm$ 0.852 & 0.01 \\
IRAS23436+5257 & 1122216 & 0.36 $\pm$ 0.036 & 3.04 $\pm$ 0.219 & 0.62 $\pm$ 0.093 & 0.20 $\pm$ 0.034 & 3.08 $\pm$ 0.526 & 0.03 \\
IRASF06076-2139 & 5200268 & 0.36 $\pm$ 0.030 & 1.84 $\pm$ 0.111 & 0.18 $\pm$ 0.037 & 0.10 $\pm$ 0.021 & 1.54 $\pm$ 0.327 & 0.02 \\
IRASF10565+2448 & 3051019 & 1.04 $\pm$ 0.047 & 7.02 $\pm$ 0.146 & 0.95 $\pm$ 0.140 & 0.13 $\pm$ 0.020 & 2.06 $\pm$ 0.314 & 0.02 \\
IRASF16330-6820 & 3370010 & 0.69 $\pm$ 0.039 & 11.25 $\pm$ 0.247 & 2.60 $\pm$ 0.473 & 0.23 $\pm$ 0.042 & 3.48 $\pm$ 0.657 & 0.02 \\
IRASF17207-0014 & 3370001 & 0.64 $\pm$ 0.080 & 7.36 $\pm$ 0.232 & 1.15 $\pm$ 0.218 & 0.16 $\pm$ 0.030 & 2.38 $\pm$ 0.468 & 0.03 \\
IRASF18293-3413 & 3370016 & 0.71 $\pm$ 0.050 & 48.86 $\pm$ 0.709 & 10.85 $\pm$ 1.967 & 0.22 $\pm$ 0.040 & 3.35 $\pm$ 0.627 & 0.01 \\
LEDA19076 & 3370015 & 0.56 $\pm$ 0.072 & 11.56 $\pm$ 0.160 & 3.58 $\pm$ 0.491 & 0.31 $\pm$ 0.043 & 4.62 $\pm$ 0.663 & 0.02 \\
LEDA59176 & 1920241 & 0.37 $\pm$ 0.068 & 3.49 $\pm$ 0.123 & 0.60 $\pm$ 0.184 & 0.17 $\pm$ 0.053 & 2.63 $\pm$ 0.822 & 0.03 \\
MCG+04-48-002 & 1122037 & 0.48 $\pm$ 0.027 & 15.54 $\pm$ 0.297 & 3.74 $\pm$ 0.542 & 0.24 $\pm$ 0.035 & 3.62 $\pm$ 0.547 & 0.02 \\
MCG+07-23-019 & 1120122 & 0.37 $\pm$ 0.036 & 4.53 $\pm$ 0.118 & 0.99 $\pm$ 0.158 & 0.22 $\pm$ 0.035 & 3.31 $\pm$ 0.549 & 0.03 \\
MCG+08-18-013 & 1120118 & 0.37 $\pm$ 0.028 & 4.75 $\pm$ 0.105 & 1.11 $\pm$ 0.324 & 0.23 $\pm$ 0.069 & 3.53 $\pm$ 1.059 & 0.03 \\
MCG+12-02-001 & 5200048 & 0.40 $\pm$ 0.031 & 22.71 $\pm$ 0.416 & 8.18 $\pm$ 0.552 & 0.36 $\pm$ 0.025 & 5.33 $\pm$ 0.392 & 0.01 \\
MCG-01-60-022 & 1122252 & 0.34 $\pm$ 0.030 & 7.31 $\pm$ 0.231 & 1.29 $\pm$ 0.335 & 0.18 $\pm$ 0.046 & 2.69 $\pm$ 0.717 & 0.02 \\
MCG-02-01-051 & 1120101 & 0.37 $\pm$ 0.082 & 10.53 $\pm$ 0.238 & 3.02 $\pm$ 0.507 & 0.29 $\pm$ 0.049 & 4.29 $\pm$ 0.753 & 0.03 \\
MCG-03-04-014 & 1120104 & 0.31 $\pm$ 0.031 & 12.28 $\pm$ 0.172 & 3.55 $\pm$ 0.280 & 0.29 $\pm$ 0.023 & 4.32 $\pm$ 0.361 & 0.02 \\
MCG-07-03-014 & 1122223 & 0.31 $\pm$ 0.026 & 11.25 $\pm$ 0.168 & 2.64 $\pm$ 0.367 & 0.23 $\pm$ 0.033 & 3.53 $\pm$ 0.510 & 0.02 \\
Mrk18 & 1122043 & 0.00 $\pm$ 0.000 & 4.12 $\pm$ 0.110 & 1.94 $\pm$ 0.205 & 0.47 $\pm$ 0.051 & 6.87 $\pm$ 0.797 & 0.02 \\
Mrk331 & 1120150 & 0.33 $\pm$ 0.017 & 16.42 $\pm$ 0.218 & 4.50 $\pm$ 0.752 & 0.27 $\pm$ 0.046 & 4.10 $\pm$ 0.713 & 0.01 \\
NGC1614 & 1120115 & 0.30 $\pm$ 0.021 & 34.03 $\pm$ 0.477 & 9.00 $\pm$ 0.356 & 0.26 $\pm$ 0.011 & 3.97 $\pm$ 0.173 & 0.02 \\
NGC232 & 1120102 & 0.29 $\pm$ 0.015 & 8.75 $\pm$ 0.139 & 1.60 $\pm$ 0.300 & 0.18 $\pm$ 0.034 & 2.78 $\pm$ 0.535 & 0.01 \\
NGC2388 & 1120231 & 0.26 $\pm$ 0.013 & 15.67 $\pm$ 0.236 & 3.86 $\pm$ 0.731 & 0.25 $\pm$ 0.047 & 3.71 $\pm$ 0.726 & 0.01 \\
NGC2623 & 1120116 & 0.89 $\pm$ 0.046 & 5.95 $\pm$ 0.201 & 0.20 $\pm$ 0.046 & 0.03 $\pm$ 0.008 & 0.51 $\pm$ 0.122 & 0.02 \\
NGC3110 & 1120120 & 0.22 $\pm$ 0.023 & 15.96 $\pm$ 0.524 & 5.12 $\pm$ 0.450 & 0.32 $\pm$ 0.030 & 4.77 $\pm$ 0.468 & 0.01 \\
NGC3256 & 3370036 & 0.66 $\pm$ 0.044 & 132.13 $\pm$ 3.049 & 21.62 $\pm$ 3.192 & 0.16 $\pm$ 0.024 & 2.49 $\pm$ 0.381 & 0.01 \\
NGC34 & 1120100 & 0.53 $\pm$ 0.031 & 12.81 $\pm$ 0.199 & 1.72 $\pm$ 0.321 & 0.13 $\pm$ 0.025 & 2.06 $\pm$ 0.392 & 0.01 \\
NGC4102 & 1120232 & 0.21 $\pm$ 0.014 & 29.25 $\pm$ 0.996 & 10.74 $\pm$ 2.278 & 0.37 $\pm$ 0.079 & 5.43 $\pm$ 1.217 & 0.01 \\
NGC5010 & 1122217 & 0.26 $\pm$ 0.013 & 9.16 $\pm$ 0.239 & 2.91 $\pm$ 0.897 & 0.32 $\pm$ 0.098 & 4.72 $\pm$ 1.514 & 0.01 \\
NGC5104 & 1120131 & 0.13 $\pm$ 0.011 & 6.44 $\pm$ 0.121 & 2.33 $\pm$ 0.132 & 0.36 $\pm$ 0.022 & 5.36 $\pm$ 0.335 & 0.01 \\
NGC5135 & 1120132 & 0.25 $\pm$ 0.018 & 13.32 $\pm$ 0.299 & 3.40 $\pm$ 0.597 & 0.25 $\pm$ 0.045 & 3.83 $\pm$ 0.701 & 0.01 \\
NGC520 & 1120229 & 0.39 $\pm$ 0.016 & 21.17 $\pm$ 0.407 & 3.87 $\pm$ 0.915 & 0.18 $\pm$ 0.043 & 2.77 $\pm$ 0.673 & 0.01 \\
NGC5256 & 1120133 & 0.28 $\pm$ 0.017 & 7.05 $\pm$ 0.109 & 1.25 $\pm$ 0.289 & 0.18 $\pm$ 0.041 & 2.69 $\pm$ 0.639 & 0.02 \\
NGC5728 & 1120086 & 0.06 $\pm$ 0.038 & 8.29 $\pm$ 0.389 & 1.67 $\pm$ 0.325 & 0.20 $\pm$ 0.040 & 3.05 $\pm$ 0.626 & 0.01 \\
NGC6090 & 1120140 & 0.36 $\pm$ 0.036 & 10.23 $\pm$ 0.152 & 2.46 $\pm$ 0.450 & 0.24 $\pm$ 0.044 & 3.62 $\pm$ 0.685 & 0.02 \\
NGC6240 & 3370014 & 0.50 $\pm$ 0.062 & 21.68 $\pm$ 0.565 & 3.01 $\pm$ 0.566 & 0.14 $\pm$ 0.026 & 2.12 $\pm$ 0.410 & 0.02 \\
NGC6285 & 1122138 & 0.43 $\pm$ 0.091 & 3.09 $\pm$ 0.098 & 0.87 $\pm$ 0.212 & 0.28 $\pm$ 0.069 & 4.24 $\pm$ 1.071 & 0.03 \\
NGC6621 & 1122248 & 0.31 $\pm$ 0.030 & 5.13 $\pm$ 0.079 & 1.08 $\pm$ 0.183 & 0.21 $\pm$ 0.036 & 3.19 $\pm$ 0.556 & 0.02 \\
NGC6670b & 3370025 & 0.27 $\pm$ 0.085 & 5.97 $\pm$ 0.179 & 2.40 $\pm$ 0.406 & 0.40 $\pm$ 0.069 & 5.91 $\pm$ 1.067 & 0.03 \\
NGC6701 & 1122292 & 0.15 $\pm$ 0.016 & 11.22 $\pm$ 0.417 & 2.74 $\pm$ 0.863 & 0.24 $\pm$ 0.077 & 3.67 $\pm$ 1.196 & 0.02 \\
NGC6786 & 5200045 & 0.16 $\pm$ 0.014 & 6.28 $\pm$ 0.094 & 1.31 $\pm$ 0.094 & 0.21 $\pm$ 0.015 & 3.15 $\pm$ 0.239 & 0.01 \\
NGC695 & 1120108 & 0.38 $\pm$ 0.038 & 18.81 $\pm$ 0.506 & 3.52 $\pm$ 0.258 & 0.19 $\pm$ 0.015 & 2.84 $\pm$ 0.228 & 0.02 \\
NGC7469 & 1120055 & 0.05 $\pm$ 0.014 & 19.35 $\pm$ 0.318 & 4.24 $\pm$ 0.550 & 0.22 $\pm$ 0.029 & 3.31 $\pm$ 0.446 & 0.01 \\
NGC7552 & 1122297 & 0.49 $\pm$ 0.036 & 57.08 $\pm$ 1.037 & 10.30 $\pm$ 1.184 & 0.18 $\pm$ 0.021 & 2.74 $\pm$ 0.327 & 0.02 \\
NGC7592W & 1120147 & 0.20 $\pm$ 0.028 & 6.95 $\pm$ 0.118 & 2.16 $\pm$ 0.188 & 0.31 $\pm$ 0.028 & 4.63 $\pm$ 0.429 & 0.02 \\
NGC7679 & 1122290 & 0.30 $\pm$ 0.028 & 14.31 $\pm$ 0.261 & 3.13 $\pm$ 0.562 & 0.22 $\pm$ 0.039 & 3.31 $\pm$ 0.613 & 0.02 \\
NGC7771 & 1120149 & 0.38 $\pm$ 0.023 & 6.19 $\pm$ 0.171 & 1.20 $\pm$ 0.308 & 0.19 $\pm$ 0.050 & 2.93 $\pm$ 0.774 & 0.02 \\
NGC838 & 1122173 & 0.33 $\pm$ 0.028 & 20.82 $\pm$ 0.192 & 5.49 $\pm$ 0.568 & 0.26 $\pm$ 0.027 & 3.95 $\pm$ 0.426 & 0.02 \\
NGC992 & 3371007 & 0.17 $\pm$ 0.029 & 19.35 $\pm$ 0.478 & 6.75 $\pm$ 0.513 & 0.35 $\pm$ 0.028 & 5.17 $\pm$ 0.434 & 0.01 \\
UGC12150 & 1122239 & 0.32 $\pm$ 0.015 & 6.01 $\pm$ 0.140 & 1.07 $\pm$ 0.188 & 0.18 $\pm$ 0.032 & 2.70 $\pm$ 0.492 & 0.01 \\
UGC1385 & 3330029 & 0.12 $\pm$ 0.019 & 5.16 $\pm$ 0.090 & 1.70 $\pm$ 0.264 & 0.33 $\pm$ 0.052 & 4.90 $\pm$ 0.799 & 0.01 \\
UGC1845 & 3371006 & 0.41 $\pm$ 0.023 & 12.58 $\pm$ 0.346 & 2.15 $\pm$ 0.448 & 0.17 $\pm$ 0.036 & 2.61 $\pm$ 0.558 & 0.01 \\
UGC2238 & 1120110 & 0.46 $\pm$ 0.040 & 21.00 $\pm$ 0.413 & 4.50 $\pm$ 0.452 & 0.21 $\pm$ 0.022 & 3.24 $\pm$ 0.342 & 0.02 \\
UGC2982 & 1120113 & 0.26 $\pm$ 0.019 & 19.36 $\pm$ 0.292 & 5.66 $\pm$ 0.381 & 0.29 $\pm$ 0.020 & 4.37 $\pm$ 0.314 & 0.01 \\
UGC8335W & 1122132 & 0.74 $\pm$ 0.131 & 1.79 $\pm$ 0.051 & 0.37 $\pm$ 0.096 & 0.20 $\pm$ 0.054 & 3.09 $\pm$ 0.833 & 0.06 \\
UGC9618N & 1120136 & 0.22 $\pm$ 0.025 & 8.78 $\pm$ 0.253 & 1.04 $\pm$ 0.249 & 0.12 $\pm$ 0.029 & 1.81 $\pm$ 0.444 & 0.01 \\
VIIZw031 & 5200264 & 0.72 $\pm$ 0.056 & 10.68 $\pm$ 0.190 & 2.52 $\pm$ 0.448 & 0.24 $\pm$ 0.042 & 3.55 $\pm$ 0.654 & 0.02 \\
VV114b & 1120103 & 1.01 $\pm$ 0.021 & 18.02 $\pm$ 0.616 & 1.76 $\pm$ 0.444 & 0.10 $\pm$ 0.025 & 1.50 $\pm$ 0.387 & 0.01 \\
VV250a & 1120129 & 0.93 $\pm$ 0.089 & 9.91 $\pm$ 0.245 & 2.03 $\pm$ 0.292 & 0.21 $\pm$ 0.030 & 3.11 $\pm$ 0.465 & 0.03 \\
VV283 & 3370024 & 0.99 $\pm$ 0.121 & 5.71 $\pm$ 0.217 & 0.81 $\pm$ 0.102 & 0.14 $\pm$ 0.019 & 2.17 $\pm$ 0.290 & 0.03 \\
VV705 & 1120138 & 0.80 $\pm$ 0.062 & 8.63 $\pm$ 0.183 & 1.45 $\pm$ 0.157 & 0.17 $\pm$ 0.019 & 2.56 $\pm$ 0.289 & 0.02 \\
\end{longtable}

\begin{minipage}{\linewidth}\footnotesize
Columns: (1) Object name; (2) AKARI Observation ID; (3) Optical depth
of H$_2$O ice absorption; (4) Intensity of the 3.3$\mum$ aromatic PAH
feature; (5) Total intensity of the aliphatic PAH features;
(6) Ratio of aliphatic to aromatic PAH feature intensities; (7) Aliphatic fraction;
(8) The fitting error $\sigma$, defined as the ratio of the integrated absolute fitting
residual to the total integrated flux of the input spectrum.\\
(This table is available in its entirety in machine-readable form.)
\end{minipage}
}
%%% Table 4 %%%

%%% Table 5 %%%
\begin{landscape}
\begin{table}[p]
\centering
\scriptsize
\caption{Multiwavelength Photometry of Our Samples ($\lambda<1\mum$)}\label{tabE:photometry1}
\setlength{\tabcolsep}{3pt}
\begin{tabular}{cccccccccccc}
\hline
Name & $E(B-V)$ & FUV & NUV & $u$ & $v$ & $g$ & $r$ & $i$ & $z$ & $y$ & flag\\
 &  & 152.8 nm & 354.3 nm &  & 477.0 nm & 623.1 nm & 762.5 nm & 152.8 nm & 913.4 nm &  & SD\\
 &  &  &  &  &  & 481.1 nm & 615.6 nm & 750.4 nm & 866.9 nm & 961.3 nm & PS1\\
 &  &  &  & 350.0 nm & 387.9 nm & 501.6 nm & 607.7 nm & 773.3 nm & 912.0 nm &  & SM\\
\hline
AM0702-601 & 0.11276 & $\cdots$ & $\cdots$ & 0.48 $\pm$ 0.14 & 0.80 $\pm$ 0.11 & 3.75 $\pm$ 0.13 & 5.78 $\pm$ 0.05 & 7.88 $\pm$ 0.24 & 10.26 $\pm$ 0.29 & $\cdots$ & SM\\
Arp193 & 0.01276 & 0.280 $\pm$ 0.017 & 0.637 $\pm$ 0.017 & 1.89 $\pm$ 0.03 & $\cdots$ & 6.19 $\pm$ 0.01 & 10.73 $\pm$ 0.03 & 14.08 $\pm$ 0.03 & 17.86 $\pm$ 0.06 & $\cdots$ & SD\\
Arp220 & 0.05153 & 0.138 $\pm$ 0.013 & 0.563 $\pm$ 0.018 & 2.50 $\pm$ 0.02 & $\cdots$ & 11.04 $\pm$ 0.03 & 23.68 $\pm$ 0.04 & 35.74 $\pm$ 0.06 & 46.76 $\pm$ 0.11 & $\cdots$ & SD\\
CGCG011-076 & 0.05476 & 0.023 $\pm$ 0.007 & 0.148 $\pm$ 0.013 & 0.86 $\pm$ 0.01 & $\cdots$ & 4.84 $\pm$ 0.01 & 12.44 $\pm$ 0.02 & 19.44 $\pm$ 0.03 & 28.09 $\pm$ 0.06 & $\cdots$ & SD\\
CGCG049-057 & 0.03952 & 0.004 $\pm$ 0.001 & 0.050 $\pm$ 0.002 & 0.34 $\pm$ 0.01 & $\cdots$ & 2.92 $\pm$ 0.01 & 6.86 $\pm$ 0.02 & 10.41 $\pm$ 0.02 & 14.65 $\pm$ 0.05 & $\cdots$ & SD\\
CGCG052-037 & 0.06113 & 0.123 $\pm$ 0.011 & 0.409 $\pm$ 0.015 & $\cdots$ & $\cdots$ & 8.53 $\pm$ 0.24 & 12.28 $\pm$ 1.08 & 20.28 $\pm$ 0.11 & 22.79 $\pm$ 0.35 & 28.28 $\pm$ 0.61 & PS1\\
CGCG436-030 & 0.03628 & 0.467 $\pm$ 0.004 & 0.764 $\pm$ 0.003 & 1.92 $\pm$ 0.09 & $\cdots$ & 4.92 $\pm$ 0.23 & 7.20 $\pm$ 0.34 & 9.39 $\pm$ 0.44 & 10.73 $\pm$ 0.51 & $\cdots$ & SD\\
CGCG453-062 & 0.10251 & 0.099 $\pm$ 0.012 & 0.388 $\pm$ 0.016 & 0.94 $\pm$ 0.02 & $\cdots$ & 3.90 $\pm$ 0.01 & 9.03 $\pm$ 0.02 & 13.89 $\pm$ 0.03 & 19.49 $\pm$ 0.06 & $\cdots$ & SD\\
ESO099-G004 & 0.57965 & $\cdots$ & $\cdots$ & $\cdots$ & $\cdots$ & $\cdots$ & 21.55 $\pm$ 6.34 & 19.35 $\pm$ 1.67 & 23.71 $\pm$ 1.30 & $\cdots$ & SM\\
\hline
\end{tabular}
\begin{minipage}{0.95\linewidth}\footnotesize
The FUV and NUV photometry is obtained from GALEX.
The sources of optical band data are indicated in the flag column:
(SD) Sloan Digital Sky Survey DR18, (SM) SkyMapper Southern Survey DR4, and (PS1) Pan-STARRS1 Survey DR2.\\
(This table is available in its entirety in machine-readable form.)
\end{minipage}
\end{table}
\end{landscape}
%%% Table 5 %%%

%%% Table 6 %%%
\begin{landscape}
\begin{table}[p]
\centering
\footnotesize
\caption{Multiwavelength Photometry of our Samples ($\lambda$\,=\,1--500$\mum$)}\label{tabE:photometry2}
\setlength{\tabcolsep}{3pt}
\begin{tabular}{cccccccccccccc}
\hline
Name & $J$ & $H$ & $K_s$ & W1 & W2 & W3 & W4 & Blue & Green & Red & PMW & PMW & PLW\\
 & 1.235$\mum$ & 1.662$\mum$ & 2.159$\mum$ & 3.353$\mum$ & 4.603$\mum$ & 11.56$\mum$ & 22.09$\mum$ & 70$\mum$ & 100$\mum$ & 160$\mum$ & 250$\mum$ & 350$\mum$ & 500$\mum$\\
\hline
AM0702-601 & 13.53 & 22.30 & 33.62 & 57.05 & 89.43 & 234.20 & 709.69 & 2554 & 2700 & 2081 & 874 & 344 & 96\\
 & 0.46 & 0.71 & 0.97 & 1.17 & 1.58 & 3.26 & 9.21 & 114 & 120 & 89 & 54 & 24 & 12 \\
Arp193 & 20.90 & 24.07 & 27.16 & 18.90 & 20.22 & 214.92 & 842.72 & 21740 & 25820 & 18080 & 6356 & 2345 & 646 \\
 & 0.37 & 0.55 & 0.81 & 0.40 & 0.38 & 2.99 & 11.72 & 1000 & 1180 & 810 & 393 & 146 & 41 \\
Arp220 & 54.12 & 71.55 & 70.38 & 28.36 & 31.63 & 383.08 & 4119.76 & 139200 & $\cdots$ & 86320 & 30870 & 12800 & 3865 \\
 & 0.96 & 1.26 & 1.51 & 0.28 & 0.56 & 5.33 & 30.47 & 6600 & $\cdots$ & 3990 & 1920 & 800 & 24 \\
CGCG011-076 & 34.98 & 46.09 & 43.00 & 25.72 & 25.85 & 192.61 & 545.11 & 6742 & 9474 & 8312 & 3795 & 1653 & 562 \\
 & 0.55 & 0.94 & 1.24 & 0.55 & 0.48 & 2.86 & 11.16 & 295 & 406 & 350 & 252 & 112 & 42 \\
CGCG049-057 & 19.91 & 26.59 & 22.51 & 11.79 & 8.75 & 62.15 & 413.51 & 27650 & 31670 & 21830 & 8430 & 3225 & 1059 \\
 & 0.46 & 0.75 & 1.06 & 0.24 & 0.16 & 0.86 & 5.75 & 1280 & 1450 & 990 & 528 & 204 & 68 \\
CGCG052-037 & 34.88 & 46.34 & 41.94 & 22.97 & 18.95 & 200.20 & 607.69 & 9151 & 12460 & 10040 & 3618 & 1423 & 492 \\
 & 0.61 & 0.94 & 1.25 & 0.47 & 0.37 & 2.60 & 11.87 & 406 & 550.0 & 440 & 240 & 96 & 36 \\
CGCG247-020 & 14.78 & 19.16 & 19.43 & 12.49 & 10.71 & 108.17 & 593.30 & 6514 & 7450 & 4904 & 1789 & 577 & 175 \\
 & 0.39 & 0.63 & 0.82 & 0.27 & 0.20 & 1.50 & 9.92 & 326 & 373 & 246 & 120 & 48 & 15 \\
CGCG436-030 & 13.80 & 17.58 & 19.04 & 14.09 & 22.54 & 198.73 & 958.70 & 11710 & 11330 & 6907 & 2125 & 770 & 230 \\
 & 0.40 & 0.54 & 1.10 & 0.29 & 0.42 & 2.76 & 17.82 & 530 & 510 & 306 & 128 & 48 & 15 \\
CGCG453-062 & 31.02 & 39.05 & 37.68 & 15.42 & 12.45 & 108.91 & 321.57 & 9269 & 11420 & 9233 & 4418 & 1749 & 609 \\
 & 0.58 & 0.87 & 1.27 & 0.32 & 0.22 & 1.41 & 5.98 & 422 & 520 & 409 & 272 & 111 & 38 \\
ESO099-G004 & 31.79 & 44.52 & 43.10 & 24.69 & 25.26 & 241.14 & 953.41 & 12540 & 14610 & 10870 & 3970 & 1674 & 505 \\
 & 1.68 & 2.01 & 2.11 & 0.53 & 0.47 & 23.13 & 17.73 & 1570 & 660 & 470 & 242 & 109 & 39 \\
\hline
\end{tabular}

\begin{minipage}{0.95\linewidth}\footnotesize
The data are sourced from 2MASS, WISE, and Herschel.
Uncertainties are listed in the rows beneath the data.\\
(This table is available in its entirety in machine-readable form.)
\end{minipage}
\end{table}
\end{landscape}
%%% Table 6 %%%

\clearpage

%\bibliography{ms}{}
%\bibliographystyle{aasjournal}

\end{document}